\documentclass[]{bytedance_seed}

\usepackage[toc,page,header]
{appendix}

\usepackage{minitoc}
\usepackage[ruled]{algorithm2e}
\usepackage{amsmath,amssymb,textcomp,mathtools,amsthm,gensymb,graphicx,enumitem}
\usepackage{color,xcolor}        
\usepackage{transparent}   
\usepackage{booktabs}
\usepackage{multirow}
\usepackage{makecell}
\usepackage{array}

\usepackage{pifont}

\SetKwComment{Comment}{\# }{}

\SetCommentSty{commentfont}
\SetKwFor{For}{for }{:}{}

\title{Stable Autoregressive Speech Generation with Low-Frame-Rate High-Dimensional Continuous Tokens}

\author[]{Yi Luo}
\author[]{Rongzhi Gu}
\author[*]{Jixun Yao}

\affiliation[]{ByteDance Seed}

\contribution[*]{Work done during an internship at ByteDance Seed.}

\abstract{
Balancing sequence length, representational capacity, and long-horizon stability is a central problem in autoregressive (AR) speech and audio generation. Representations with higher frame rates or greater capacity can preserve more signal detail, but they also make streaming generation more vulnerable to distribution drift and AR error accumulation. Conversely, shorter and more compressed representations simplify AR modeling, but their limited bandwidth may discard important components and constrain the upper bound of reconstruction fidelity and generation quality. We ask whether a low-frame-rate, high-dimensional, high-bandwidth continuous representation can be co-designed with a streaming generation framework to support robust high-fidelity reconstruction, strong single-token predictability, and superior long-horizon stability. We decompose this goal into two coupled problems: what geometric and statistical properties a high-dimensional representation space should have, and how an AR continuous-token generator should be structured to resist error accumulation. Accordingly, we propose \emph{Locodec}, a locally encoded codec that shapes its representation space to improve the interpolatability of a lower-dimensional core manifold and the identifiability of the native high-dimensional coordinates, thereby improving the predictability of high-dimensional high-bandwidth tokens. We also propose \emph{MP-ELD}, a single-token AR flow-matching framework that uses multi-path information routing and residual classifier-free guidance to mitigate error accumulation. Experiments with \(8\)-Hz, \(768\)-dimensional tokens show that our design preserves reconstruction quality, improves single-token predictability, achieves competitive WER, and maintains stable long-form synthesis, without using external SSL/ASR models, pretrained text language models, or post-training stages.
}

\date{\today}
\correspondence{Yi Luo at \email{yl3364@columbia.edu}}

\begin{document}
\maketitle


\section{Introduction}

Audio is one of the most natural interfaces for human--machine interaction, and the ability to perform stable streaming understanding and generation has become increasingly important for modern audio and multimodal foundation models~\cite{wu2023next, hurst2024gpt, gemmateam2026gemma4technicalreport}. From the input side of such systems, a representation that preserves more information generally provides a higher ceiling for downstream understanding. If the representation has already discarded task-relevant cues through aggressive lossy compression, no subsequent model can recover them reliably. In this sense, rich or near-complete representations allow the model to learn which factors are useful for a given task, rather than forcing it to operate on a representation that may not contain those factors in the first place. From the output side, however, the same preference for high-information representations creates a different difficulty. A higher-bandwidth or higher-dimensional target space is usually harder to predict, and prediction errors in an autoregressive (AR) system are fed back as future context. Small local prediction errors may therefore accumulate over time, leading to drift in loudness, timbre, speaking rate, or spectral quality, and in severe cases to unstable or collapsed audio outputs. This creates a basic tension: representations that are desirable for general understanding and high-fidelity reconstruction can be difficult to use as stable generation targets. Balancing representation capacity, AR stability, and computational efficiency is therefore a central design problem in streaming audio systems.

As text language models and audio--text datasets continue to improve, transcription, captioning, and other understanding-oriented capabilities have made steady progress in large audio models~\cite{chu2023qwen, defossez2024moshi, ding2025kimi, zhang2025mimo, wu2025step, xu2025qwen3, ghosh2026audio, yang2026moss}. By contrast, stable and efficient high-fidelity streaming generation remains a bottleneck: it directly determines the perceptual quality of human-facing interaction, and it also affects subsequent machine-side understanding in closed-loop systems. The problem becomes more pronounced as interactive scenarios move from second-scale responses to minute- or hour-scale sessions, while practical context length and computation budgets remain finite. Ideally, one would like to use a representation that is informative enough to support high-fidelity reconstruction, short enough to reduce the AR horizon, and simple enough to model without expensive local predictors or post-processing modules. These requirements, however, seem to form an unfavorable triangle: high information content, low AR error accumulation, and low model complexity are difficult to satisfy simultaneously.

This paper explores whether this triangle can be relaxed by jointly designing the representation space and the AR generative model. Motivated by recent progress in raw signal-space prediction~\cite{wang2025pixnerd, li2026back}, representation-space analysis~\cite{yu2024representation, zheng2025diffusion}, and AR flow-matching systems~\cite{li2024autoregressive, jia2025ditar}, we study a low-frame-rate, high-dimensional continuous-token formulation together with its corresponding AR generation framework. Instead of relying on externally pretrained self-supervised learning (SSL) or automatic speech recognition (ASR) models to define a ``semantic'' space as a supervision signal for shaping the representation space, we train a reconstruction-first tokenizer whose latent geometry is shaped directly. The resulting tokenizer, \emph{Locodec}, produces spherical continuous tokens whose high-dimensional space is organized around a lower-dimensional \emph{interpolatable} core manifold, while its native coordinates are encouraged to develop an energy hierarchy that improves per-token \emph{identifiability}. This design aims to keep the bandwidth and reconstruction ceiling of a high-capacity representation, while making single-token prediction substantially easier for the generative model.

To address the remaining long-horizon instability of high-information AR generation, we further examine how error accumulation arises during guided streaming synthesis. Our empirical observations suggest that a major source of accumulated drift is instability under classifier-free guidance (CFG), especially when partially overlapping acoustic cues are carried by different guidance paths and are amplified inconsistently. Rather than suppressing guidance globally, we aim to reduce such conflicts by encouraging different types of information to be routed through functionally distinct pathways. We therefore propose \emph{MP-ELD}, a multi-path encoder--LM--decoder framework for AR flow matching. Through explicit information routing and training-time path dropout, MP-ELD encourages different aspects of the audio state to be represented by different conditioning pathways. This allows guidance to be applied as structured residual corrections, so that different factors, such as acoustic consistency and external-condition alignment, can be strengthened separately. In practice, even without pretrained text language models providing a strong semantic inductive bias, this design allows a generative model trained from scratch to form distinguishable acoustic-state and alignment-related pathways. This separation makes it possible to control their CFG scales independently, thereby mitigating accumulated acoustic drift and improving stability in long-form synthesis.

The rest of this paper is organized as follows. Section~\ref{sec:problem} discusses the main assumptions and design principles behind our model design, including the roles of bitrate, representation-space geometry, and AR stability. Section~\ref{sec:methodology} presents the methodology used to shape the tokenizer latent space and to construct the flow-matching bridge and training objective. Section~\ref{sec:architecture} describes the concrete model architectures of the tokenizer and the AR generative model. Section~\ref{sec:experiments} reports the experimental results on tokenizer reconstruction quality and generative model performance on the Seed-TTS-eval dataset. Section~\ref{sec:conclusion} concludes the paper.
\section{Main Assumptions and Design Principles}
\label{sec:problem}

In this section, we revisit several existing design choices for tokenizers and generative models, discuss their advantages and limitations, and then motivate the principles that guide our overall design.

\subsection{Frame rate, bitrate, and reconstruction fidelity}

Due to the nature of audio signals, audio processing has long been confronted with the challenge of modeling extremely long sequences~\cite{van2016wavenet, mehri2016samplernn, huang2018music, narayanan2019recognizing, dhariwal2020jukebox, luo2020dual, mao2020speech, battenberg2020location, shi2021emformer, raj2021integration, han2021continuous, evans2024long, ning2025diffrhythm, yuan2025yue}. For a fixed audio duration, the token sequence length is determined by the token frame rate, which directly controls the temporal span and granularity covered by each token. Together with the information capacity of each token, it also determines the amount of information that can be allocated per unit time, i.e., the bandwidth. While tokenizer and generative model designs have been extensively explored for moderate bitrate configurations~\cite{defossez2022high, kumar2023high, wang2023neural, chen2024vall, zhang2024speechtokenizer, defossez2024moshi, ji2025wavtokenizer, liao2026fish}, a major line of research in recent years has focused on achieving the highest possible reconstruction fidelity under aggressively reduced frame rate and bitrate settings~\cite{du2024cosyvoice, guo2024fireredtts, ju2024naturalspeech, jiang2025diffrhythm, parker2025scaling, wang2026scaling}. Lowering the frame rate is one of the most direct ways to reduce the complexity of the downstream sequence model, while lowering the bitrate can further simplify the generative modeling problem and reduce computational cost. For example, replacing residual vector quantization (RVQ)~\cite{zeghidour2021soundstream} with single-stage vector quantization (VQ)~\cite{van2017neural} removes the need to model a residual hierarchy, thereby reducing both the required computation and the effective parameter budget.

However, from the perspective of lossless compression, low bitrate and high reconstruction fidelity are fundamentally difficult, and often impossible, to achieve simultaneously. As a result, existing tokenizers typically preserve only a subset of the most important signal attributes, such as content, timbre, or pitch, while (intentionally or unintentionally) discarding other factors that are more difficult to retain faithfully. Under such a training objective, many low-bitrate tokenizers are often closer to a \emph{resynthesis} system than to a strict \emph{codec}. Consequently, in certain regimes or tasks, such as high-fidelity generation or local editing, the upper bound of performance may be directly constrained by the tokenizer bitrate and by the corresponding training paradigm.

Conversely, when one increases bitrate in order to improve the information capacity of discrete tokens, for instance via deeper RVQ hierarchies or larger finite scalar quantization (FSQ) bitrates~\cite{mentzer2024finite}, the resulting latent space gradually becomes closer to a continuous one. From the modeling perspective, when a discrete tokenization scheme such as RVQ uses many residual levels, and the generative model must unroll these levels hierarchically during prediction, e.g., via delay-pattern modeling~\cite{copet2023simple} or residual-quantization transformers (RQTransformer)~\cite{lee2022autoregressive}, the model complexity increases substantially. Recent work has therefore increasingly explored continuous representations as a means of enabling higher-bandwidth modeling~\cite{eskimez2024e2, liu2024autoregressive, peng2025vibevoice, meng2025autoregressive, chen2025f5, xin2026longcat}. On the one hand, under the same model architecture, frame rate, and token dimensionality, one can often significantly improve reconstruction fidelity simply by removing the quantization step altogether. On the other hand, from the perspective of iterative prediction, the residual hierarchy in high-bitrate discrete tokenization is, to some extent, analogous to the iterative denoising or function evaluation process in flow-based continuous generation. In this sense, once bitrate is increased sufficiently, continuous tokenization becomes a natural modeling direction.

At the same time, both high-bitrate discrete tokens and continuous tokens have substantially more target-space degrees of freedom than low-bitrate tokens, which in practice can make AR systems more vulnerable to accumulated prediction errors and exposure bias~\cite{schmidt2019generalization}. Moreover, when the tokenizer frame rate is high and the AR sequence is correspondingly long, this accumulation can be further amplified. Therefore, for high-bandwidth tokenization, reducing the frame rate, or more generally reducing the effective frame rate seen by the AR model, becomes a necessary means of controlling AR error accumulation.

\subsection{Single high-dimensional space versus product-structured space}

When discussing low-frame-rate, high-bandwidth continuous tokens, a natural design question arises: should one directly construct tokens that are natively low in frame rate but higher in dimensionality, so that the AR model operates directly at the token rate; or should one instead construct tokens that are higher in frame rate but lower in dimensionality, and then group nearby tokens so that the AR model effectively operates at the grouped frame rate?

Suppose that the overall tokenizer compression ratio is fixed across these two choices. Then, from the perspective of the AR model, the main difference is whether the model sees a \emph{single high-dimensional space} or a \emph{product-structured space} composed of multiple lower-dimensional spaces. In most existing approaches, the latter is more common: each local group of low-dimensional tokens is treated as a short local sequence, which is first compressed into a single embedding before entering the language model, and is then decoded back into the next local token sequence at prediction time~\cite{zhou2025voxcpm, jia2025ditar, jiang2025diffrhythm, zhou2026voxcpm2}. This strategy is related to what prior work on efficient speech enhancement and separation referred to as a \emph{context codec}, namely a local compression/decompression mechanism designed to shorten the effective input length of the sequence model~\cite{luo2021group}. By contrast, among systems that attempt to operate at a natively low frame rate, one rarely observes continuous token configurations that are simultaneously very high-dimensional (e.g., $256$ dimensions or above) and sufficiently high in bandwidth to support high reconstruction fidelity. One possible reason is the geometric and statistical difficulty of modeling such a high-dimensional space, together with the well-known ``curse of dimensionality''~\cite{altman2018curse}. As a consequence, a grouped low-dimensional product-structured representation appears, at least superficially, to be a more practical way of constructing high-capacity tokens.

Nevertheless, this observation immediately raises another question. If one must already introduce an additional encoder to compress a group of low-dimensional tokens into a single representation for the language model, and then introduce a corresponding decoder to reconstruct the next token group from the language-model output, why should this compression--decompression process be placed inside the generative model rather than inside the tokenizer itself? In other words, if the generative model must reshape a high-frame-rate, low-dimensional token sequence into a low-frame-rate, high-dimensional latent space, and then invert that reshaping during prediction, does this not suggest that one may instead train a tokenizer that directly produces such a native low-frame-rate, high-dimensional token space? Ideally, such a tokenizer would preserve sufficient bandwidth to maintain strong reconstruction fidelity, while eliminating the need for additional local encoding and decoding modules inside the generative model, especially prediction-time modules such as local DiTs that substantially increase computational cost and architectural complexity.

From the viewpoint of the generative model input, compressing a group of low-dimensional tokens can be implemented either inside the tokenizer or inside the generative model, and these two choices are not fundamentally different. However, from the viewpoint of the generative model output, the difference is substantial. A local token group, being itself a short sequence, can be processed by a local sequence model that has sufficient capacity to progressively lift and reshape the local representation, thereby making the corresponding product-structured representation considerably easier to model. In contrast, a single high-dimensional token is not itself a sequence and therefore cannot benefit from such sequential lifting; its representational expansion inside the model is much more limited, since excessively large hidden dimensions would directly increase model size, training difficulty, and computational cost. Therefore, the key challenge is not simply to use a high-dimensional space, but to construct such a space so that the generative model can perform stable and efficient \emph{single high-dimensional token prediction}, ideally reaching performance comparable to, or better than, sequential prediction over grouped low-dimensional tokens, at significantly lower computational cost.

\subsection{Disentangled representation space}

One of the most common strategies in prior work for reducing generative modeling difficulty, applicable to both discrete and continuous tokens, is to explicitly construct a coarse-to-fine representation space, most often in the form of semantic--acoustic disentanglement~\cite{borsos2023audiolm, mousavi2024should, du2024cosyvoice, liu2024semanticodec, bai2024seed, ye2025codec, wang2025spark, chen2026sac}. In such representations, the semantic component typically serves as a compact core representation that preserves the main attributes of the signal, and may come from the hidden space of a pretrained SSL model~\cite{baevski2020wav2vec, chen2022wavlm, hsu2021hubert, chiu2022self} or an ASR model~\cite{radford2023robust, chen2025minmo}. The acoustic component then complements this semantic representation by providing additional detail, often through a residual or hierarchical structure. By feeding the language model with a more clustered, more constrained, or otherwise lower-freedom semantic representation, AR error accumulation can be effectively reduced on the input side; and on the output side, the semantic embedding itself can simplify and stabilize the prediction. Similar trends have also appeared increasingly in image generation, where large-scale pretrained SSL embeddings have been used either as alignment supervision during tokenizer training or directly as tokens themselves, improving the generative performance of continuous, especially high-dimensional continuous, representations~\cite{leng2025repa, zheng2025diffusion, tong2026scaling}.

However, relying on such predefined or externally trained semantic embeddings is not a free lunch. On the one hand, semantic embeddings derived from models trained for specific domains or attributes, such as ASR models, inevitably inherit strong biases toward the corresponding target attributes, for example, word or phoneme structure, or even language-dependent biases. Such biases may reduce the capacity of the tokenizer to encode other types of information, such as non-semantic or non-vocal content, forcing those properties to be represented only in the acoustic component. If the generative model then relies heavily on the semantic component in order to reduce AR error accumulation, other types of information may be underrepresented in its input, limiting the model's understanding and generation capabilities. On the other hand, semantic embeddings derived from SSL models are heavily shaped by the SSL training data, model scale, and pretraining objective. Different generative tasks, e.g., speech synthesis versus music generation, may prefer different representation-space properties. If one wishes to build a single task-general representation space, the cost of training a sufficiently large and sufficiently universal SSL model may even exceed that of training the generative model itself.

Therefore, from the perspectives of simplicity and generality, we prefer not to rely on explicitly disentangled semantic--acoustic representations, nor on external large-scale SSL training to shape or constrain the high-dimensional token space. Instead, we aim to let the tokenizer optimize primarily for high-fidelity reconstruction, while making its decoder sufficiently robust to cover the types of prediction errors produced by the generative model, and to use lower-cost mechanisms to induce a high-dimensional space that is itself easier for the generative model to predict.

\subsection{Raw input space modeling and the manifold hypothesis}

Another line of work that has recently attracted increasing attention is to perform generative modeling directly in the high-dimensional raw signal space, without any separately trained tokenizer. Such approaches are often motivated by the \emph{manifold hypothesis}, namely the assumption that many real-world high-dimensional signals in fact lie near a lower-dimensional latent manifold~\cite{chapelle2006discussion, gorban2018blessing}. Under this view, directly modeling such signals may be easier than modeling an arbitrary high-dimensional vector that does not inherit such structure, such as a freely learned high-dimensional token. This line of work has shown strong potential in image generation~\cite{hoogeboom2025simpler, yu2026pixeldit, li2026back}, and has also begun to appear in audio generation~\cite{zhou2026wavflow, chen2026wavtts, fan2026barewave}.

However, unlike direct modeling in image pixel space, direct modeling in waveform space might not necessarily be a natural choice for audio. In waveform space, an isotropic Gaussian corruption process, when viewed under an orthonormal time--frequency transform, corresponds to broadly full-band white noise in the frequency domain. Real-world audio signals, by contrast, tend to exhibit substantially lower natural energy in the mid- and high-frequency bands. As a consequence, if the generative model fails to remove the injected noise completely, the residual error in these bands may become perceptually salient under human auditory perception~\cite{miller1947sensitivity, zwicker2013psychoacoustics}. This may differ from the image case where human sensitivity and tolerance to such noises might have a different mechanism~\cite{campbell1968application, barten1999contrast, chen2023importance}, and where the masking effects of additive noise might also be qualitatively different from those in audio~\cite{watson1997model}. Therefore, the same type of prediction error may have significantly different perceptual consequences in audio and image generation. In addition, although pixel-space modeling in image generation has been empirically shown to achieve performance comparable to, and in some cases better than, latent-space modeling, it may still exhibit artifacts or noise due to the lack of a noise-robust tokenizer decoder that can refine or compensate for prediction errors. Therefore, what we seek is a high-dimensional space that still benefits from the manifold hypothesis, but at the same time possesses a meaningful degree of reconstruction robustness, so that the generative model can learn it more easily due to the presence of a structured and interpolatable low-dimensional manifold, while also retaining enough robustness to prevent generative prediction errors from directly becoming perceptually severe distortions.

\subsection{Representational interpolatability and identifiability}

Another attribute that affects whether a representation space can be effectively modeled by a generative model is its \emph{interpolatability}~\cite{arvanitidis2017latent, wessels2025grounding}. An interpolatable representation space is often associated with a smoother manifold and better clustering properties, and prior work has observed that such spaces are more likely to lead to better generation performance~\cite{arvanitidis2020geometrically, daly2022variational, de2024pullback, sun2024geometry, yue2026matters, xu2026making}. However, directly constructing sufficiently strong interpolatability over an entire high-dimensional space is mathematically highly impractical. The capacity of a high-dimensional space grows so rapidly that neither finite training data nor local noise augmentation can densely cover it in any meaningful sense.

We illustrate this point using a simplified spherical-cap covering problem. We consider a high-dimensional sphere and spherical tokens lying on it. We consider the noise robustness of a tokenizer by assuming a ``reconstruction-stable basin'', where for each token there exists a spherical cap that allows the tokenizer decoder to reconstruct a near-identical raw signal from any token within the cap. We thus ask: how many non-overlapping basins can the sphere $\mathbb{S}^{N-1}$ contain? In an interpolatable space, one would expect reconstruction-stable basins to be sufficiently connected or overlapping at a moderate basin density, so that moving between nearby valid regions does not frequently pass through invalid regions on the sphere. On the other hand, if the number of non-overlapping basins is enormously large and far exceeds the number of training samples, it means that the space has enough capacity for training samples to occupy mutually isolated basins without forcing substantial basin overlap. As a result, the valid regions on the sphere only cover a minor portion of the total capacity, and global interpolatability in this case is nearly impossible.

We formulate this problem via mathematical arguments. Consider the unit sphere
\[
\mathbb{S}^{N-1}
=
\{x\in\mathbb{R}^{N}:\|x\|_2=1\}.
\]
For a reference point $e_1=(1,0,\dots,0)$, define the spherical cap of geodesic half-angle $\theta\in(0,\pi/2)$ as
\[
\mathcal{C}_{\theta}
=
\left\{
x\in\mathbb{S}^{N-1}:
\arccos(\langle x,e_1\rangle)\le \theta
\right\}.
\]
Let
\[
\mu_N(\theta)
\]
denote the normalized surface measure of this cap. Equivalently, if
$x$ is uniformly distributed on $\mathbb{S}^{N-1}$, then
\[
\mu_N(\theta)
=
\Pr\big(\langle x,e_1\rangle\ge \cos\theta\big).
\]
The exact expression is
\begin{equation}
\mu_N(\theta)
=
\frac{\int_0^\theta \sin^{N-2}\phi\,d\phi}
     {\int_0^\pi \sin^{N-2}\phi\,d\phi}.
\label{eq:cap_measure_exact_rewrite}
\end{equation}
Equivalently, the first coordinate of a uniformly sampled point on the sphere has density
\[
p_N(u)
=
\frac{\Gamma(N/2)}
{\sqrt{\pi}\Gamma((N-1)/2)}
(1-u^2)^{(N-3)/2},
\qquad u\in[-1,1],
\]
and therefore
\begin{equation}
\mu_N(\theta)
=
\int_{\cos\theta}^{1}
\frac{\Gamma(N/2)}
{\sqrt{\pi}\Gamma((N-1)/2)}
(1-u^2)^{(N-3)/2}\,du .
\label{eq:cap_measure_density_rewrite}
\end{equation}

Applying a standard endpoint Laplace approximation, for any fixed $\theta\in(0,\pi/2)$ we obtain
\begin{equation}
\mu_N(\theta)
\sim
\frac{(\sin\theta)^{N-1}}
{\cos\theta\sqrt{2\pi N}},
\qquad
N\to\infty.
\label{eq:cap_measure_endpoint_asymptotic}
\end{equation}
Thus the reciprocal cap measure satisfies
\begin{equation}
K_{\mathrm{area}}(N,\theta)
:=
\frac{1}{\mu_N(\theta)}
\sim
\cos\theta\sqrt{2\pi N}\,
(\sin\theta)^{-(N-1)}.
\label{eq:k_area_asymptotic_rewrite}
\end{equation}
The essential point is that for every fixed $\theta<90^\circ$, this quantity grows exponentially in $N$. Let $\mathcal{N}_{\mathrm{cov}}(N,\theta)$ denote the minimum number of spherical caps of radius $\theta$ required to cover $\mathbb{S}^{N-1}$. Since each cap occupies surface fraction $\mu_N(\theta)$, one necessarily has
\begin{equation}
\mathcal{N}_{\mathrm{cov}}(N,\theta)
\ge
\frac{1}{\mu_N(\theta)}
=
K_{\mathrm{area}}(N,\theta).
\label{eq:covering_lower_bound_rewrite}
\end{equation}
Therefore, if one wants every point on the high-dimensional sphere to lie within angle $\theta$ of some reconstruction-stable region, the required number of such regions is already at least exponential in $N$. For non-overlapping basins, let $\mathcal{N}_{\mathrm{pack}}(N,\theta)$ denote the maximum number of disjoint spherical caps of radius $\theta$. The simple volume argument gives
\[
\mathcal{N}_{\mathrm{pack}}(N,\theta)
\le
\frac{1}{\mu_N(\theta)}.
\]
Conversely, a standard maximal-packing argument gives an area-scale lower bound. For \(\theta<\pi/4\), choose a maximal collection of disjoint \(\theta\)-caps. By maximality, the caps with doubled radius \(2\theta\) must cover the sphere. If the number of selected caps is \(M\), then
\[
M\,\mu_N(2\theta)\ge 1.
\]
Since \(\mathcal{N}_{\mathrm{pack}}(N,\theta)\) is the maximum number of disjoint \(\theta\)-caps, we obtain
\begin{equation}
\mathcal{N}_{\mathrm{pack}}(N,\theta)
\ge
\frac{1}{\mu_N(2\theta)}.
\label{eq:packing_lower_bound_rewrite}
\end{equation}
Thus, even conservative packing lower bounds can be exponential in dimension. For example, the scale \(K_{\mathrm{area}}(N,60^\circ)\) gives a lower-bound scale for the number of disjoint \(30^\circ\) basins obtainable by such a maximal-packing argument.

\begin{table}[!t]
\centering
\caption{Area scale $K_{\mathrm{area}}(N,\theta)=1/\mu_N(\theta)$ for spherical caps on $\mathbb{S}^{N-1}$. This quantity is a lower bound on the number of 
$\theta$-caps required to cover the sphere, and also indicates the exponential area scale associated with angular regions. Values are approximate.}
\label{tab:cap_covering}
\begin{tabular}{c|ccc}
\toprule
$N$ & $\theta=30^\circ$ & $\theta=45^\circ$ & $\theta=60^\circ$ \\
\midrule
$4$   & $\approx 3.5\times 10^{1}$   & $\approx 1.1\times 10^{1}$  & $\approx 5.1$ \\
$8$   & $\approx 7.9\times 10^{2}$   & $\approx 6.0\times 10^{1}$  & $\approx 1.2\times 10^{1}$ \\
$16$  & $\approx 2.5\times 10^{5}$   & $\approx 1.3\times 10^{3}$  & $\approx 4.9\times 10^{1}$ \\
$32$  & $\approx 2.6\times 10^{10}$  & $\approx 4.6\times 10^{5}$  & $\approx 6.1\times 10^{2}$ \\
$64$  & $\approx 1.6\times 10^{20}$  & $\approx 4.3\times 10^{10}$ & $\approx 8.6\times 10^{4}$ \\
$128$ & $\approx 4.2\times 10^{39}$  & $\approx 2.6\times 10^{20}$ & $\approx 1.2\times 10^{9}$ \\
$256$ & $\approx 2.0\times 10^{78}$  & $\approx 6.8\times 10^{39}$ & $\approx 1.7\times 10^{17}$ \\
$512$ & $\approx 3.3\times 10^{155}$ & $\approx 3.3\times 10^{78}$ & $\approx 2.4\times 10^{33}$ \\
\bottomrule
\end{tabular}
\end{table}

To make the scale concrete, Table~\ref{tab:cap_covering} reports $K_{\mathrm{area}}(N,\theta) = 1/\mu_N(\theta)$ for representative dimensions and angles. The values should be interpreted as area/coverage scales rather than exact packing numbers. They are computed from the exact cap-measure expression when numerically convenient and from the asymptotic expression in Eq.~\eqref{eq:k_area_asymptotic_rewrite} in the large-$N$ regime. Several observations follow. First, even for a relatively large angular tolerance such as $60^\circ$, the area scale reaches roughly $10^5$ at $N=64$ and $10^{17}$ at $N=256$. For smaller angular tolerances, the numbers become enormous much earlier. Second, these estimates are already sufficient to show that direct dense interpolatability in a high-dimensional spherical token space is not a practical goal. Local noise injection around training samples can improve decoder robustness near the data manifold, but it cannot make the entire high-dimensional sphere densely connected in any global sense. Therefore, if one wishes to induce interpolatability, it is far more plausible to construct it on a lower-dimensional manifold embedded within the high-dimensional space, and then use the geometry of this lower-dimensional manifold to shape the high-dimensional token space.

At the same time, the dimension of the lower-dimensional manifold should not be too small, as interpolatability is only a valid objective when the low-dimensional space itself contains sufficient information for coarse reconstruction. A representation space can be highly interpolatable simply because it has collapsed most information: if the decoder cannot reconstruct the signal meaningfully, then moving smoothly in that space is of little value. More formally, let $X$ denote the data signal and let an encoder--decoder pair with a latent representation of token dimension $d$, denoted by $\mathcal{M}_d$, induce reconstructions
\[
\widehat X = D(E(X)),
\qquad E(X)\in \mathcal{M}_d .
\]
For a reconstruction loss $\ell$, define the best achievable distortion at dimension $d$ as
\begin{equation}
\mathcal{D}^{\star}(d)
=
\inf_{E,D}
\mathbb{E}\big[\ell(X,D(E(X)))\big],
\label{eq:best_distortion_dim_d}
\end{equation}
under the architectural, smoothness, regularization, and bandwidth constraints of interest. A meaningful low-dimensional manifold must satisfy
\[
\mathcal{D}^{\star}(d)
\le
\varepsilon_{\mathrm{rec}}
\]
for a reconstruction tolerance $\varepsilon_{\mathrm{rec}}$ relevant to the target task. If $d$ is too small, this condition fails regardless of how smooth or interpolatable the latent space appears.

The same issue can be expressed in terms of over-clustering. Suppose that the latent space is approximately spherical with token dimension \(d\). At an angular resolution \(\rho\), the number of disjoint stable regions is upper-bounded by the cap-area scale
\[
K_{\mathrm{area}}(d,\rho)=\frac{1}{\mu_d(\rho)}.
\]
Let \(M_{\mathrm{data}}(\varepsilon)\) denote the effective number of perceptually distinguishable signal states at tolerance \(\varepsilon\). If
\[
K_{\mathrm{area}}(d,\rho)
\ll
M_{\mathrm{data}}(\varepsilon),
\]
then even this optimistic upper bound is insufficient to assign distinct stable regions to all perceptually distinguishable signal states, and many distinct signals must therefore be mapped into the same or nearby latent regions. Under a smooth or noise-robust decoder, nearby latent points tend to decode to nearby reconstructions, and excessive clustering may further lead to averaged or over-smoothed outputs. Consequently, an excessively low-dimensional manifold may force unrelated or only weakly related signals to be clustered together, leading to over-smoothing, loss of diversity, or mode merging. Moreover, if the high-dimensional token space is explicitly aligned with this low-dimensional manifold, such over-clustering can also bias the information layout of the high-dimensional space and make the generative model inherit the same smoothing or diversity-loss tendency. In practice, this suggests choosing a moderate dimension: large enough to support
meaningful reconstruction and sufficient latent capacity, but small enough that the induced geometry remains
substantially more interpolatable than the original high-dimensional token sphere.

Besides interpolatability, \emph{identifiability} is another factor that may affect modeling difficulty. In many audio processing systems, time--frequency representations play a central role partly because they introduce an energy-based inductive bias: perceptually important or structurally dominant components often occupy higher-energy regions, while finer details occupy lower-energy regions. Under an additive isotropic corruption model
\[
y_i = x_i + \epsilon_i,
\qquad
\epsilon_i \sim \mathcal{N}(0,\sigma^2),
\]
a feature coordinate with energy $e_i=\mathbb{E}[x_i^2]$ has an effective coordinate-wise signal-to-noise ratio that scales as
\[
\mathrm{SNR}_i
\propto
\frac{e_i}{\sigma^2}.
\]
Higher-energy components therefore remain identifiable under stronger corruption. Motivated by this observation, we define identifiability in the present context as an inductive bias that highlights core content through energy allocation in feature space. If a high-dimensional token space can be shaped so that more reconstruction-critical information tends to occupy higher-energy components, then the generative model may more easily infer and preserve the core content of each token even under prediction noise.

In summary, interpolatability and identifiability impose complementary requirements. Interpolatability suggests that the high-dimensional token space should be constrained by a lower-dimensional manifold, while identifiability suggests that the high-dimensional coordinates should develop an energy hierarchy that makes important information more robust and easier to recover. Together, these two attributes may make the high-dimensional space easier for a generative model to learn and predict.

\subsection{Core objectives of tokenizer design}

The discussion above suggests a set of core objectives for the type of low-frame-rate, high-dimensional token space that we would like to construct:
\begin{itemize}
    \item it should be constrained by, or organized around, a lower-dimensional manifold embedded within the high-dimensional space, where this manifold is interpolatable and well-clustered while still having sufficient dimensionality to support meaningful reconstruction and avoid excessive over-clustering or smoothing;
    \item it should have sufficiently strong reconstruction robustness under noise;
    \item it should not rely on semantic or SSL spaces defined by external models;
    \item it should have sufficiently strong per-token identifiability.
\end{itemize}
\section{Methodology}
\label{sec:methodology}

In this section, we describe the concrete methods used to design our tokenizer and generative model, following the assumptions and design principles discussed above.

\subsection{Concentration of statistics in high-dimensional spaces}

A standard variational autoencoder regularizes its approximate posterior toward a standard Gaussian prior through the KL term. Let
\[
g\sim\mathcal{N}(0,I_N),
\qquad
g\in\mathbb{R}^N.
\]
Then its squared norm satisfies
\[
\|g\|_2^2=\sum_{i=1}^N g_i^2\sim\chi_N^2,
\]
with
\[
\mathbb{E}\big[\|g\|_2^2\big] = N,
\qquad
\mathrm{Var}\big(\|g\|_2^2\big) = 2N.
\]
As a consequence,
\[
\frac{\|g\|_2}{\sqrt{N}} \to 1
\quad\text{in probability as } N \to \infty,
\]
which implies that high-dimensional Gaussian vectors concentrate near the sphere of radius $\sqrt{N}$. Therefore, in sufficiently high dimensions, a standard Gaussian latent can be viewed, to a good approximation, as living on a thin spherical shell. This naturally motivates learning a spherical token space, i.e., a token space with approximately fixed norm. From the perspective of prediction, for any target token \(x\in\mathbb{R}^N\) and prediction \(\hat{x}\in\mathbb{R}^N\), the prediction error can be decomposed in polar form into radial and angular components. If both the target and the prediction are constrained or normalized to the same sphere, the radial degree of freedom is removed, and the remaining prediction error is determined by angular discrepancy. This is attractive for two reasons. First, the tokenizer decoder only needs to learn robustness with respect to angular perturbations. Second, the generative model can in principle remove one source of exposure bias by predicting within a fixed-norm space, rather than having to model both the norm and the direction of the token.

For these reasons, in the remainder of this paper we assume a high-dimensional \emph{spherical} token space as the default setting.

\subsection{Shaping decoder noise robustness}

Prior work on spherical VAEs often adopts the von Mises--Fisher (vMF) distribution as the prior over spherical latent variables~\cite{davidson2018hyperspherical}. However, in practice, the KL term and reparameterization for the vMF distribution are typically more cumbersome to implement, and often require additional approximations or specialized estimators~\cite{ke2025hyperspherical}. Here we instead consider a simpler alternative: rather than imposing an explicit prior through a KL term, we shape the information bottleneck and decoder robustness by injecting stronger noise directly into the token space without constraining the token prior distribution.

For a spherical token space, the most natural corruption mechanism is a random rotation on the sphere. Let
\[
x \in \mathbb{S}^{N-1}(R)
=
\{u \in \mathbb{R}^{N} : \|u\|_2 = R\}
\]
be a clean token. To construct a noisy token on the same sphere, we first sample a random tangent direction. Concretely, let
\[
\xi \sim \mathcal{N}(0, I_N),
\]
and project it onto the tangent space of the sphere at \(x\):
\begin{equation}
u
=
\xi
-
\frac{\langle \xi, x\rangle}{\|x\|_2^2} x.
\label{eq:tangent_projection}
\end{equation}
Then
\[
u \in T_x \mathbb{S}^{N-1}(R),
\qquad
\langle u, x\rangle = 0.
\]
We obtain a unit tangent direction at \(x\) by normalizing
\[
\bar{u}=\frac{u}{\|u\|_2}.
\]
We then sample a rotation angle \(\theta \in [0, \pi/2]\) and define the rotated noisy token through the spherical exponential map:
\begin{equation}
x^{\mathrm{rot}}
=
\cos\theta \, x
+
\sin\theta \, R \bar{u}.
\label{eq:spherical_noisy_token}
\end{equation}
Since $x$ and $\bar{u}$ are orthogonal, it follows directly that
\[
\|x^{\mathrm{rot}}\|_2^2
=
\cos^2\theta \, \|x\|_2^2
+
\sin^2\theta \, R^2
=
R^2,
\]
so $x^{\mathrm{rot}}$ remains exactly on the same sphere. Moreover, the geodesic angle between $x$ and $x^{\mathrm{rot}}$ is precisely $\theta$.

In practice, we sample a scalar
\[
s\sim\mathrm{Beta}(1,2)
\]
and set
\[
\theta=\frac{\pi}{2}s.
\]
This produces a distribution over rotation angles that is biased toward smaller perturbations while still allowing large-angle corruption up to \(90^\circ\), which corresponds to orthogonal tokens on the sphere. The motivation is twofold. First, a stronger mass near small angles makes the decoder spend more capacity on the local robustness regime that is most relevant to the prediction errors encountered during generation. Second, allowing occasional large-angle perturbations still provides a meaningful information bottleneck and prevents the decoder from overfitting to an excessively narrow neighborhood around the clean token. Under this corruption scheme, the tokenizer decoder is trained to reconstruct from randomly rotated noisy tokens. As a result, the decoder is explicitly encouraged to learn angular robustness on the spherical token space, without the need for an explicit spherical prior. In this sense, noise injection here serves both as a training-time robustness mechanism and as an implicit bottleneck that shapes the geometry of the learned token space.

\subsection{Shaping per-token identifiability}

As discussed above, the core idea behind identifiability is to emphasize more important content through higher-energy components. Here we aim to induce such a structure in the high-dimensional token space by explicitly introducing an \emph{availability bias} over token dimensions and combining it with sufficiently strong spherical corruption. The resulting training dynamics encourage more frequently available dimensions to carry larger energy, thereby increasing their effective signal-to-noise ratio under corruption and making the token more identifiable.

Concretely, inspired by residual dropout strategies commonly used in RVQ-based systems~\cite{kumar2023high}, we introduce a \emph{postfix dimension dropout} mechanism over the token dimensions. Let
\[
x=(x_1,\dots,x_N)\in\mathbb{S}^{N-1}(R)\subset\mathbb{R}^N
\]
be the rotated noisy spherical token. For each token independently, with probability $p$ we keep all dimensions unchanged. Otherwise, we sample an integer
\[
K \sim \mathrm{Unif}\{1,2,\dots,N-1\},
\]
and keep only the prefix dimensions $1,\dots,K$, while dropping the postfix dimensions $K+1,\dots,N$. Equivalently, if we define a random effective prefix length
\[
K_{\mathrm{eff}}
=
\begin{cases}
N, & \text{with probability } p,\\
K, & \text{with probability } 1-p,
\end{cases}
\]
then the dropout mask $m \in \{0,1\}^N$ is
\[
m_i = \mathbf{1}[i \le K_{\mathrm{eff}}],
\qquad i=1,\dots,N,
\]
and the corrupted token is
\begin{equation}
x^{\mathrm{drop}} = x \odot m,
\label{eq:postfix_dropout}
\end{equation}
where $\odot$ denotes elementwise multiplication.

Under this sampling rule, when \(p<1\), lower-indexed dimensions have strictly higher probabilities of being retained. For any \(i\in\{1,\dots,N\}\), we have
\begin{equation}
\Pr(m_i=1)
=
p+(1-p)\Pr(K\ge i)
=
p+(1-p)\frac{N-i}{N-1},
\label{eq:dim_keep_prob}
\end{equation}
where the last expression also gives \(\Pr(m_1=1)=1\) and \(\Pr(m_N=1)=p\). Therefore, this operation induces a clear prefix-to-postfix availability ordering over all dimensions. Note that the resulting token no longer lies on the high-dimensional sphere as we do not renormalize it before sending it to the decoder. We set $p=0.5$ by default.

The training-dynamics intuition is straightforward. Because prefix dimensions are more likely to be preserved, the model is repeatedly required to reconstruct the signal from partially observed and corrupted tokens in which only a prefix, sometimes a short one, remains available. At the same time, the clean token has a fixed total energy budget due to the spherical token geometry. Under this budget, if the model is to maximize reconstruction robustness under such corruption, one natural optimization direction is to allocate larger energy to the more frequently available dimensions. In other words, the availability bias is transformed through training into an energy bias:
\[
\text{higher availability}
\;\Longrightarrow\;
\text{higher learned energy}
\;\Longrightarrow\;
\text{higher corruption-time identifiability}.
\]

\subsection{Shaping a low-dimensional manifold}

We would like to construct, at low cost, a low-dimensional manifold embedded in the high-dimensional token space that is both well-clustered and strongly interpolatable. Since the high-dimensional token lies on a sphere, a natural design is to introduce a low-dimensional spherical space and to align the two spaces using orthogonal projection and lifting operations. As discussed above, due to the spherical-cap covering behavior, different tokens overlap much more strongly under large rotational corruption in lower dimensions than in higher dimensions. Therefore, if we impose strong perturbations on the low-dimensional sphere, it naturally creates a much stronger information bottleneck, which in turn encourages stronger cross-token interpolatability through larger overlap regions. This is also reminiscent of the feature disentanglement and clustering effects induced by stronger bottlenecks such as those used in $\beta$-VAE~\cite{higgins2017betavae}.

To preserve geometric structure as much as possible, we use a learnable row-orthogonal linear projection to map the clean high-dimensional spherical token into a lower-dimensional spherical space. Let
\[
x \in \mathbb{S}^{D_{\mathrm{tok}}-1}(R_H)
\]
be a native high-dimensional token, and let \(d_{\mathrm{core}}\ll D_{\mathrm{tok}}\) denote the dimension of the core manifold. We use a learnable row-orthogonal projection matrix
\[
W_{\downarrow} \in \mathbb{R}^{d_{\mathrm{core}} \times D_{\mathrm{tok}}},
\qquad
W_{\downarrow} W_{\downarrow}^{\top} = I_{d_{\mathrm{core}}},
\]
to map the high-dimensional token into the low-dimensional spherical space. In implementation, we maintain an unconstrained matrix
\[
A_{\downarrow}\in\mathbb{R}^{D_{\mathrm{tok}}\times d_{\mathrm{core}}},
\]
compute its thin QR decomposition
\[
A_{\downarrow}=Q_{\downarrow}R_{\downarrow},
\qquad
Q_{\downarrow}^{\top}Q_{\downarrow}=I_{d_{\mathrm{core}}},
\]
and set
\[
W_{\downarrow}=Q_{\downarrow}^{\top}.
\]
We define the low-dimensional token as
\begin{equation}
z
=
R_L
\frac{W_{\downarrow}x}{\|W_{\downarrow}x\|_2},
\qquad
z \in \mathbb{S}^{d_{\mathrm{core}}-1}(R_L).
\label{eq:lowdim_projection}
\end{equation}
The lifting map is defined by the same QR-based orthogonalization-and-renormalization template, but in the reverse direction. Specifically, we maintain an unconstrained matrix
\[
A_{\uparrow}\in\mathbb{R}^{D_{\mathrm{tok}}\times d_{\mathrm{core}}},
\]
compute its thin QR decomposition
\[
A_{\uparrow}=Q_{\uparrow}R_{\uparrow},
\qquad
Q_{\uparrow}^{\top}Q_{\uparrow}=I_{d_{\mathrm{core}}},
\]
and set
\[
W_{\uparrow}=Q_{\uparrow}.
\]
The lifted high-dimensional token is then defined as
\begin{equation}
x^{\mathrm{lift}}
=
R_H
\frac{W_{\uparrow}z}{\|W_{\uparrow}z\|_2},
\qquad
x^{\mathrm{lift}}\in\mathbb{S}^{D_{\mathrm{tok}}-1}(R_H).
\label{eq:lowdim_lifting}
\end{equation}
Since \(W_{\uparrow}\) is column-orthogonal, \(\|W_{\uparrow}z\|_2=\|z\|_2=R_L\) in exact arithmetic, so the above is equivalently
\[
x^{\mathrm{lift}}=\frac{R_H}{R_L}W_{\uparrow}z.
\]
We keep the explicit renormalization in Eq.~\eqref{eq:lowdim_lifting} for numerical consistency.

The motivation for using orthogonal mappings is geometric. The row-orthogonal projection \(W_{\downarrow}\) has orthonormal rows and therefore acts as a partial isometry on its retained subspace; in particular, it does not introduce anisotropic scaling among the retained directions. The subsequent normalization maps the projected vector back onto the low-dimensional sphere. For the lifting step, the mapping is strictly angle-preserving on the low-dimensional sphere: for any two low-dimensional tokens $z_1,z_2 \in \mathbb{S}^{d-1}(R_L)$,
\begin{equation}
\frac{\langle W_{\uparrow} z_1, W_{\uparrow} z_2\rangle}
{\|W_{\uparrow} z_1\|_2 \, \|W_{\uparrow} z_2\|_2}
=
\frac{\langle z_1, z_2\rangle}{\|z_1\|_2 \, \|z_2\|_2}.
\label{eq:strict_conformal_lift}
\end{equation}
Hence the angular similarity between low-dimensional tokens is preserved exactly after lifting.

To align the two spherical spaces, we employ a bidirectional commitment loss in the spirit of VQ-based discrete tokenizers. Let \(x\) denote the clean native high-dimensional token and \(x^{\mathrm{lift}}\) the lifted low-dimensional token. We penalize their angular discrepancy in both directions using cosine similarity. Denoting stop-gradient by \(\operatorname{sg}[\cdot]\), we use the following bidirectional cosine commitment loss:
\begin{equation}
\mathcal{L}_{\mathrm{commit}}
=
\bigl(1-\cos(x,\operatorname{sg}[x^{\mathrm{lift}}])\bigr)
+
\bigl(1-\cos(\operatorname{sg}[x],x^{\mathrm{lift}})\bigr).
\label{eq:bidirectional_commitment}
\end{equation}
This encourages the low-dimensional token to remain geometrically aligned with the high-dimensional token, while also encouraging the lifted low-dimensional representation to occupy a direction close to the original high-dimensional token.

We apply the strong information bottleneck and corruption to the low-dimensional sphere by constructing a noisy low-dimensional reconstruction path. That is, we apply the same spherical corruption process to the low-dimensional token $z$, lift the noisy low-dimensional token back to the high-dimensional sphere, and require the decoder to reconstruct the signal from this lifted noisy low-dimensional representation as well. If \(z^{\mathrm{rot}}\) denotes the spherically corrupted low-dimensional token and \(x^{\mathrm{low}}\) denotes its lifted high-dimensional version, then the decoder is trained on both reconstruction paths:
\[
\text{corrupted native high-dimensional token} \;\to\; \text{decoder} \;\to\; \text{signal},
\]
and
\[
\text{lifted corrupted low-dimensional token} \;\to\; \text{decoder} \;\to\; \text{signal}.
\]
As a result, the low-dimensional space is not merely an auxiliary projection of the high-dimensional token, but is explicitly required to support signal reconstruction under corruption, thereby encouraging it to encode the most reconstruction-critical information. Combined with the bidirectional commitment loss, this dual-path noisy reconstruction objective shapes a low-dimensional manifold with a strong information bottleneck and encourages the high-dimensional token space to be organized around the lifted low-dimensional geometry induced by the orthogonal mapping pair.

\subsection{Bridge construction and training target in the generative model}

For spherical tokens, the most natural evolution path is the geodesic path on the sphere, and the corresponding bridge construction is spherical flow matching (SFM), i.e., the spherical special case of Riemannian flow matching~\cite{chen2023flow}. Several recent works have explored the use of SFM for improving generative modeling on spherical token spaces~\cite{lee2026geometry, meral2026aligning}. We therefore use SFM as the default bridge and discuss several properties that motivate this choice.

\paragraph{Connection to VP-path flow.}
In high dimensions, SFM is closely related to the standard trigonometric interpolant bridge used in variance-preserving (VP) paths~\cite{song2020score, albergo2022building, albergo2025stochastic}. Let \(x_1\in\mathbb{S}^{D_{\mathrm{tok}}-1}(R)\) denote a data token sampled from the tokenizer distribution. For the source endpoint, we sample an isotropic Gaussian vector $g\sim\mathcal{N}(0,I_{D_{\mathrm{tok}}})$ and project it onto the same sphere:
\begin{equation}
x_0
=
R\frac{g}{\|g\|_2},
\qquad
x_0\in\mathbb{S}^{D_{\mathrm{tok}}-1}(R).
\label{eq:sfm_source_sample}
\end{equation}
The VP-style trigonometric interpolation between the source sample \(x_0\) and the target data token \(x_1\) is
\begin{equation}
x_t^{\mathrm{VP}}
=
\cos\!\left(\frac{\pi t}{2}\right)x_0
+
\sin\!\left(\frac{\pi t}{2}\right)x_1,
\qquad t\in[0,1].
\label{eq:vp_trig_path}
\end{equation}
Its squared norm is
\begin{equation}
\|x_t^{\mathrm{VP}}\|_2^2
=
R^2\cos^2\!\left(\frac{\pi t}{2}\right)
+
R^2\sin^2\!\left(\frac{\pi t}{2}\right)
+
2\cos\!\left(\frac{\pi t}{2}\right)\sin\!\left(\frac{\pi t}{2}\right)
\langle x_0,x_1\rangle .
\label{eq:vp_trig_norm}
\end{equation}
Equivalently,
\[
\|x_t^{\mathrm{VP}}\|_2^2
=
R^2
+
2\cos\!\left(\frac{\pi t}{2}\right)\sin\!\left(\frac{\pi t}{2}\right)
\langle x_0,x_1\rangle .
\]
Conditioned on any fixed data token \(x_1\), the source direction \(x_0/R\) is uniformly distributed on the unit sphere. Therefore,
\[
\mathbb{E}\left[
\frac{\langle x_0,x_1\rangle}{R^2}
\,\middle|\,x_1
\right]
=0,
\qquad
\frac{\langle x_0,x_1\rangle}{R^2}
=
O_p\!\left(D_{\mathrm{tok}}^{-1/2}\right).
\]
Hence, in high dimensions,
\[
\frac{\langle x_0,x_1\rangle}{R^2}\approx 0.
\]
It follows that
\[
\|x_t^{\mathrm{VP}}\|_2^2 \approx R^2,
\]
which shows that the VP trigonometric path is approximately norm-preserving. Under the same high-dimensional near-orthogonality condition, the geodesic angle
\[
\Omega
=
\arccos\!\left(\frac{\langle x_0,x_1\rangle}{R^2}\right)
\]
concentrates near \(\pi/2\), and the SFM geodesic interpolation
\begin{equation}
x_t^{\mathrm{SFM}}
=
\frac{\sin((1-t)\Omega)}{\sin\Omega}x_0
+
\frac{\sin(t\Omega)}{\sin\Omega}x_1
\label{eq:sfm_path_general}
\end{equation}
reduces approximately to
\begin{equation}
x_t^{\mathrm{SFM}}
\approx
\cos\!\left(\frac{\pi t}{2}\right)x_0
+
\sin\!\left(\frac{\pi t}{2}\right)x_1.
\label{eq:sfm_path_highdim}
\end{equation}
Therefore, in the high-dimensional near-orthogonal regime, SFM is nearly equivalent to the VP-style trigonometric interpolant on the same radius-\(R\) sphere. 

\paragraph{Norm stability of \(x_t\).}
A further property of SFM is that the norm of \(x_t\) remains constant along the bridge. By contrast, under a linear interpolation bridge \(x_t=(1-t)x_0+t x_1\), the norm of \(x_t\) varies with \(t\). In the high-dimensional near-orthogonal case, for example, we have \(\|x_t\|_2^2 \approx R^2\big((1-t)^2+t^2\big)\), which is strongly coupled with the bridge time. Since the model typically receives an explicit time embedding, such a bridge introduces an additional and unnecessary coupling between time and radial scale. At inference time, once the generated trajectory deviates from the ideal bridge, the radial scale of the current state may become inconsistent with the explicit time embedding, increasing the risk of exposure bias.

\paragraph{Equivalence of \(x\)-pred and \(v\)-pred.}
As discussed above, in raw signal-space generative modeling, the manifold hypothesis often motivates direct target prediction, which can be more effective than predicting noise or velocity in certain regimes. It is useful to first contrast the role of endpoint prediction in Euclidean rectified flow and in spherical flow matching. For the standard Euclidean rectified-flow bridge
\[
x_t=(1-t)x_0+t x_1,
\qquad t\in[0,1],
\]
the oracle velocity is $v_t=x_1-x_0$, which lives in the full ambient Euclidean space and is unconstrained. Therefore, a direct \(v\)-pred parameterization requires the model output head to predict a vector containing the high-entropy source term \(x_0\). By contrast, an \(x\)-pred parameterization lets the model predict \(\hat x_1\) and converts it into a velocity by
\begin{equation}
\hat v_t^{\mathrm{RF}}
=
\frac{\hat x_1-x_t}{1-t}.
\label{eq:rf_xpred_to_v}
\end{equation}
Algebraically, \(x\)-pred and \(v\)-pred can be converted into each other in the Euclidean bridge. However, as discussed in recent works~\cite{li2026back}, they are not equivalent as output-layer parameterizations: \(x\)-pred asks the model to predict the data endpoint, which may benefit from the manifold structure of the data distribution, whereas \(v\)-pred directly asks the model to predict the unconstrained displacement \(x_1-x_0\). 

For spherical flow matching, the situation is different. A valid velocity at the bridge state \(x_t\in\mathbb{S}^{D_{\mathrm{tok}}-1}(R)\) must lie in the tangent space
\[
T_{x_t}\mathbb{S}^{D_{\mathrm{tok}}-1}(R)
=
\left\{
u\in\mathbb{R}^{D_{\mathrm{tok}}}:
\langle u,x_t\rangle=0
\right\}.
\]
Thus, unlike Euclidean rectified flow, the velocity is not an unconstrained ambient vector, and any direct velocity parameterization must either explicitly predict a tangent vector or project an ambient prediction onto the tangent space. For any \(y\in\mathbb{S}^{D_{\mathrm{tok}}-1}(R)\), define the geodesic angle between \(x_t\) and \(y\) as
\[
\alpha_t(y)
=
\arccos\!\left(
\frac{\langle x_t,y\rangle}{R^2}
\right).
\]
The tangent projection of \(y\) at \(x_t\) is
\begin{equation}
\Pi_{x_t}^{\perp}(y)
=
y
-
\frac{\langle y,x_t\rangle}{R^2}x_t
=
y-\cos\!\big(\alpha_t(y)\big)x_t .
\label{eq:tangent_projector_xt}
\end{equation}
The Riemannian logarithm map from \(x_t\) to \(y\) is
\begin{equation}
\log_{x_t}(y)
=
\frac{\alpha_t(y)}{\sin\alpha_t(y)}
\Pi_{x_t}^{\perp}(y),
\qquad
\log_{x_t}(y)\in T_{x_t}\mathbb{S}^{D_{\mathrm{tok}}-1}(R),
\label{eq:sphere_log_map}
\end{equation}
with
\[
\|\log_{x_t}(y)\|_2
=
R\,\alpha_t(y).
\]

Therefore, if the model predicts a spherical endpoint \(\hat x_1\), the corresponding valid SFM velocity over the remaining interval \([t,1]\) is
\begin{equation}
\hat v_t^{\mathrm{full}}
=
\frac{1}{1-t}\log_{x_t}(\hat x_1).
\label{eq:xpred_full_velocity}
\end{equation}
The oracle SFM velocity is obtained by setting \(y=x_1\). Let
\[
\Omega
=
\arccos\!\left(
\frac{\langle x_0,x_1\rangle}{R^2}
\right),
\]
and let the SFM bridge be
\begin{equation}
x_t
=
\frac{\sin((1-t)\Omega)}{\sin\Omega}x_0
+
\frac{\sin(t\Omega)}{\sin\Omega}x_1 .
\label{eq:sfm_bridge}
\end{equation}
Then the oracle tangent velocity is
\begin{equation}
v_t
=
\dot{x}_t
=
\frac{\Omega}{\sin\Omega}
\left(
-\cos((1-t)\Omega)x_0
+
\cos(t\Omega)x_1
\right),
\label{eq:sfm_velocity}
\end{equation}
or equivalently
\[
v_t=\frac{1}{1-t}\log_{x_t}(x_1).
\]
Thus, for SFM, estimating a geometrically valid velocity is essentially equivalent to estimating a spherical endpoint together with its induced tangent direction and scale. In this sense, the distinction between \(x\)-pred and \(v\)-pred is weaker than in Euclidean rectified flow: the tangent-space constraint naturally turns a valid velocity parameterization into an endpoint-induced parameterization.

\paragraph{Under-stepping and direction-only modeling.}
A conventional velocity objective would minimize the full velocity MSE,
\begin{equation}
\mathcal{L}_{\mathrm{v}}
=
\left\|
\hat v_t
-
v_t
\right\|_2^2 .
\label{eq:full_velocity_loss}
\end{equation}
However, in the present high-dimensional spherical setting, full velocity MSE introduces a potential \emph{under-stepping} effect. To see this, suppose the oracle velocity can be written as
\[
v_t=s_t d_t,
\qquad
s_t=\|v_t\|_2,
\qquad
\|d_t\|_2=1,
\]
and suppose the model predicts a tangent velocity but with a direction error. Write its prediction as
\[
\hat v_t=a_t \hat d_t,
\qquad
\|\hat d_t\|_2=1,
\]
where \(a_t\ge 0\) is the predicted speed. Let \(\psi_t\) denote the angle between the predicted and oracle directions:
\[
\cos\psi_t
=
\langle \hat d_t,d_t\rangle .
\]
For a fixed predicted direction \(\hat d_t\), the velocity MSE as a function of the predicted magnitude \(a_t\) is
\begin{align}
\left\|
a_t\hat d_t-s_t d_t
\right\|_2^2
&=
a_t^2
-
2a_t s_t\langle \hat d_t,d_t\rangle
+
s_t^2
\nonumber\\
&=
a_t^2
-
2a_t s_t\cos\psi_t
+
s_t^2 .
\label{eq:mse_speed_derivation}
\end{align}
Minimizing this quadratic over \(a_t\) gives
\begin{equation}
a_t^{\star}
=
s_t\cos\psi_t,
\label{eq:mse_optimal_speed}
\end{equation}
Therefore, unless the predicted direction is perfectly aligned with the oracle direction, the nonnegative MSE-optimal velocity magnitude is smaller than the oracle magnitude. In other words, full velocity MSE allows the model to explain directional uncertainty by reducing the step size.

For the SFM geodesic, the oracle speed is constant along the bridge and satisfies exactly
\begin{equation}
\|v_t\|_2
=
R\Omega .
\label{eq:sfm_velocity_norm_exact}
\end{equation}
In high dimensions, when \(x_0/R\) and \(x_1/R\) are weakly correlated random unit directions, their normalized inner product concentrates near zero, and hence
\[
\Omega\approx \frac{\pi}{2}.
\]
Thus the oracle velocity norm concentrates near
\begin{equation}
\|v_t\|_2
\approx
\frac{\pi}{2}R .
\label{eq:vt_norm_concentrated}
\end{equation}
Under full velocity MSE, however, if the model has high uncertainty in endpoint or direction estimation, Eq.~\eqref{eq:mse_optimal_speed} encourages a smaller predicted speed. During inference, this can lead to a trajectory whose accumulated path length is systematically shorter than the typical SFM geodesic length. In the high-dimensional regime, where the target endpoint is typically close to \(90^\circ\) away from the source endpoint, such under-stepping creates the risk that the generated endpoint remains at an angle substantially smaller than \(90^\circ\) from the source sample. To avoid this failure mode, we use a direction-only parameterization: the model still predicts a spherical endpoint \(\hat x_1\), but this endpoint is used only to define the tangent direction at the current bridge state. Specifically, we define the predicted and oracle unit tangent directions as
\begin{equation}
\hat d_t
=
\frac{\log_{x_t}(\hat x_1)}
{\|\log_{x_t}(\hat x_1)\|_2},
\qquad
d_t
=
\frac{v_t}{\|v_t\|_2}
=
\frac{\log_{x_t}(x_1)}
{\|\log_{x_t}(x_1)\|_2}.
\label{eq:unit_tangent_directions}
\end{equation}
We then supervise only the tangent direction using the cosine direction loss
\begin{equation}
\mathcal{L}_{\mathrm{dir}}
=
1-\langle \hat d_t,d_t\rangle .
\label{eq:direction_loss}
\end{equation}
At inference time, before any additional guidance combination, we fix the predicted velocity magnitude to its high-dimensional SFM limit:
\begin{equation}
\hat v_t
=
\frac{\pi R}{2}\hat d_t .
\label{eq:fixed_norm_predicted_velocity}
\end{equation}
In this way, the model-predicted endpoint serves as a direction parameterization rather than as an exact endpoint estimate or a full velocity estimate. This removes the need for the model to predict the velocity norm, avoids MSE-induced under-stepping, and keeps the generated trajectory length compatible with the typical high-dimensional SFM geometry.
\section{Model Architecture Design}
\label{sec:architecture}

In this section, we provide concrete architecture instantiations of a minimalist tokenizer model and a minimalist AR flow-matching model. For the tokenizer, we intentionally restrict ourselves to simple and standard building blocks, in order to show that once the token space is shaped appropriately, neither the overall model structure nor its basic units need to be overly complicated to achieve a good balance between fidelity and predictability. For the AR flow-matching model, we adopt a modular design that is tailored to the error-accumulation and exposure-bias issues that arise easily in AR flow systems. Through explicit functional disentanglement, the system can, via training dynamics, automatically optimize different information extraction paths, while at inference time appropriate multi-path CFG can be used to selectively enhance different functional components, thereby mitigating sequential error accumulation during AR generation.

\subsection{Locodec: a minimalist, locally encoded tokenizer architecture}

\begin{figure}[!t]
    \centering
    \includegraphics[width=1.0\textwidth]{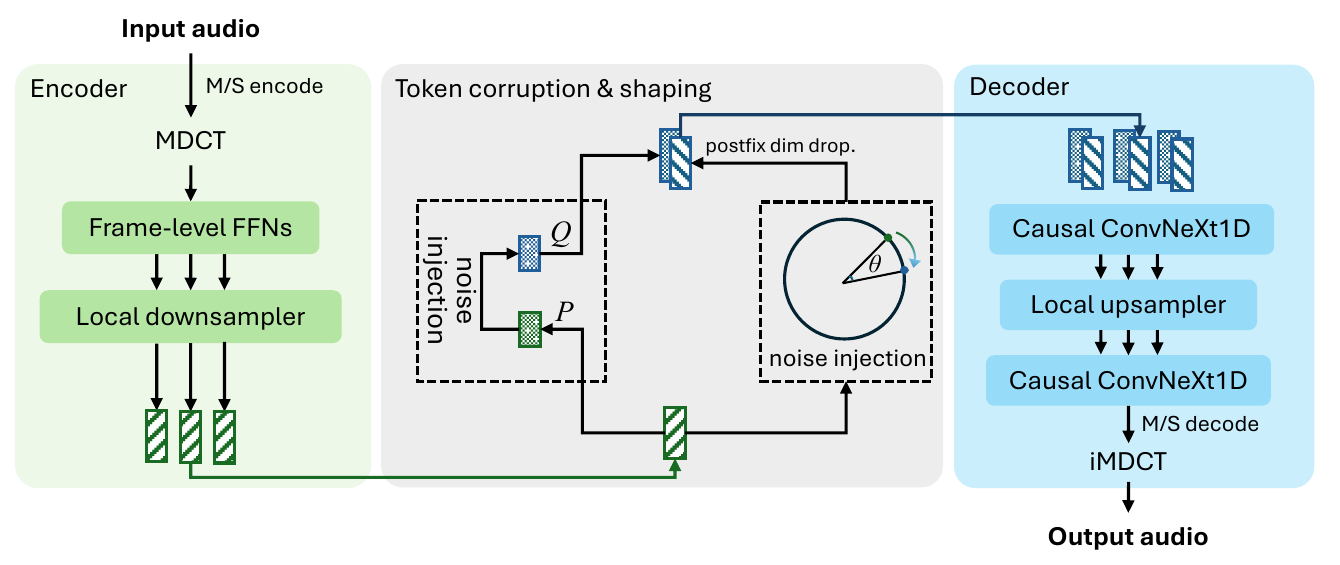}
    \caption{Illustration of the Locodec tokenizer architecture, consisting of MDCT coefficient extraction, a fully local encoder, a token-grid refiner, a coefficient-grid predictor, and an inverse-MDCT stage for waveform reconstruction.}
    \label{fig:tokenizer_design}
\end{figure}

We propose \emph{Locodec}, a \emph{lo}cally encoded \emph{codec}. Our goal here is deliberately minimalist: the architecture is built from simple, standard components, so that the emphasis remains on token-space shaping rather than on architectural complexity. Moreover, although the experiments in this paper mainly focus on single-channel TTS, the tokenizer design and training losses described here are formulated to support both monaural and stereo audio in a unified manner. Figure~\ref{fig:tokenizer_design} shows the Locodec pipeline.

\paragraph{Mid/side input formulation and MDCT front-end.}
We adopt a mid/side formulation at the waveform level. For the left and right channels $x_L$ and $x_R$, we define
\begin{equation}
x_{\mathrm{mid}} = \frac{x_L + x_R}{2},
\qquad
x_{\mathrm{side}} = \frac{x_L - x_R}{2}.
\label{eq:mid_side_encode_wave}
\end{equation}
For monaural input, we simply set $x_L = x_R$ so that $x_{\mathrm{side}} = 0$. This allows monaural and stereo samples to be mixed naturally during training while sharing the same tokenizer architecture and loss functions. 

Motivated by conventional audio codecs~\cite{atal1970adaptive, schroeder1985code, bessette2002adaptive}, we apply an MDCT transform to obtain real-valued time--frequency representations, i.e., MDCT coefficients, for both the mid and side waveforms. For an MDCT window size of \(k\), the corresponding hop size is \(k/2\), which determines the native coefficient-frame rate seen by the tokenizer encoder. We further apply a simple signed dynamic-range compression to both the mid and side coefficient sequences before feeding them into the encoder:
\begin{equation}
c \;\leftarrow\; \operatorname{sign}(c)\,|c|^{1/3}.
\label{eq:mdct_compress_audio}
\end{equation}
This transformation preserves coefficient sign while compressing amplitude range, reducing the tendency of the model to over-emphasize large low-frequency coefficients and underfit smaller but still perceptually important higher-frequency components. The compressed mid and side coefficient sequences are then concatenated along the feature dimension and used as the input to the tokenizer encoder.

\paragraph{Fully local encoder.}

On top of this MDCT coefficient sequence, we use a fully local encoder to map the coefficients to a hidden space. The encoder first applies a stack of frame-level local feed-forward networks (FFNs) at the native MDCT frame rate, and then uses a single-step linear downsampling layer that groups \(s\) neighboring frame-level embeddings, concatenates them along the feature dimension, and projects the concatenated vector directly into the target token dimension. More precisely, if the frame-level embedding dimension is $D_{\mathrm{enc}}$ and the temporal downsampling factor is $s$, then the downsampling layer is implemented as a linear map
\[
\mathbb{R}^{D_{\mathrm{enc}} \cdot s} \rightarrow \mathbb{R}^{D_{\mathrm{tok}}}.
\]
The token sequence is strictly local in the sense that each token is formed only from its corresponding local coefficient frames, without cross-token information exchange. We intentionally keep the encoder fully local to make the tokenization closer in spirit to raw input-space modeling with non-overlapping chunking or patchification: each token corresponds to a local region of the input signal, and the corresponding manifold assumption is imposed at the level of local signal chunks. Moreover, this locality also reduces the risk of reconstruction conflicts during generation. If the encoder introduces strong cross-token interactions, then the information required to reconstruct a given signal segment may be distributed across multiple neighboring tokens. While the tokenizer decoder can exploit such distributed information during reconstruction training, an AR generative model predicts these tokens separately at inference time, and the prediction errors or inconsistencies among neighboring tokens may then provide conflicting evidence about the same underlying signal segment, making reconstruction less stable. We therefore isolate the information scope of each token at the encoder level, so that each token primarily models its own local signal region.

Each token is then normalized onto the high-dimensional sphere. As described in the previous section, all token-space shaping mechanisms, including spherical corruption, postfix dimension dropout, low-dimensional projection and lifting, and dual-path noisy reconstruction, are applied on top of this token sequence.

\paragraph{Stacked causal convolutional decoder.}

The decoder starts from either a noisy high-dimensional token sequence or a lifted noisy low-dimensional token sequence. It first applies a token-grid refinement module implemented as stacked causal ConvNeXt1D blocks~\cite{liu2022convnet}. This stage operates directly on the low-frame-rate token sequence and is responsible for reconciling the corrupted token representation before coefficient-frame reconstruction. The refined token sequence is then mapped back to the native MDCT coefficient-frame rate through a single-step linear upsampling layer, which serves as the decoder-side counterpart of the single-step linear downsampling layer. Another stack of causal ConvNeXt1D blocks then performs coefficient-grid prediction at the native MDCT frame rate to recover frame-level detail. The mid/side hidden representations are then decoded via a channel-separation layer: let $h \in \mathbb{R}^{T_{\mathrm{coef}} \times D}$ denote the refined hidden sequence produced by the coefficient-grid ConvNeXt1D predictor, where $T_{\mathrm{coef}}$ is the native MDCT coefficient-frame length and $D$ is the hidden dimension. We pass $h$ through two separate FFNs to obtain a mid-channel hidden sequence and a side-channel hidden sequence:
\begin{equation}
h_{\mathrm{mid}} = \mathrm{FFN}_{\mathrm{mid}}(h),
\qquad
h_{\mathrm{side}} = \mathrm{FFN}_{\mathrm{side}}(h).
\label{eq:mid_side_hidden}
\end{equation}
These two hidden sequences are then mapped to coefficient predictions using a \emph{shared} gated output layer: let $\mathcal{O}_{\mathrm{coef}}:\mathbb{R}^{D}\rightarrow \mathbb{R}^{2F}$ denote the shared coefficient output layer, where $F$ is the number of MDCT frequency bins per frame. Applying \(\mathcal{O}_{\mathrm{coef}}\) to either \(h_{\mathrm{mid}}\) or \(h_{\mathrm{side}}\) produces two tensors, corresponding to an amplitude branch and a signed-gate branch. Writing
\begin{equation}
(h_{\mathrm{amp}}^{(b)},\,h_{\mathrm{sgn}}^{(b)})
=
\mathcal{O}_{\mathrm{coef}}(h_b),
\qquad
b\in\{\mathrm{mid},\mathrm{side}\},
\label{eq:shared_coef_head}
\end{equation}
the predicted MDCT coefficients for branch \(b\) are parameterized as
\begin{equation}
\widehat{c}_b
\;=\;
\exp\!\big(h_{\mathrm{amp}}^{(b)}\big)
\cdot
\tanh\!\big(h_{\mathrm{sgn}}^{(b)}\big),
\qquad
b\in\{\mathrm{mid},\mathrm{side}\}.
\label{eq:coef_param_audio_rewrite}
\end{equation}
Here \(\exp(h_{\mathrm{amp}}^{(b)})\) controls a nonnegative amplitude scale, while \(\tanh(h_{\mathrm{sgn}}^{(b)})\) acts as a bounded signed gate. The left and right MDCT coefficients are recovered by
\begin{equation}
\widehat{c}_L
=
\widehat{c}_{\mathrm{mid}}
+
\widehat{c}_{\mathrm{side}},
\qquad
\widehat{c}_R
=
\widehat{c}_{\mathrm{mid}}
-
\widehat{c}_{\mathrm{side}}.
\label{eq:mid_side_decode_coef_rewrite}
\end{equation}
Finally, inverse MDCT is applied to \(\widehat{c}_L\) and \(\widehat{c}_R\) to reconstruct the left and right waveforms \(\hat{x}_L\) and \(\hat{x}_R\) in the original time domain.

\paragraph{Training objectives.}

The training objective consists of three parts: a noisy reconstruction loss, a perceptual loss, and the high--low-dimensional commitment loss. 

The noisy reconstruction loss is applied to both high-dimensional and low-dimensional reconstruction paths. Following prior works~\cite{kumar2023high, luo2024gull}, we compute STFT magnitudes at window sizes
\[
\mathcal{W}_{\mathrm{STFT}}
=
\{2^5,2^6,\dots,2^{12}\}.
\]
For each \(w\in\mathcal{W}_{\mathrm{STFT}}\), define
\[
A_w(y)=\left|\mathrm{STFT}_w(y)\right|.
\]
Here and below, \(y\) denotes a waveform and \(\hat y\) denotes its reconstruction. The normalized magnitude loss is defined channel-wise. For the left, right, and mid channels, we use the target magnitude of the same channel as the normalization factor:
\begin{equation}
r_w^{b}(y)
=
\operatorname{mean}\!\left(A_w(y_b)\right),
\qquad
b\in\{L,R,\mathrm{mid}\}.
\label{eq:mag_loss_denominator_main}
\end{equation}
Here \(\operatorname{mean}(\cdot)\) averages over the time--frequency bins of a single example. For the side channel, however, we normalize by the target mid-channel magnitude:
\begin{equation}
r_w^{\mathrm{side}}(y)
=
\operatorname{mean}\!\left(A_w(y_{\mathrm{mid}})\right).
\label{eq:mag_loss_denominator_side}
\end{equation}
This avoids an ill-conditioned normalization in the monaural case, where the target side channel is identically zero. The channel-wise normalized magnitude loss is then
\begin{equation}
\ell_{w}^{\mathrm{mag},b}(\hat y,y)
=
\mathbb{E}
\left[
\frac{
\operatorname{mean}\!\left(
\left|A_w(\hat y_b)-A_w(y_b)\right|
\right)
}{
r_w^{b}(y)+\varepsilon
}
\right],
\qquad
b\in\{L,R,\mathrm{mid},\mathrm{side}\},
\label{eq:mag_loss}
\end{equation}
where \(\mathbb{E}[\cdot]\) denotes the empirical average over the training batch. The log-magnitude loss is applied only to the left and right channels:
\begin{equation}
\ell_{w}^{\mathrm{lmag},b}(\hat y,y)
=
\mathbb{E}
\left[
\operatorname{mean}\!\left(
\left|
\log_{10}\!\big(A_w(\hat y_b)+\varepsilon\big)
-
\log_{10}\!\big(A_w(y_b)+\varepsilon\big)
\right|
\right)
\right],
\qquad
b\in\{L,R\}.
\label{eq:log_mag_loss}
\end{equation}

For a reconstructed stereo waveform, we define
\[
\hat y_{\mathrm{mid}}
=
\frac{\hat y_L+\hat y_R}{2},
\qquad
\hat y_{\mathrm{side}}
=
\frac{\hat y_L-\hat y_R}{2},
\]
and analogously for the target waveform \(y\). Let \(\hat y^{(H)}\) and \(\hat y^{(L)}\) denote the reconstructions from the high-dimensional and low-dimensional paths, respectively. We define the reconstruction loss for each path as
\begin{align}
\mathcal{L}_{\mathrm{rec}}^{(m)}
=
\sum_{w\in\mathcal{W}_{\mathrm{STFT}}}
\Bigg[
&
\sum_{b\in\{L,R\}}
\left(
\ell_w^{\mathrm{mag},b}(\hat y^{(m)},y)
+
\lambda_{\mathrm{lmag}}\ell_w^{\mathrm{lmag},b}(\hat y^{(m)},y)
\right)
\nonumber\\
&
+
\ell_w^{\mathrm{mag},\mathrm{mid}}(\hat y^{(m)},y)
+
\ell_w^{\mathrm{mag},\mathrm{side}}(\hat y^{(m)},y)
\Bigg],
\qquad m\in\{H,L\}.
\label{eq:rec_loss}
\end{align}

In the current implementation, we use the same reconstruction-loss form for both paths, while the low-dimensional path is assigned a smaller weight in the final objective.

The perceptual loss used to train the tokenizer decoder is composed of an adversarial generator loss and a feature-matching loss. The GAN discriminator operates on multi-resolution uncompressed MDCT coefficients computed directly from the waveform, and is built from stacked Conv1d blocks~\cite{luo2024gull}. We use MDCT coefficient matrices computed with four window sizes $256,\ 512,\ 1024,\ 2048$, and for each resolution, we instantiate both a full-band discriminator and a sub-band discriminator. The full-band discriminator takes the entire coefficient matrix as input, and the sub-band discriminator uniformly partitions the spectrum into sub-bands of \(4\)-kHz bandwidth. In implementation, the sub-band discriminator is realized with grouped convolutions: each spectral sub-band is processed by its own convolutional group, while all sub-bands at the same MDCT resolution are treated as one discriminator module. Therefore, each resolution contributes one full-band discriminator and one grouped sub-band discriminator, yielding a total of eight discriminator modules across the four resolutions.

We use the least-squares GAN (LSGAN) objective~\cite{mao2017least}. Let \(\mathcal{J}\) denote the set of all discriminators. For each \(j\in\mathcal{J}\), let \(C_j(\cdot)\) denote the MDCT representation associated with discriminator \(D_j\). The discriminator loss is
\begin{equation}
\mathcal{L}_{D}^{(j)}
=
\mathbb{E}_{y}
\left[
\left(D_j(C_j(y))-1\right)^2
\right]
+
\mathbb{E}_{\hat{y}^{(H)}}
\left[
D_j\!\left(C_j(\operatorname{sg}[\hat{y}^{(H)}])\right)^2
\right],
\label{eq:LSGAN_disc_loss}
\end{equation}
and the adversarial generator loss for the tokenizer is
\begin{equation}
\mathcal{L}_{\mathrm{adv}}^{(m, j)}
=
\mathbb{E}_{\hat{y}^{(m)}}
\left[
\left(D_j(C_j(\hat{y}^{(m)}))-1\right)^2
\right], \quad m\in\{H, L\}.
\label{eq:LSGAN_gen_loss}
\end{equation}
Note that the discriminators are trained using only the high-dimensional reconstructions as negative samples. 

A feature-matching loss is also computed as a layer-wise normalized MAE between the discriminator hidden activations of the reconstructed and target waveforms. Let \(D_j^{(\ell)}(\cdot)\) denote the hidden activation of the \(\ell\)-th layer of discriminator \(D_j\), and let \(L_j\) be the number of hidden layers used for feature matching. We define
\begin{equation}
\mathcal{L}_{\mathrm{FM}}^{(m,j)}
=
\frac{1}{L_j}
\sum_{\ell=1}^{L_j}
\mathbb{E}
\left[
\frac{
\operatorname{mean}\!\left(
\left|
D_j^{(\ell)}(C_j(\hat{y}^{(m)}))
-
\operatorname{sg}\!\left[D_j^{(\ell)}(C_j(y))\right]
\right|
\right)
}{
\operatorname{mean}\!\left(
\left|
\operatorname{sg}\!\left[D_j^{(\ell)}(C_j(y))\right]
\right|
\right)
+\epsilon
}
\right], \quad m\in\{H, L\}.
\label{eq:feature_matching_loss}
\end{equation}

The overall perceptual loss used to train the tokenizer is
\begin{equation}
\mathcal{L}_{\mathrm{perc}}^{(m)}
=
\sum_{j\in\mathcal{J}}
\left(
\mathcal{L}_{\mathrm{adv}}^{(m,j)}
+
\mathcal{L}_{\mathrm{FM}}^{(m,j)}
\right), \quad m\in\{H, L\},
\label{eq:perceptual_loss}
\end{equation}
and the discriminator objective is
\begin{equation}
\mathcal{L}_{D}
=
\sum_{j\in\mathcal{J}}
\mathcal{L}_{D}^{(j)}.
\label{eq:total_discriminator_loss}
\end{equation}

Putting everything together, the overall tokenizer objective takes the form
\begin{equation}
\mathcal{L}_{\mathrm{tok}}
=
\mathcal{L}_{\mathrm{rec}}^{(H)}
+
\mathcal{L}_{\mathrm{perc}}^{(H)}
+
\lambda_{\mathrm{low}}\left(\mathcal{L}_{\mathrm{rec}}^{(L)}
+
\mathcal{L}_{\mathrm{perc}}^{(L)}
\right)
+
\lambda_{\mathrm{commit}}
\mathcal{L}_{\mathrm{commit}},
\label{eq:tokenizer_total_loss_arch_rewrite}
\end{equation}
where $\mathcal{L}_{\mathrm{commit}}$ is the bidirectional high--low-dimensional commitment loss defined previously. We set $\lambda_{\mathrm{lmag}}=0.2$, $\lambda_{\mathrm{low}}=0.2$, and $\lambda_{\mathrm{commit}}=0.1$ by default.

\subsection{MP-ELD: a multi-path information-routing encoder-LM-decoder architecture}
\label{sec:mpeld}

We find that constraining tokens to a high-dimensional sphere or explicitly learning a low-dimensional manifold does not automatically solve the AR error-accumulation problem. Based on empirical observations, we hypothesize that one important remaining source of AR error accumulation is \emph{information-pathway conflict under CFG}. Specifically, conditioning signals with partially overlapping functionality may appear in multiple CFG paths but be combined inconsistently across paths; the resulting conflicts are then fed back through the AR loop and manifest as gradual drift of the corresponding attributes. In practice, degradation in a TTS system often emerges as drift in \emph{acoustic attributes}---e.g., spectral distortion, gradually drifting loudness, or changes in speaking rate---whereas \emph{content consistency} is typically much more stable. This asymmetry suggests that acoustic-state drift, rather than content inconsistency, is the dominant mode of error accumulation. We attribute it to partially overlapping or redundant acoustic cues across CFG paths that can become mutually inconsistent under guidance, thereby inducing systematic drift in the AR inference process. Motivated by this hypothesis, we explicitly decouple information pathways by routing conditioning information into functionally distinct channels and applying multi-path CFG as a structured residual correction, so that each guidance component targets a specific role rather than implicitly entangling overlapping information across paths.

\paragraph{The ELD framework.}

We start from the \emph{ELD} (Encoder--LM--Decoder) framework, which is a common recipe for in-context conditional AR generation~\cite{lee2023sequential, ruhe2024rolling, jia2025ditar, yin2025slow, jiang2025diffrhythm}. Given a user-specified or task-provided control signal \(c\) (e.g., text, labels, or reference inputs), ELD aggregates the control signal \(c\) and the native-token prefix \(x_{1:i}\) into a step-wise conditioning signal \(c'_i\), and the decoder predicts the next native high-dimensional token conditioned on \(c'_i\):
\begin{equation}
p_\theta(x_{i+1}\mid x_{1:i},c)
=
p_\theta(x_{i+1}\mid c'_i),
\qquad
c'_i=f_\theta(x_{1:i},c).
\label{eq:eld_conditional_parameterization}
\end{equation}
ELD constructs $c'_i$ using an encoder, a sequence modeling module (LM), and a flow-matching decoder. A token encoder \(E\) maps a clean native token to a hidden representation:
\begin{equation}
h_i = E(x_i),
\qquad
h_i \in \mathbb{R}^{D_{\mathrm{model}}}.
\end{equation}
This serves as a token adapter that maps the token into a hidden space better aligned with sequential modeling and conditioning. The control signal \(c\) is embedded as a sequence \(\phi(c)\) in the same hidden space and concatenated with the encoded prefix:
\begin{equation}
s_i = [\,\phi(c)\,;\, h_{1:i}\,],
\end{equation}
and an LM then produces step-wise conditioning vectors
\begin{equation}
c'_{1:i} = \mathrm{LM}(s_i),
\qquad
c'_j \in \mathbb{R}^{D_{\mathrm{model}}}.
\end{equation}
Given \(c'_i\) and the bridge state at position \(i+1\), the decoder predicts a spherical endpoint estimate \(\hat{x}_{i+1}\). This endpoint estimate is then converted, using the logarithm-map construction described in the previous section, into a pathwise fixed-norm tangent velocity estimate before CFG. In addition to the step-wise conditioning vector, the decoder also receives the bridge time $\tau \in [0,1]$ and its time embedding.

To mitigate exposure bias, during training the ground-truth prefix tokens are perturbed with small noise before being fed to the encoder. We denote the resulting noisy teacher-forced native token by
\[
x_i^{\mathrm{ctx}}=\operatorname{Corrupt}(x_i).
\]
This simulates mild inference-time distribution drift while staying within a locally reconstruction-stable neighborhood and is a standard technique~\cite{ning2023input, chen2024diffusion}. At inference time, \(x_i^{\mathrm{ctx}}\) denotes the previously generated native token at step \(i\). Moreover, CFG is applied only at the decoder level, while the encoder and LM are evaluated once to produce the step-wise conditioning vectors. This is computationally cheaper than settings in which the LM or backbone must participate in CFG, since guidance does not require multiple forward passes through the encoder or LM.

\begin{figure}[!t]
    \centering
    \includegraphics[width=1.0\textwidth]{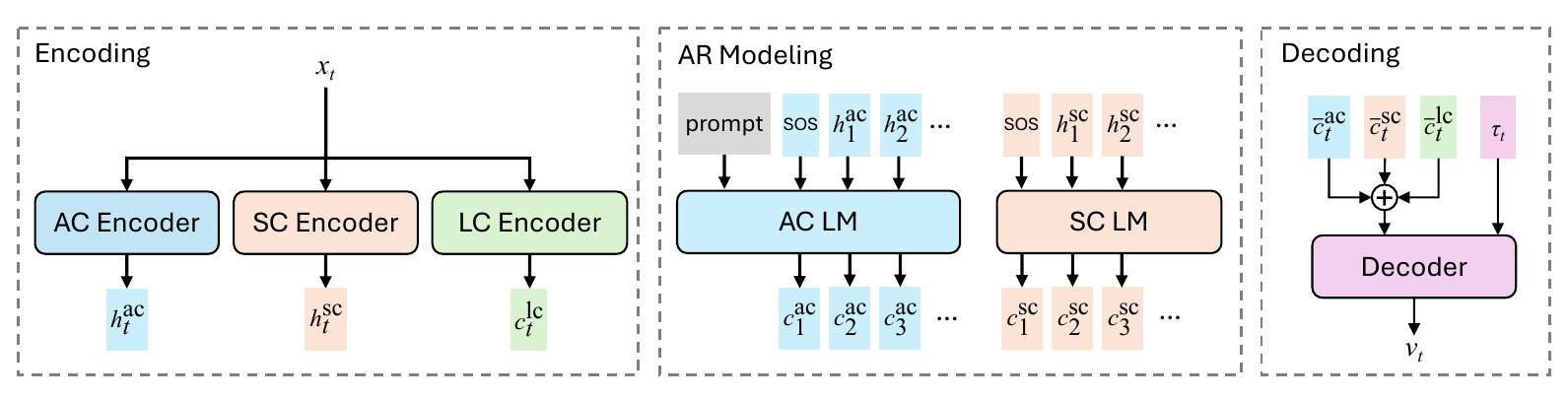}
    \caption{Illustration of the MP-ELD framework. Multiple local encoders map the same input token into distinct hidden spaces, whose functional roles are shaped by different AR information paths through training dynamics. A flow-matching decoder predicts pathwise next-token velocity fields conditioned on the corresponding path embeddings, which are combined by multi-path residual CFG during inference.}
    \label{fig:generative_model}
\end{figure}

However, in practice we find that the ELD framework can become fragile when stronger CFG is applied in the decoder. Empirically, this failure mode is closely tied to two design ambiguities: (i) what information the step-wise conditioning signal is expected to preserve and how it should be used or amplified by the decoder, and (ii) how to define the ``conditional'' and ``unconditional'' paths in CFG for the decoder. In particular, a fully null unconditional path can be problematic in this setting, since it provides no local continuation anchor and may yield a high-variance early-time velocity estimate, making training statistically difficult. Based on these observations, we introduce \emph{multi-path ELD (MP-ELD)}, a modification of ELD that makes the roles of conditioning signals explicit. The core idea is \emph{information routing}: we construct multiple conditioning paths that intentionally receive different information, so that each path naturally specializes to a different role. Figure~\ref{fig:generative_model} illustrates the MP-ELD framework.

At AR step $i$, MP-ELD constructs three step-wise conditioning vectors:
\begin{itemize}
\item \emph{Local-continuity conditioning} $c^{\mathrm{lc}}_i$: a local continuation anchor derived from the current token, mainly responsible for enforcing short-range continuity.
\item \emph{Self-consistency conditioning} $c^{\mathrm{sc}}_i$: an in-context summary of modality-internal evolution over the prefix, mainly responsible for long-horizon stability and internal attribute consistency.
\item \emph{Alignment-consistency conditioning} $c^{\mathrm{ac}}_i$: a condition-aware signal mainly responsible for aligning the generated output with the external control signal.
\end{itemize}
All three are represented in a shared \(D_{\mathrm{model}}\)-dimensional conditioning space. For notational clarity, we use \(\bar c_i^{(\cdot)}\) to denote the decoder-side version of each conditioning vector, and write a path configuration as a tuple only to indicate which components are present.

\paragraph{Information modeling paths.}
We construct three conditioning sources using three lightweight token encoders and two LMs:
\begin{itemize}
\item A \emph{local encoder} $E_{\mathrm{lc}}$ produces the local-continuity signal from the current token:
\begin{equation}
c^{\mathrm{lc}}_i = E_{\mathrm{lc}}(x_i^{\mathrm{ctx}}) \in \mathbb{R}^{D_{\mathrm{model}}}.
\end{equation}
This can be viewed as a local-continuation path with minimal self-conditioning.

\item A \emph{self-consistency encoder} $E_{\mathrm{sc}}$ and a \emph{self-consistency LM} $\mathrm{LM}_{\mathrm{sc}}$ produce the self-consistency signal from the history tokens:
\begin{equation}
h^{\mathrm{sc}}_{i}=E_{\mathrm{sc}}(x_i^{\mathrm{ctx}}),
\qquad
c^{\mathrm{sc}}_{1:i}=\mathrm{LM}_{\mathrm{sc}}(h^{\mathrm{sc}}_{1:i}),
\qquad
c^{\mathrm{sc}}_i\in\mathbb{R}^{D_{\mathrm{model}}}.
\end{equation}
This can be viewed as a global-continuation path with full self-conditioning over the token history.

\item An \emph{alignment-consistency encoder} $E_{\mathrm{ac}}$ and an \emph{alignment-consistency LM} $\mathrm{LM}_{\mathrm{ac}}$ produce the alignment signal by integrating the external condition with the history tokens:
\begin{equation}
h^{\mathrm{ac}}_{i} = E_{\mathrm{ac}}(x_i^{\mathrm{ctx}}),
\qquad
c^{\mathrm{ac}}_{1:i}
=
\mathrm{LM}_{\mathrm{ac}}\big([\phi(c);\, h^{\mathrm{ac}}_{1:i}]\big),
\qquad
c^{\mathrm{ac}}_i\in\mathbb{R}^{D_{\mathrm{model}}}.
\end{equation}
This can be viewed as a cross-modal, externally conditioned global-continuation path.
\end{itemize}

These three paths can be intuitively interpreted as follows:
\begin{itemize}
    \item $c^{\mathrm{lc}}_i$ encourages smooth local continuation of short-range acoustic attributes, such as pitch, phase, and energy, and helps prevent abrupt discontinuities.
    \item $c^{\mathrm{sc}}_i$ captures slowly varying acoustic attributes that should remain stable over an utterance, such as timbre, overall energy profile, and accent or style.
    \item $c^{\mathrm{ac}}_i$ indicates where the model is in the conditioned content and therefore what should be generated next, e.g., which part of the text should be spoken at the current step.
\end{itemize}

We augment the alignment-consistency LM with an additional scalar head for stop detection. Concretely, in addition to the step-wise alignment vector $c^{\mathrm{ac}}_i$, the alignment-consistency LM outputs a scalar
\[
\pi_i\in[0,1]
\]
representing the probability of continuing generation after step \(i\). During training, we supervise \(\pi_i\) with a binary cross-entropy loss against the ground-truth continuation label derived from the sequence length. During inference, we terminate decoding when \(\pi_i<0.5\).

\paragraph{Orthogonal residual conditioning.}
To encourage a residual, non-overlapping decomposition of conditioning information, we orthogonalize the three condition vectors via a Gram--Schmidt transform and normalize each component to a fixed radius. For a target radius \(r=\sqrt{D_{\mathrm{model}}}\), let
\[
\operatorname{Norm}_{r}(u)
=
r\frac{u}{\|u\|_2+\epsilon}.
\]
At AR step \(i\), we construct the decoder-side conditioning components as
\begin{align}
\bar c^{\mathrm{lc}}_i
&=
\operatorname{Norm}_{\sqrt{D_{\mathrm{model}}}}
\left(c^{\mathrm{lc}}_i\right),
\\
\bar c^{\mathrm{sc}}_i
&=
\operatorname{Norm}_{\sqrt{D_{\mathrm{model}}}}
\left(
c^{\mathrm{sc}}_i
-
\frac{
\langle c^{\mathrm{sc}}_i,\bar c^{\mathrm{lc}}_i\rangle
}{
\|\bar c^{\mathrm{lc}}_i\|_2^2+\epsilon
}
\bar c^{\mathrm{lc}}_i
\right),
\\
\bar c^{\mathrm{ac}}_i
&=
\operatorname{Norm}_{\sqrt{D_{\mathrm{model}}}}
\left(
c^{\mathrm{ac}}_i
-
\frac{
\langle c^{\mathrm{ac}}_i,\bar c^{\mathrm{lc}}_i\rangle
}{
\|\bar c^{\mathrm{lc}}_i\|_2^2+\epsilon
}
\bar c^{\mathrm{lc}}_i
-
\frac{
\langle c^{\mathrm{ac}}_i,\bar c^{\mathrm{sc}}_i\rangle
}{
\|\bar c^{\mathrm{sc}}_i\|_2^2+\epsilon
}
\bar c^{\mathrm{sc}}_i
\right).
\end{align}
Intuitively, this orthogonalization operation encourages each condition vector to encode \emph{residual information} that the preceding condition vectors do not contain. This allows us to define the full-path condition as the \emph{summation} of the orthogonalized components instead of the \emph{concatenation} of the original vectors, which also controls the overall conditioning dimension and the related model complexity when \(D_{\mathrm{model}}\) is large. 

\paragraph{Multi-path residual CFG.}
Using the orthogonalized components defined above, we apply CFG dropout only to the \emph{non-local} branches, while always keeping the local-continuity anchor \(\bar c_i^{\mathrm{lc}}\) present. This preserves an always-available local continuation signal throughout training. For a given path configuration, we form the additive path condition
\begin{equation}
\tilde c_i
=
\bar c_i^{\mathrm{lc}}
+
\delta_i^{\mathrm{sc}}\bar c_i^{\mathrm{sc}}
+
\delta_i^{\mathrm{ac}}\bar c_i^{\mathrm{ac}},
\qquad
\delta_i^{\mathrm{sc}},\delta_i^{\mathrm{ac}}\in\{0,1\}.
\label{eq:mpeld_additive_path_condition}
\end{equation}
The decoder receives the concatenation of the additive path condition and the bridge-time embedding \(e_{\mathrm{time}}(\tau)\), denoted by \(q_i(\tau)\). We use three path configurations in this work:
\begin{itemize}
\item \(L\): \emph{l}ocal-continuity only,
\[
\tilde c_i^{L}
=
\bar c_i^{\mathrm{lc}},
\qquad
q_i^{L}(\tau)
=
\left[
\tilde c_i^{L}\ ;\
e_{\mathrm{time}}(\tau)
\right];
\]
\item \(LS\): \emph{l}ocal-continuity and \emph{s}elf-consistency,
\[
\tilde c_i^{LS}
=
\bar c_i^{\mathrm{lc}}
+
\bar c_i^{\mathrm{sc}},
\qquad
q_i^{LS}(\tau)
=
\left[
\tilde c_i^{LS}\ ;\
e_{\mathrm{time}}(\tau)
\right];
\]
\item \(LSA\): \emph{l}ocal-continuity, \emph{s}elf-consistency, and \emph{a}lignment-consistency,
\[
\tilde c_i^{LSA}
=
\bar c_i^{\mathrm{lc}}
+
\bar c_i^{\mathrm{sc}}
+
\bar c_i^{\mathrm{ac}},
\qquad
q_i^{LSA}(\tau)
=
\left[
\tilde c_i^{LSA}\ ;\
e_{\mathrm{time}}(\tau)
\right].
\]
\end{itemize}
During training, we sample the three paths with probabilities
\[
\Pr(L)=0.1,\qquad
\Pr(LS)=0.1,\qquad
\Pr(LSA)=0.8.
\]
At inference time, for each conditioning path, we run the decoder with the corresponding concatenated condition \(q_i^{L}(\tau)\), \(q_i^{LS}(\tau)\), or \(q_i^{LSA}(\tau)\), convert its endpoint prediction into a tangent velocity estimate using the endpoint-to-velocity construction described earlier, and then combine these tangent velocity estimates linearly. Following the residual decomposition of these three conditioning paths, we perform \emph{multi-path residual CFG}:
\begin{equation}
v_{\tau}
=
v_{\tau}^{L}
+
\lambda_{\mathrm{sc}}\big(v_{\tau}^{LS}-v_{\tau}^{L}\big)
+
\lambda_{\mathrm{ac}}\big(v_{\tau}^{LSA}-v_{\tau}^{LS}\big),
\label{eq:multipath_cfg_mpeld}
\end{equation}
where \(v_{\tau}^{LS}-v_{\tau}^{L}\) is the self-consistency residual relative to the local-continuity path, and \(v_{\tau}^{LSA}-v_{\tau}^{LS}\) is the alignment residual relative to the self-consistent path. When
\(\lambda_{\mathrm{sc}}=\lambda_{\mathrm{ac}}=1\), the inference reduces to the no-guidance case
\(v_{\tau}=v_{\tau}^{LSA}\), which matches the standard full-condition path.

Although each pathwise velocity is obtained from the endpoint-parameterized direction-only construction described earlier, we do not re-normalize the CFG-combined velocity \(v_\tau\) back to the fixed magnitude \(\pi R/2\). Since all pathwise velocities are tangent vectors at the same bridge state \(x_\tau\), their linear combination remains in the same tangent space:
\[
v_{\tau}^{L},\,v_{\tau}^{LS},\,v_{\tau}^{LSA}
\in
T_{x_\tau}\mathbb{S}^{D_{\mathrm{tok}}-1}(R)
\quad\Longrightarrow\quad
v_\tau\in T_{x_\tau}\mathbb{S}^{D_{\mathrm{tok}}-1}(R).
\]
Therefore, keeping the CFG-combined magnitude does not violate the spherical geometry. The reason for not fixing the magnitude after CFG is that a fixed velocity norm also fixes the total path length. If one re-normalizes the velocity to
\(\|v_\tau\|_2=\pi R/2\) for all \(\tau\in[0,1]\), then the generated trajectory has total length
\[
\int_0^1 \|v_\tau\|_2\,d\tau
=
\frac{\pi R}{2}.
\]
This is appropriate when the trajectory is close to the high-dimensional SFM geodesic, whose typical endpoint distance is approximately \(\pi R/2\). However, after CFG, the guided vector field may no longer follow the geodesic direction exactly; the resulting trajectory can be curved. A curved trajectory connecting the same conceptual endpoints generally requires a different path length, often larger than the geodesic length. Enforcing a fixed total length in this case can unnecessarily constrain the guided ODE and may prevent the trajectory from reaching the desired endpoint under the learned vector field. By allowing the CFG residuals to change the velocity magnitude, guidance can adjust not only the tangent direction but also the effective integration speed. This gives the guided trajectory additional flexibility while still preserving tangency to the sphere.

\paragraph{Time-dependent CFG guidance.}
Empirically, we find that extrapolating the self-consistency residual, i.e., using $\lambda_{\mathrm{sc}}>1$, is important for improving in-context and in-domain consistency, such as speaker-timbre consistency. At the same time, it is also the main source of distribution drift and long-horizon error accumulation. By contrast, extrapolating the alignment residual, i.e., using $\lambda_{\mathrm{ac}}>1$, is important for strengthening conditional controllability, and we do not observe comparable drift induced by this term. We further hypothesize that the instability associated with self-consistency extrapolation is mainly caused by an unreliable estimate of $v_{\tau}^{L}$ in the small-$\tau$ regime: the $L$ path contains only local continuation information and is therefore substantially less informative than the $LS$ path for estimating the flow output near the early-time bridge regime. As a result, the base point $v_{\tau}^{L}$ on which the self-consistency extrapolation is applied can be unreliable, and extrapolation from it may produce off-manifold intermediate states whose errors are then amplified through AR iteration. By contrast, since the alignment residual is defined relative to $v_{\tau}^{LS}$ rather than $v_{\tau}^{L}$, extrapolation along $v_{\tau}^{LSA}-v_{\tau}^{LS}$ is typically stable across the full bridge time range. Motivated by this observation, we use a time-dependent self-consistency guidance weight \(\lambda_{\mathrm{sc}}(\tau)\) to avoid extrapolating from an unreliable base in the early-time regime. Concretely, we start from \(\lambda_{\mathrm{sc}}(0)=1\) and gradually increase it to a predefined maximum value \(\lambda_{\mathrm{sc}}^{\max}\):
\begin{equation}
\lambda_{\mathrm{sc}}(\tau)
=
1+\big(\lambda_{\mathrm{sc}}^{\max}-1\big)s(\tau),
\qquad
s(0)=0,\quad s(1)=1,
\label{eq:lambda_sc_schedule}
\end{equation}
where \(s(\tau)\) is a non-decreasing schedule, e.g.,
\[
s(\tau)=\tau^\gamma,\qquad \gamma>0.
\]
By default, we keep \(\lambda_{\mathrm{ac}}\) constant.

\paragraph{Riemannian integration on the sphere.}
After CFG, the combined velocity remains a tangent vector at the current bridge state:
\[
v_{\tau}
\in
T_{x_\tau}\mathbb{S}^{D_{\mathrm{tok}}-1}(R).
\]
To preserve the spherical constraint during numerical integration, we update the bridge state using the Riemannian exponential map. For a step size \(\Delta\tau\), the update is
\begin{equation}
x_{\tau+\Delta\tau}
=
\operatorname{Exp}_{x_\tau}\!\left(\Delta\tau\,v_\tau\right),
\label{eq:riemannian_exp_update}
\end{equation}
where, for any tangent vector \(u\in T_x\mathbb{S}^{D_{\mathrm{tok}}-1}(R)\),
\begin{equation}
\operatorname{Exp}_{x}(u)
=
\cos\!\left(\frac{\|u\|_2}{R}\right)x
+
R\sin\!\left(\frac{\|u\|_2}{R}\right)
\frac{u}{\|u\|_2}.
\label{eq:sphere_exp_map}
\end{equation}

\paragraph{Bridge-time sampling.}
During training, the bridge time is sampled from a simple mixture distribution. With probability \(0.75\), we sample
\[
a \sim \mathcal{N}(-1,1),
\qquad
\tau = \sigma(a)=\frac{1}{1+\exp(-a)},
\]
i.e., \(\tau\) follows a logit-normal distribution biased toward the early, high-noise part of the bridge. With the remaining probability \(0.25\), we sample
\[
\tau \sim \mathrm{Unif}(0,1).
\]
This sampling rule allocates more training probability to small-\(\tau\) states, where the bridge state is closer to the noisy source and the denoising problem is more ambiguous, while the uniform component guarantees nonzero coverage over the entire bridge interval, including the moderate- and large-\(\tau\) low-noise regimes.
\section{Experiments and Results}
\label{sec:experiments}

In this section, we evaluate Locodec and MP-ELD from two complementary perspectives. First, we study whether the proposed token-space shaping mechanisms preserve reconstruction quality while changing the geometry and statistics of the latent space. Second, we evaluate whether these shaped representations are easier to predict, and whether MP-ELD maintains short-form generation quality while improving long-horizon stability.

\subsection{Experimental setup}

A central motivation of this work is to examine how much can be gained from representation-space and generation-framework design under moderate model size and restrained budgets relative to large industrial systems. We do not attempt to establish a general scaling law or to claim that design choices universally dominate data scale or model scale. Instead, by using comparatively restrained data and model sizes, and by avoiding external pretrained components and post-training stages, we aim to make the effects of token-space shaping and information routing easier to isolate. This setting allows us to test whether strong reconstruction quality and stable AR generation can be obtained without relying solely on larger datasets, larger models, or additional pretrained inductive biases.

\paragraph{Datasets and evaluation.}
We train Locodec on an internal bilingual speech dataset of second-scale utterances, and MP-ELD on an internal bilingual dataset consisting of both second-scale and minute-scale utterances, both from real-world recordings. We evaluate both tokenizer reconstruction and generative synthesis on the Seed-TTS-eval~\cite{anastassiou2024seed}, where the ZH subset contains \(2020\) utterances from DiDiSpeech 2~\cite{guo2021didispeech} and the EN subset contains \(1088\) utterances from Common Voice~\cite{ardila2020common}. In addition, we construct a smaller medium-length test set from real-world recordings (ZH only) for efficient CFG grid search, and a long-form test set from real-world recordings (ZH only) to analyze long-horizon AR error accumulation under different CFG configurations.

For tokenizer reconstruction, we report fidelity-oriented metrics, including Mel-cepstral distortion (MCD)\footnote{\url{https://github.com/chenqi008/pymcd}}~\cite{kubichek1993mel}, STOI\footnote{\url{https://github.com/mpariente/pystoi}}~\cite{taal2011algorithm}, and ViSQOL\footnote{\url{https://github.com/google/visqol}}~\cite{chinen2020visqol}. For both reconstruction and generation, we report task-oriented metrics, including word error rate (WER) and speaker similarity (SIM). The evaluation models and configurations for WER and SIM follow DiTAR~\cite{jia2025ditar}. On the long-form test set, we additionally split each generated utterance into non-overlapping \(10\)-second segments and report the segment-level SIM. We also provide spectral visualizations to illustrate the specific acoustic manifestations of AR error accumulation.

\paragraph{Locodec configurations.}
For \(24\)-kHz stereo audio, Locodec uses a \(512\)-point MDCT window as the signal front-end, and we group \(11\) adjacent MDCT frames and map them into one token. The resulting frame rate is approximately \(8.5\) Hz, which we refer to as \(8\)-Hz tokens for simplicity. The native high-dimensional token dimension is fixed to \(D_{\mathrm{tok}}=768\). The encoder consists of \(6\) FFN blocks followed by a single-step linear downsampling layer. The decoder consists of \(6\) causal ConvNeXt1D blocks on the token grid and \(12\) causal ConvNeXt1D blocks on the MDCT coefficient grid. All ConvNeXt1D blocks use kernel size \(7\) with causal left zero padding, and all FFNs in both the encoder and decoder are SwiGLU FFNs~\cite{shazeer2020glu}. Under this configuration, the encoder has \(59.5\)M parameters and \(5.1\)G MACs for a one-second input, while the decoder has \(180.9\)M parameters and \(12.3\)G MACs for a one-second input. For higher sampling rates or non-speech audio such as stereo music, the MDCT window size, token rate, and model size can be adjusted accordingly. We leave such extensions outside the scope of this paper.

We evaluate five core-manifold dimensions $d_{\mathrm{core}}\in\{768,256,64,32,16\}$, where \(d_{\mathrm{core}}=768\) is equivalent to not imposing a lower-dimensional bottleneck. For \(d_{\mathrm{core}}\in\{64,32,16\}\), we additionally train variants with postfix dimension dropout, abbreviated as PDD hereafter. This gives eight Locodec configurations in total. All tokenizers are trained on \(1\)-second audio clips with a global batch size of \(256\) seconds for \(250\)k iterations.

\paragraph{MP-ELD configurations.}
For each Locodec configuration, we train one corresponding MP-ELD model. The MP-ELD architecture contains three token encoders, each consisting of \(3\) FFN blocks, corresponding to the local-continuity, self-consistency, and alignment-consistency paths. The alignment-consistency LM is a \(12\)-layer RoFormer~\cite{su2024roformer}, the self-consistency LM is a \(3\)-layer RoFormer, and the flow-matching decoder is a \(3\)-block FFN network with adaLN-Zero conditioning~\cite{peebles2023scalable}. All FFNs are also SwiGLU FFNs. The hidden size of the encoders and LMs is \(1536\), the hidden size of the decoder is \(2048\), and the dimension of time embedding is \(256\). The full model has \(0.74\)B parameters. The three token encoders together contain \(180.6\)M parameters and require \(1.4\)G MACs for a one-second (\(8\)-token) input. The decoder contains \(127.7\)M parameters and requires \(1.0\)G MACs for a one-second input. All models are trained with a global token budget of \(200\)k tokens (including phonemes and audio embeddings) per iteration for \(400\)k iterations. We maintain an exponential moving average (EMA) version of the model for inference, with the EMA decay set to \(0.9995\) by default. Inference uses \(20\) NFEs with Riemannian exponential-map integration on the token sphere.

Following DiTAR, we use phonemes as the text front-end and adopt the same in-context learning (ICL) formulation. During training, each training example is serialized as a phoneme sequence followed by its corresponding audio-token sequence, i.e., \((\text{phoneme}, \text{audio})\). The flow-matching next-token prediction loss is applied only to the audio tokens. During inference, the model takes \((\text{prompt phoneme}, \text{target phoneme}, \text{prompt audio})\) as input and autoregressively continues from the prompt-audio tokens to generates the target-audio tokens.

In practical systems, pretrained components could be introduced at multiple stages of the MP-ELD pipeline. For example, the alignment-consistency path could use a pretrained text LM or a pretrained cross-modal alignment model, while the token encoders or condition encoders could be initialized from strong SSL or ASR models to provide additional semantic or acoustic inductive biases. These extensions are compatible with the MP-ELD formulation and may further improve performance. In this work, however, we deliberately avoid using pretrained models in any MP-ELD component and train all encoders, LMs, and decoders from scratch. This prevents external pretrained representations from dominating the information layout and allows us to test whether the proposed local-continuity, self-consistency, and alignment-consistency routing behavior can emerge from the training objective and architecture alone.

\subsection{Results on reconstruction}

We first evaluate whether the proposed token-space shaping mechanisms affect tokenizer reconstruction quality. Table~\ref{tab:locodec_reconstruction_main} reports full-dimensional token reconstruction quality on Seed-TTS-eval, where reconstruction is performed using the native \(768\)-dimensional high-dimensional token. Under our training configuration, imposing a low-dimensional core manifold does not degrade full-dimensional token reconstruction quality. Across \(d_{\mathrm{core}}\in\{768,256,64,32,16\}\) without PDD, all metrics remain nearly unchanged. In fact, the small variations across core dimensions are comparable to normal training variation, and no consistent loss of reconstruction fidelity is observed as the core dimension is reduced. This indicates that the low-dimensional reconstruction path can reshape the geometry of the high-dimensional token space without noticeably reducing the information preserved by the full native token. 

PDD has a slightly different effect. Compared with the corresponding non-PDD configurations, PDD leads to a mild degradation in MCD, suggesting a small increase in spectral mismatch. However, its effect on the remaining metrics is minimal: STOI and ViSQOL remain almost unchanged, and the task-oriented WER and SIM scores show only negligible differences. As all metrics except MCD are computed at \(16\) kHz, the fact that PDD mainly affects MCD while leaving STOI, ViSQOL, WER, and SIM essentially unchanged suggests that the induced coordinate-wise energy hierarchy primarily introduces a mild mismatch in less critical spectral details, especially in the middle- and high-frequency regions, while preserving the core low- and mid-frequency information needed for intelligibility, perceptual quality, and speaker identity.

\begin{figure}[!t]
    \centering
    \includegraphics[width=1.0\textwidth]{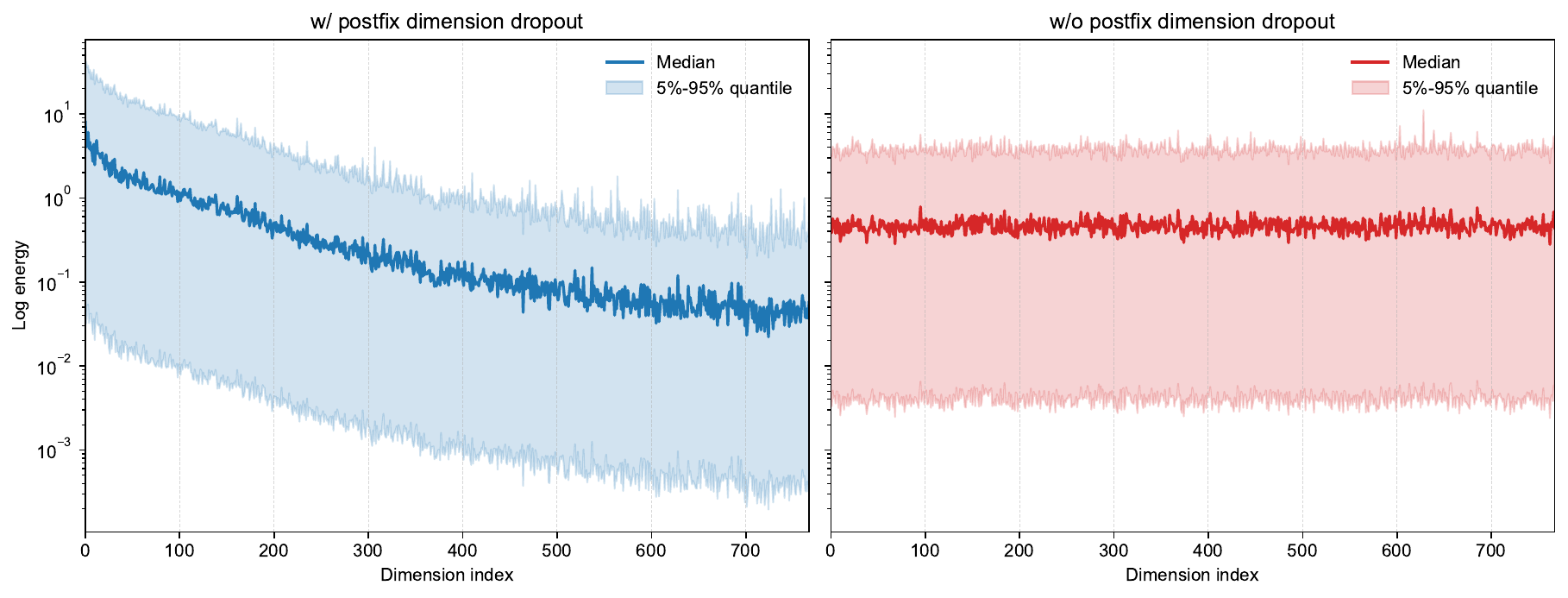}
    \caption{Coordinate-wise energy profiles of Locodec tokens. When combined with rotation noise injection, PDD automatically induces a stable prefix-to-postfix energy hierarchy, while models without postfix dropout keep an almost uniform energy distribution across dimensions.}
    \label{fig:locodec_energy_profile}
\end{figure}

To provide an intuitive view of the energy bias induced by PDD, Figure~\ref{fig:locodec_energy_profile} visualizes the average coordinate-wise token energy. Without PDD, the native high-dimensional token energy is nearly uniform across dimensions. With PDD, prefix dimensions acquire substantially larger energy, postfix dimensions acquire substantially smaller energy, and the logarithm of the per-dimension energy decays approximately linearly with the dimension index. This indicates that the combination of noise injection and PDD effectively induces a stable coordinate-wise energy bias through training dynamics, without explicitly prescribing a target energy profile.

\begin{table}[!t]
\centering
\small
\setlength{\tabcolsep}{3pt}
\caption{Full-dimensional token reconstruction quality of different Locodec configurations on Seed-TTS-eval. Reconstruction is performed using the native \(768\)-dimensional high-dimensional token.}
\label{tab:locodec_reconstruction_main}
\begin{tabular}{cc|ccccc|ccccc}
\toprule
\multirow{2}{*}{\(d_{\mathrm{core}}\)}
& \multirow{2}{*}{PDD}
& \multicolumn{5}{c|}{ZH}
& \multicolumn{5}{c}{EN} \\
\cline{3-12}
& 
& MCD \(\downarrow\)
& STOI \(\uparrow\)
& ViSQOL \(\uparrow\)
& WER (\%) \(\downarrow\)
& SIM \(\uparrow\)
& MCD \(\downarrow\)
& STOI \(\uparrow\)
& ViSQOL \(\uparrow\)
& WER (\%) \(\downarrow\)
& SIM \(\uparrow\) \\
\midrule
768 & \(\times\)
    & 2.32 & 0.98 & 4.65 & 1.34 & 0.740
    & 2.56 & 0.98 & 4.65 & 2.18 & 0.714 \\
256 & \(\times\)
    & 2.38 & 0.98 & 4.64 & 1.32 & 0.740
    & 2.64 & 0.98 & 4.64 & 2.16 & 0.713 \\
\multirow{2}{*}{64}
    & \(\times\)
    & 2.27 & 0.98 & 4.68 & 1.33 & 0.740
    & 2.51 & 0.98 & 4.68 & 2.20 & 0.715 \\
    & \(\checkmark\)
    & 2.62 & 0.97 & 4.61 & 1.37 & 0.736
    & 2.85 & 0.98 & 4.63 & 2.26 & 0.711 \\
\multirow{2}{*}{32}
    & \(\times\)
    & 2.22 & 0.98 & 4.68 & 1.32 & 0.741
    & 2.44 & 0.98 & 4.67 & 2.15 & 0.716 \\
    & \(\checkmark\)
    & 2.60 & 0.97 & 4.63 & 1.36 & 0.737
    & 2.84 & 0.98 & 4.63 & 2.14 & 0.712 \\
\multirow{2}{*}{16}
    & \(\times\)
    & 2.21 & 0.98 & 4.68 & 1.33 & 0.742
    & 2.42 & 0.98 & 4.68 & 2.13 & 0.717 \\
    & \(\checkmark\)
    & 2.51 & 0.98 & 4.63 & 1.34 & 0.738
    & 2.72 & 0.98 & 4.64 & 2.19 & 0.713 \\
\bottomrule
\end{tabular}
\end{table}

\begin{table}[!t]
\centering
\small
\setlength{\tabcolsep}{3pt}
\caption{Reconstruction from restricted representations for PDD-enabled Locodec models. ``Core'' denotes reconstruction from the lifted low-dimensional token. ``Prefix-\(K\)'' denotes reconstruction after keeping only the first \(K\) dimensions of the native high-dimensional token and zeroing out the remaining dimensions.}
\label{tab:locodec_low_prefix_reconstruction}
\begin{tabular}{cc|ccccc|ccccc}
\toprule
\multirow{2}{*}{\(d_{\mathrm{core}}\)}
& \multirow{2}{*}{Repr.}
& \multicolumn{5}{c|}{ZH}
& \multicolumn{5}{c}{EN} \\
\cline{3-12}
&
& MCD \(\downarrow\)
& STOI \(\uparrow\)
& ViSQOL \(\uparrow\)
& WER (\%) \(\downarrow\)
& SIM \(\uparrow\)
& MCD \(\downarrow\)
& STOI \(\uparrow\)
& ViSQOL \(\uparrow\)
& WER (\%) \(\downarrow\)
& SIM \(\uparrow\) \\
\midrule
\multirow{5}{*}{16}
& Core
    & 6.53 & 0.83 & 3.15 & 15.81 & 0.396
    & 6.63 & 0.83 & 3.29 & 23.00 & 0.284 \\
& Prefix-16
    & 7.93 & 0.75 & 2.78 & 62.90 & 0.334
    & 8.09 & 0.76 & 2.86 & 72.37 & 0.164 \\
& Prefix-32
    & 6.45 & 0.84 & 3.23 & 18.63 & 0.411
    & 6.62 & 0.84 & 3.36 & 24.65 & 0.295 \\
& Prefix-64
    & 5.11 & 0.90 & 3.71 & 3.86 & 0.540
    & 5.35 & 0.90 & 3.83 & 6.46 & 0.482 \\
& Full & 2.51 & 0.98 & 4.63 & 1.34 & 0.738
    & 2.72 & 0.98 & 4.64 & 2.19 & 0.713 \\
\midrule
\multirow{5}{*}{32}
& Core
    & 5.24 & 0.89 & 3.60 & 4.92 & 0.484
    & 5.47 & 0.89 & 3.74 & 7.69 & 0.417 \\
& Prefix-16
    & 8.03 & 0.74 & 2.72 & 60.58 & 0.335
    & 8.17 & 0.75 & 2.85 & 72.11 & 0.160 \\
& Prefix-32
    & 6.35 & 0.83 & 3.26 & 18.30 & 0.426
    & 6.47 & 0.84 & 3.41 & 24.42 & 0.304 \\
& Prefix-64
    & 5.09 & 0.89 & 3.70 & 4.14 & 0.540
    & 5.30 & 0.90 & 3.82 & 6.21 & 0.477 \\
& Full & 2.60 & 0.97 & 4.63 & 1.36 & 0.737
    & 2.84 & 0.98 & 4.63 & 2.14 & 0.712 \\
\midrule
\multirow{5}{*}{64}
& Core
    & 4.50 & 0.92 & 3.96 & 2.47 & 0.595
    & 4.77 & 0.92 & 4.06 & 3.87 & 0.553 \\
& Prefix-16
    & 7.93 & 0.74 & 2.77 & 61.94 & 0.344
    & 7.96 & 0.75 & 2.91 & 68.41 & 0.187 \\
& Prefix-32
    & 6.31 & 0.83 & 3.27 & 16.92 & 0.426
    & 6.44 & 0.84 & 3.42 & 26.25 & 0.306 \\
& Prefix-64
    & 5.00 & 0.90 & 3.73 & 4.10 & 0.548
    & 5.21 & 0.90 & 3.85 & 5.98 & 0.483 \\
& Full & 2.62 & 0.97 & 4.61 & 1.37 & 0.736
    & 2.85 & 0.98 & 4.63 & 2.26 & 0.711 \\
\bottomrule
\end{tabular}
\end{table}

We further examine the relationship between the explicitly learned low-dimensional core and the prefix subspace induced by PDD. This analysis is meaningful only for PDD-enabled tokenizers, because without PDD the coordinate order is not trained to carry a prefix-to-postfix availability hierarchy. We therefore evaluate the three PDD-enabled configurations with \(d_{\mathrm{core}}\in\{16,32,64\}\). For each configuration, we compare reconstruction from three restricted representations: the lifted low-dimensional core token, the prefix-\(K\) subspace of the native high-dimensional token, and the full native token. For prefix-\(K\) reconstruction, we keep only the first \(K\) coordinates of the native token and set all remaining coordinates to zero before feeding the token to the decoder, without renormalization. This matches the form of the training-time postfix dropout corruption.

The results are reported in Table~\ref{tab:locodec_low_prefix_reconstruction}. They show that the prefix subspaces indeed form a functional hierarchy: within each PDD-enabled tokenizer, reconstruction quality improves consistently from Prefix-\(16\) to Prefix-\(32\) and then to Prefix-\(64\). This confirms that the energy profile induced by PDD is not merely a statistical artifact, but corresponds to an actual ordering of decodable information, in a manner loosely analogous to the coarse-to-fine information hierarchy in RVQ. At the same time, the core dimension is not equivalent to the prefix dimension: for the same nominal dimensionality, Core-\(d\) consistently outperforms Prefix-\(d\), even though the low-dimensional reconstruction path is down-weighted during training. This indicates that the low-dimensional path maintains an effective shaping pressure on the token space. We also observe that the reconstruction quality of a fixed prefix length is largely stable across different choices of \(d_{\mathrm{core}}\). For example, Prefix-\(64\) gives very similar reconstruction metrics for \(d_{\mathrm{core}}=16,32,\) and \(64\). This suggests that the native coordinate hierarchy is mainly determined by the PDD availability bias, whereas the choice of \(d_{\mathrm{core}}\) primarily affects the quality of the explicit core representation. Finally, Prefix-\(K\) is not constrained to have a fixed norm within its active \(K\)-dimensional subspace: after masking, its norm varies and its coordinate energies are highly non-uniform. In this sense, it is geometrically less constrained than the fixed-radius core representation and in principle has an additional radial degree of freedom. Nevertheless, Prefix-\(K\) remains weaker than Core-\(K\) at the same nominal dimension, including at \(K=64\). A different dropout schedule or substantially longer training might improve fixed-prefix reconstruction, but we leave this direction outside the scope of the present study. Taken together, these results suggest that the low-dimensional constraint and PDD play complementary roles, and that they do not introduce severe conflicts that would compromise full-dimensional token reconstruction quality.

\begin{figure}[!t]
    \centering
    \includegraphics[width=1.0\textwidth]{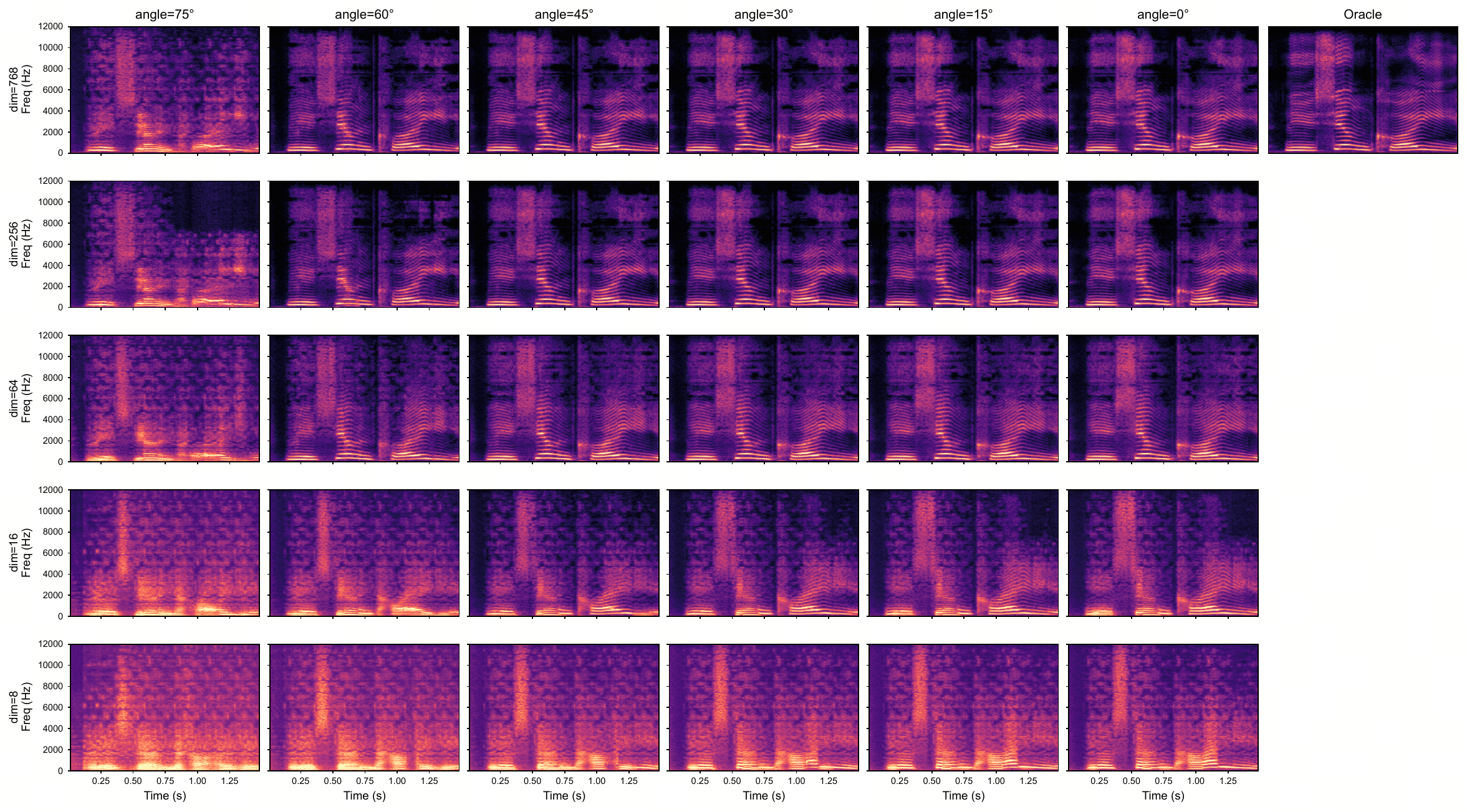}
    \caption{Spectral visualization of Locodec reconstructions under angular token corruption and prefix restriction. Columns correspond to different rotation angles and rows correspond to different retained prefix dimensions.}
    \label{fig:locodec_rotation_prefix_spectrogram}
\end{figure}

As discussed in Section~\ref{sec:methodology}, PDD is not used as an isolated masking trick: its availability bias is combined with spherical noise injection, so that dimensions that remain available more frequently are also encouraged to become more identifiable under angular corruption. To provide an intuitive view of the reconstruction behavior induced by this joint training scheme, Figure~\ref{fig:locodec_rotation_prefix_spectrogram} visualizes reconstructions from the PDD-enabled tokenizer with \(d_{\mathrm{core}}=16\) under different rotation angles and retained prefix dimensions. We rotate the high-dimensional token by angles \(\theta\in\{0^\circ,15^\circ,30^\circ,45^\circ,60^\circ,75^\circ\}\), and decode from prefix dimensions \(K\in\{8,16,64,256,768\}\). The visualization shows how reconstruction quality changes as angular corruption becomes stronger and as fewer prefix dimensions are retained, providing a qualitative view of both decoder robustness and the PDD-induced information hierarchy. Notably, the full-dimensional token still preserves reasonably good spectral quality even under large angular perturbations, such as \(60^\circ\) in the visualization. This behavior is encouraged by the training-time spherical corruption, whose maximum rotation angle is \(90^\circ\), and suggests that the decoder learns a relatively large reconstruction-stable basin around each token. However, as discussed earlier, learning such large basins in an unconstrained high-dimensional space can encourage different tokens to become widely separated, or even nearly orthogonal, thereby weakening interpolatability. The low-dimensional core constraint is therefore important for preventing robustness from relying purely on high-dimensional angular separation. We return to this point in the generation results below.

\subsection{Results on generation}

\begin{figure}[!t]
    \centering
    \includegraphics[width=1.0\textwidth]{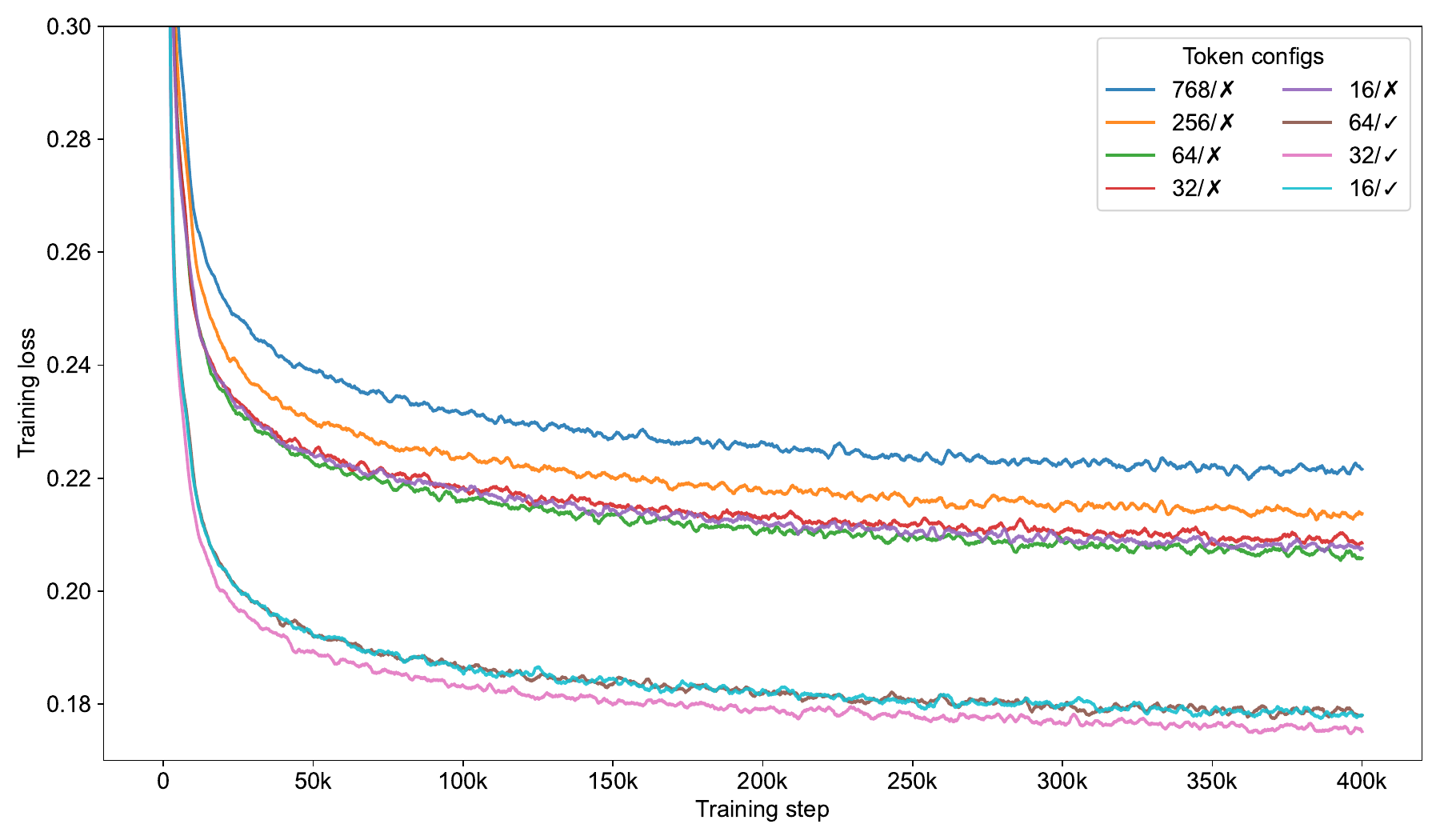}
    \caption{Training curves of the MP-ELD cosine direction loss for different Locodec configurations. The notation \(d/\mathrm{PDD}\) denotes the core dimension and whether PDD is used. All MP-ELD models use the same architecture and training setup. Faster convergence and lower final loss indicate that the corresponding token representation is easier to predict under the fixed generator configuration.}
    \label{fig:generation_loss_curves}
\end{figure}

We next evaluate how different Locodec representations affect AR continuous-token generation. For each of the eight tokenizer configurations, we train an MP-ELD model with the same architecture, training data, optimization setup, and training budget. Under this controlled setting, the convergence behavior and absolute value of the generative-model training loss provide an empirical proxy for token predictability: if the same generator converges faster and reaches a lower training loss on one token representation than on another, then the former representation can be regarded as easier to model under the fixed generator configuration. 

Figure~\ref{fig:generation_loss_curves} shows the MP-ELD cosine direction loss curves for the eight Locodec configurations. The first clear trend is that imposing a low-dimensional core manifold substantially improves predictability. The no-bottleneck configuration \(768/\times\) converges the slowest and reaches the highest final loss. Reducing the core dimension to \(256\) already lowers the loss noticeably, and further reducing it to \(64\), \(32\), or \(16\) gives an additional improvement. However, this improvement is not monotonic without limit. Once the core dimension becomes sufficiently small, the gain largely saturates: the non-PDD configurations with \(d_{\mathrm{core}}\in\{64,32,16\}\) converge to similar losses. This suggests that the low-dimensional constraint helps organize the high-dimensional token space into a more predictable geometry, but making the core dimension arbitrarily small does not continue to provide proportional benefits. 

The second and more pronounced trend is the effect of PDD. For all core dimensions, PDD-enabled models converge faster and reach substantially lower losses than their non-PDD counterparts. This indicates that the coordinate-wise energy bias induced by PDD provides an additional and strong improvement in single-token predictability beyond the low-dimensional manifold constraint. Among the PDD-enabled configurations, \(d_{\mathrm{core}}=32\) achieves the lowest final loss, while \(d_{\mathrm{core}}=16\) and \(64\) are slightly worse. Thus, from the perspective of generator-side predictability, the non-PDD and PDD comparisons together provide empirical support for our assumption that a moderate low-dimensional core manifold can make a high-dimensional token space easier to model, and that the additional identifiability induced by PDD can further improve this predictability.

\begin{figure}[!t]
    \centering
    \includegraphics[width=1.0\textwidth]{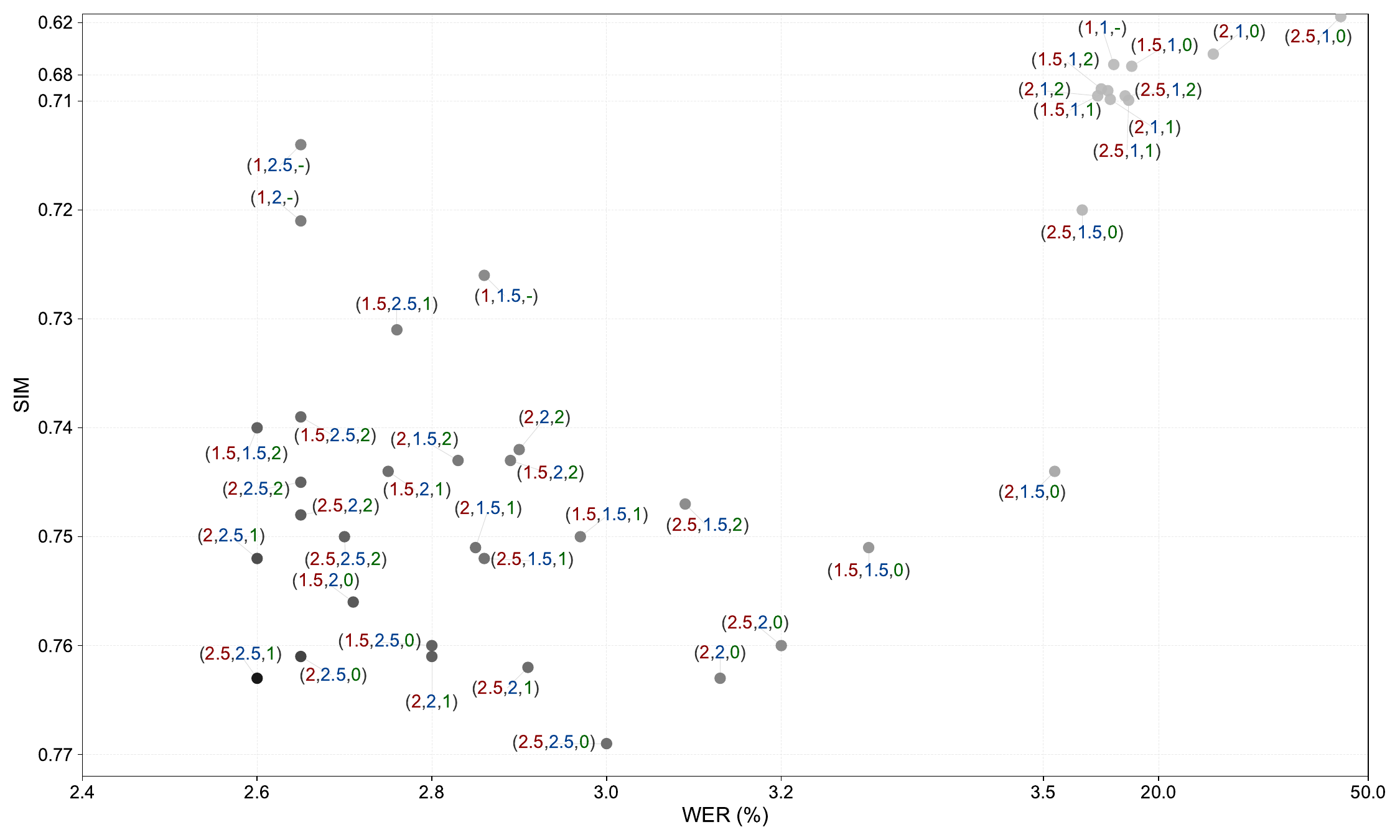}
    \caption{WER/SIM trade-off under different CFG configurations on the internal CFG-selection set. Each point corresponds to one setting from the sweep over \((\lambda_{\mathrm{sc}}^{\max}, \lambda_{\mathrm{ac}}, \gamma)\). The scatter shows that alignment-consistency guidance mainly affects WER, while self-consistency guidance mainly affects SIM, and that the bridge-time schedule controls the trade-off induced by self-consistency extrapolation. Darker points indicate configurations closer to the preferred low-WER, high-SIM region.}
    \label{fig:cfg_sweep}
\end{figure}

Since the \(d_{\mathrm{core}}=32\), PDD-enabled tokenizer achieves the lowest training loss among all tested configurations, we use it as the default representation to study the effect of CFG configurations on the balance between WER and SIM. We sweep CFG hyperparameters on a smaller internal evaluation set over a predefined grid:
\[
\lambda_{\mathrm{sc}}^{\max}\in\{1,1.5,2,2.5\},
\qquad
\lambda_{\mathrm{ac}}\in\{1,1.5,2,2.5\},
\qquad
\gamma\in\{0,1,2\}.
\]
Here \(\gamma=0\) denotes constant self-consistency guidance, i.e., \(s(\tau)\equiv 1\), while \(\gamma>0\) uses the delayed schedule \(s(\tau)=\tau^\gamma\). The setting
\(\lambda_{\mathrm{sc}}^{\max}=\lambda_{\mathrm{ac}}=1\) corresponds to the standard full-condition path \(v_\tau=v_\tau^{LSA}\) without CFG. When \(\lambda_{\mathrm{sc}}^{\max}=1\), the choice of \(\gamma\) has no effect. This gives a total of \(40\) configurations.

Figure~\ref{fig:cfg_sweep} shows the resulting WER/SIM trade-off. The first clear trend is that the alignment-consistency guidance scale is the main factor controlling whether the model enters a content-aligned generation regime. When \(\lambda_{\mathrm{sc}}^{\max}=1\), increasing \(\lambda_{\mathrm{ac}}\) from \(1\) to \(1.5\), \(2\), and \(2.5\) moves the model from a high-WER region to a much lower-WER region, with WER decreasing from above \(10\%\) to roughly the \(2.6\)--\(2.9\%\) range. In this process, SIM first improves and then degrades, indicating that stronger alignment guidance mainly improves content following, but excessive alignment extrapolation may start to trade off against acoustic similarity. This behavior is consistent with the intended role of the alignment-consistency residual \(v_\tau^{LSA}-v_\tau^{LS}\): it primarily acts as an external-condition alignment correction rather than a general acoustic-consistency correction.

The effect of self-consistency guidance is different. In the valid low-WER region, increasing \(\lambda_{\mathrm{sc}}^{\max}\) tends to improve SIM more directly than WER. For example, at comparable alignment strength, increasing the self-consistency scale from \(1.5\) to \(2\) or \(2.5\) can move SIM from the mid-\(0.74\) range to around \(0.76\), while WER changes more mildly and remains within the same general low-WER regime. This supports the interpretation that the self-consistency residual \(v_\tau^{LS}-v_\tau^{L}\) mainly strengthens modality-internal acoustic consistency, including speaker identity and slowly varying acoustic state.

However, stronger self-consistency guidance does not imply a monotonic SIM improvement. When the alignment-consistency scale is insufficient, such as \(\lambda_{\mathrm{ac}}=1\), increasing \(\lambda_{\mathrm{sc}}^{\max}\) does not rescue the high WER. Moreover, SIM also becomes non-monotonic and can drop substantially under some configurations. This suggests that if the model is not properly aligned to the external condition, amplifying self-consistency residuals may instead reinforce an unstable acoustic trajectory. In such cases, accumulated acoustic distortion can degrade not only intelligibility but also the speaker-similarity metric itself.

The schedule parameter \(\gamma\) further controls how aggressively self-consistency guidance is applied over the bridge time. For the same \((\lambda_{\mathrm{sc}}^{\max},\lambda_{\mathrm{ac}})\), constant guidance with \(\gamma=0\) often gives stronger SIM, but can also move the point away from the best WER region. For example, under strong guidance scales such as \((\lambda_{\mathrm{sc}}^{\max},\lambda_{\mathrm{ac}})=(2.5,2.5)\), using \(\gamma=0\) yields very high SIM but a higher WER, while delayed schedules such as \(\gamma=1\) or \(\gamma=2\) reduce early-time self-consistency extrapolation and give a more balanced WER/SIM trade-off. This is consistent with the motivation of the time-dependent guidance schedule: self-consistency guidance is useful, but applying it too strongly in the early high-noise part of the bridge can make the residual extrapolation less reliable.

Taken together, the sweep provides empirical evidence that MP-ELD learns a meaningful degree of information routing from scratch. Nevertheless, changing \(\lambda_{\mathrm{ac}}\) and \(\lambda_{\mathrm{sc}}^{\max}\) produces different and interpretable movements in the WER/SIM plane: the alignment-consistency residual mainly affects content alignment, while the self-consistency residual mainly affects acoustic and speaker consistency. The grid search also identifies a set of Pareto-competitive CFG configurations, from which we select the best overall WER/SIM trade-off for comparison with prior systems.

\begin{figure}[!t]
    \centering
    \includegraphics[width=1.0\textwidth]{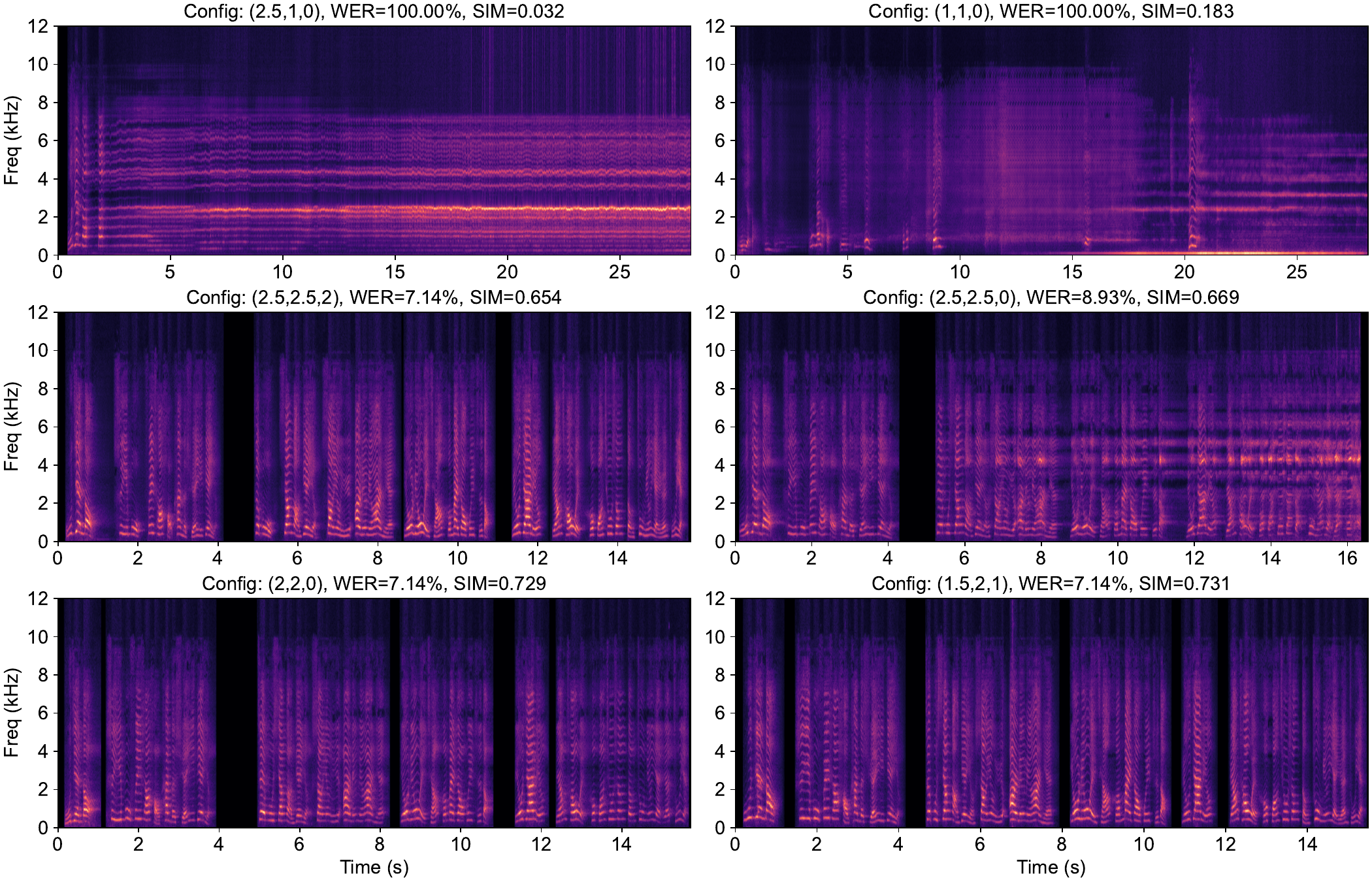}
    \caption{Spectrograms of the same utterance generated under different CFG configurations. Each title reports \((\lambda_{\mathrm{sc}}^{\max},\lambda_{\mathrm{ac}},\gamma)\), WER, and SIM for the generated sample. The six examples are ordered from top-left to bottom-right by increasing SIM. The visualization shows that different CFG configurations lead to different forms of AR error accumulation, including early collapse, delayed spectral drift, harmonic repetition, and frequency-band energy drift.}
    \label{fig:ar_drift_spectrum}
\end{figure}

To better understand what these WER/SIM differences correspond to acoustically, we visualize one representative utterance generated with different CFG configurations in Figure~\ref{fig:ar_drift_spectrum}. All six spectrograms are generated from the same input sample using the same model and tokenizer described above, differing only in the CFG configuration. The examples are ordered from top-left to bottom-right by increasing SIM for this utterance: the first row shows the two configurations with the lowest SIM, the second row shows two middle-SIM configurations, and the third row shows the two configurations with the highest SIM. Compared with the WER/SIM scatter plot alone, this visualization gives a more direct view of how different guidance choices affect the acoustic structure of the generated signal.

We can first observe that severe AR error accumulation clearly affects the metrics: when strong drift or collapse is visible, at least one of WER and SIM becomes poor. At the same time, the onset time of error accumulation is not fixed: in the first row, degradation appears very early and dominates most of the generated waveform, while in other configurations such as the middle-right and bottom-left examples, the generation is initially reasonable but begins to drift after several seconds. The acoustic manifestation of drift is also not unique: we observe elongated or locally repeated harmonic structures, gradually increasing noise-like energy, and frequency bands whose energy slowly increases or decreases over time. Note that other failure modes, such as gradually increasing speaking rate or global loudness drift, are also observed in practice but are omitted here for space.

The first row illustrates the importance of sufficient alignment-consistency guidance. Both examples use \(\lambda_{\mathrm{ac}}=1\) and \(\gamma=0\), and both show severe error accumulation. The no-extrapolation configuration \((1,1,0)\) already fails to maintain a valid content-aligned trajectory on this sample, leading to WER \(=100\%\). Increasing self-consistency guidance alone, as in \((2.5,1,0)\), does not fix this failure; instead, the generation collapses into a long, acoustically self-reinforcing trajectory with even lower SIM. This supports the interpretation from the CFG sweep that the self-consistency residual cannot replace the alignment-consistency residual. Without sufficient alignment guidance, amplifying self-consistency may reinforce an incorrect acoustic continuation rather than recover the intended content.

The second row isolates the effect of the time schedule \(\gamma\). The two configurations have the same guidance scales, \((\lambda_{\mathrm{sc}}^{\max},\lambda_{\mathrm{ac}})=(2.5,2.5)\), and differ only in the self-consistency schedule. With constant self-consistency guidance, \(\gamma=0\), the spectrogram starts to show visible drift around \(7\) seconds: energy around the \(4\)--\(5\) kHz region gradually increases, and neighboring frequency bands also exhibit different degrees of energy amplification or attenuation. With delayed guidance, \(\gamma=2\), the spectrum remains more stable over the utterance. Interestingly, the \(\gamma=0\) configuration has a slightly higher utterance-level SIM despite having a higher WER and more visible spectral distortion. This indicates that utterance-level SIM alone is not sufficient to judge the actual acoustic quality or stability of generated speech.

The third row further shows that even similar WER and SIM values do not necessarily imply identical acoustic behavior. The two configurations obtain almost the same utterance-level metrics, but the bottom-left example still shows mild frequency-band drift after around \(7\) seconds, especially near the \(3\) kHz and \(6\) kHz regions, whereas the bottom-right example remains more stable throughout the utterance. This suggests that automatic metrics such as WER and SIM are useful but incomplete summaries of generation quality, as they may fail to capture subtle spectral drift or slowly accumulating acoustic artifacts even when the final utterance-level scores are similar. 

\begin{figure}[!t]
    \centering
    \includegraphics[width=1.0\textwidth]{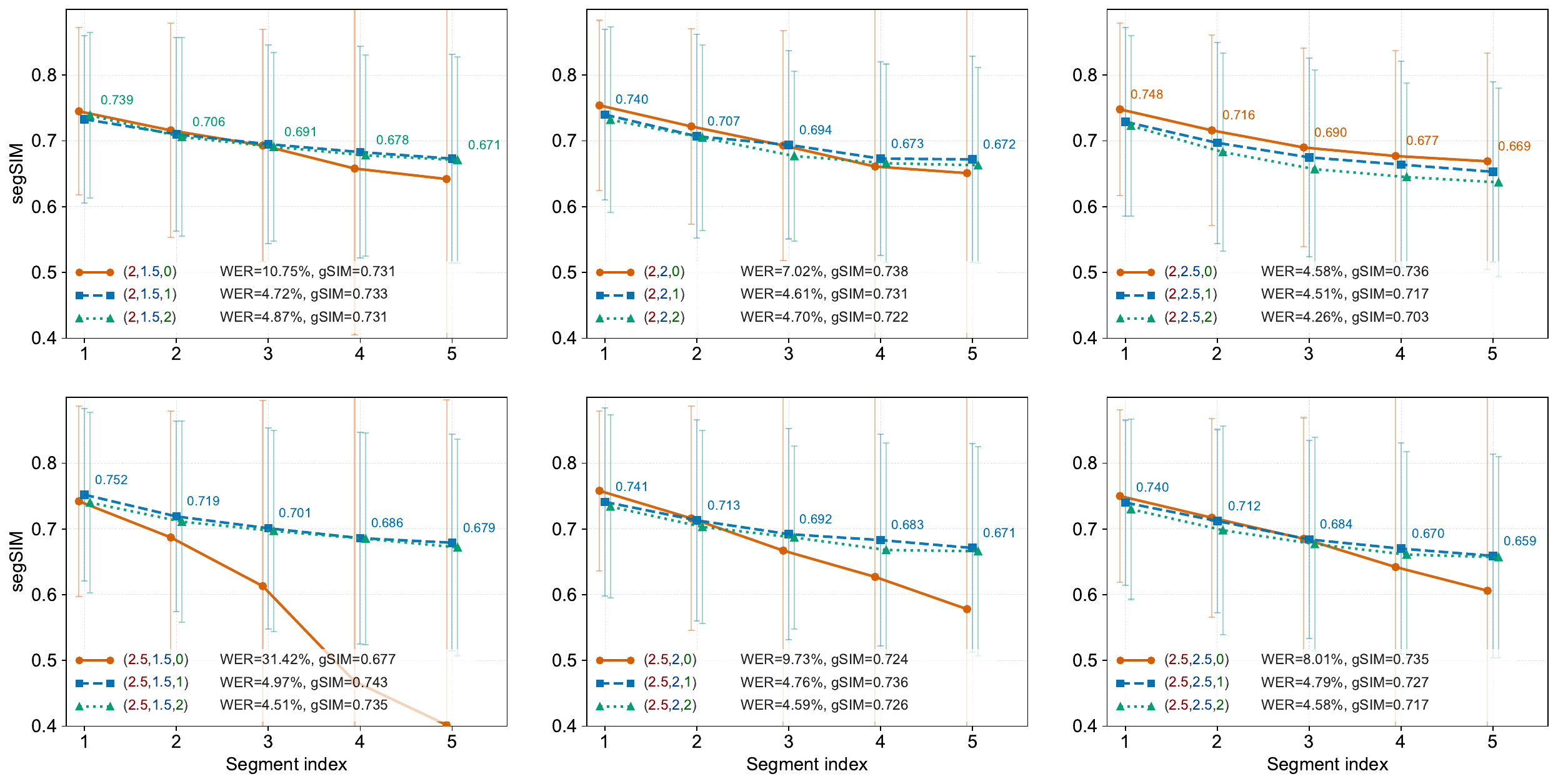}
    \caption{Long-form segment-level SIM under representative CFG configurations. Each generated utterance is split into five non-overlapping \(10\)-second segments. Each subplot fixes \((\lambda_{\mathrm{sc}}^{\max},\lambda_{\mathrm{ac}})\) and compares different \(\gamma\) values. Curves show dataset-level mean segment SIM, and error bars indicate \(95\%\) confidence intervals. The legend reports global WER and global SIM. The annotated values correspond to the configuration with the best overall stability--quality trade-off in each subplot.}
    \label{fig:longform_segment_sim}
\end{figure}

To further examine this limitation, we evaluate long-form generation using both global-level and segment-level metrics on the long-form evaluation set. Based on the CFG sweep above, we select \(18\) representative configurations from the core operating region,
\[
\lambda_{\mathrm{sc}}^{\max}\in\{2,2.5\},
\qquad
\lambda_{\mathrm{ac}}\in\{1.5,2,2.5\},
\qquad
\gamma\in\{0,1,2\}.
\]
For each generated utterance, we keep the first \(50\) seconds, split it into five non-overlapping \(10\)-second segments, and compute SIM for each segment. Figure~\ref{fig:longform_segment_sim} reports the dataset-level mean segment SIM together with \(95\%\) confidence intervals. Each subplot fixes \((\lambda_{\mathrm{sc}}^{\max},\lambda_{\mathrm{ac}})\) and compares different \(\gamma\) values. We also report the global WER and global SIM (gSIM) in the legend. For readability, in each subplot we annotate the segment-wise mean SIM values of the configuration with the best overall stability--quality trade-off.

The first observation is that constant self-consistency guidance, i.e., \(\gamma=0\), is substantially less stable in long-form generation, especially when \(\lambda_{\mathrm{sc}}^{\max}\) is large. For example, with \((\lambda_{\mathrm{sc}}^{\max},\lambda_{\mathrm{ac}},\gamma)=(2.5,1.5,0)\), the segment SIM drops from \(0.742\) in the first segment to \(0.402\) in the last segment, and the global WER increases to \(31.42\%\). Similar but less extreme degradation is also observed for \((2.5,2,0)\) and \((2.5,2.5,0)\). In contrast, delayed schedules with \(\gamma>0\) greatly reduce this long-horizon decay. For the same \((\lambda_{\mathrm{sc}}^{\max},\lambda_{\mathrm{ac}})=(2.5,1.5)\), changing \(\gamma\) from \(0\) to \(1\) improves the last-segment SIM from \(0.402\) to \(0.679\), while reducing WER from \(31.42\%\) to \(4.97\%\). This suggests that the bridge-time schedule is a key factor in preventing self-consistency guidance from becoming a source of accumulated drift. At the same time, while a larger \(\gamma\) typically improves the long-horizon stability and especially the spectrogram quality, it also delays self-consistency extrapolation more aggressively and may reduce the overall speaker-similarity level. In four of the six subplots, the best stability--quality trade-off is obtained with \(\gamma=1\), which provides a middle ground: it avoids strong self-consistency extrapolation in the early high-noise regime while still applying sufficient self-consistency guidance later in the bridge. This is consistent with the medium-length spectrogram examples above, where delayed guidance reduces spectral drift without completely suppressing the benefit of self-consistency guidance.

The curves further illustrate the different roles of \(\lambda_{\mathrm{sc}}^{\max}\) and \(\lambda_{\mathrm{ac}}\). Stronger self-consistency guidance can produce higher similarity near the beginning of an utterance, but it may also increase the risk of later drift. For example, \((2.5,2,0)\) achieves a very high first-segment SIM of \(0.758\), but its last-segment SIM drops to \(0.578\). By contrast, a more balanced scheduled configuration such as \((2,2,1)\) starts from a slightly lower first-segment SIM of \(0.740\), but decays more slowly and remains at \(0.672\) in the last segment. Thus, the highest initial similarity is not necessarily a good indicator of long-form stability. This also shows why global SIM alone can be misleading: two configurations may have very similar utterance-level SIM but very different temporal behavior. For instance, \((2.5,2.5,0)\) and \((2.5,2,1)\) have nearly identical global SIM values, \(0.735\) and \(0.736\), respectively. However, their last-segment SIM values differ substantially: \(0.606\) for \((2.5,2.5,0)\) versus \(0.671\) for \((2.5,2,1)\). The former therefore hides a much stronger long-horizon degradation behind a similar global score. Moreover, although Figure~\ref{fig:longform_segment_sim} reports only WER and segment-level SIM, our qualitative inspection shows phenomena consistent with Figure~\ref{fig:ar_drift_spectrum}: even configurations with similar segment-level SIM can exhibit noticeably different spectral quality. This further indicates that long-form audio generation should not be evaluated solely by a small set of automatic metrics, and a more complete evaluation should combine signal-level analysis, perceptual listening, and task-oriented automatic metrics. Empirically, we find that configurations with \(\lambda_{\mathrm{ac}}\ge \lambda_{\mathrm{sc}}^{\max}\) tend to be safer for long-form generation, especially in terms of spectral and temporal stability, although they often sacrifice some global or segment-level SIM. Conversely, configurations with \(\lambda_{\mathrm{sc}}^{\max}>\lambda_{\mathrm{ac}}\) can achieve stronger short-range or early-segment speaker similarity, but are more sensitive to the guidance schedule and more prone to accumulated drift. This suggests that the optimal CFG operating point may depend on the target generation length: short utterances may prefer stronger self-consistency guidance, whereas long-form generation may require more conservative or more alignment-dominant guidance. A natural extension is to use dynamic CFG schedules that vary not only with bridge time \(\tau\), but also with AR generation time. For example, one may gradually adjust the relative strength of self-consistency and alignment-consistency guidance as generation proceeds. We leave this direction for future work.

\begin{table}[!t]
\centering
\small
\setlength{\tabcolsep}{2.2pt}
\caption{Generation performance of different Locodec token configurations under three representative CFG settings. CFG-L denotes the most stable \emph{l}ong-form configuration, \((\lambda_{\mathrm{sc}}^{\max},\lambda_{\mathrm{ac}},\gamma)=(2,2,1)\). CFG-M denotes the best overall configuration selected on the \emph{m}edium-length CFG search set, \((\lambda_{\mathrm{sc}}^{\max},\lambda_{\mathrm{ac}},\gamma)=(2.5,2.5,1)\). CFG-S denotes the configuration with the highest first-segment (\emph{s}hort-horizon) SIM on the long-form set, \((\lambda_{\mathrm{sc}}^{\max},\lambda_{\mathrm{ac}},\gamma)=(2.5,2,0)\).}
\label{tab:generation_tokenizer_compare}
\begin{tabular}{cc|cccc|ccccccc}
\toprule
\multirow{3}{*}{Token}
& \multirow{3}{*}{CFG}
& \multicolumn{4}{c|}{Seed-TTS-eval}
& \multicolumn{7}{c}{Long-form set} \\
\cline{3-13}
&
& \multicolumn{2}{c|}{ZH}
& \multicolumn{2}{c|}{EN}
& \multirow{2}{*}{WER (\%) \(\downarrow\)}
& \multirow{2}{*}{gSIM \(\uparrow\)}
& \multicolumn{5}{c}{SegSIM} \\
\cline{3-6}\cline{9-13}
&
& WER (\%) \(\downarrow\)
& SIM \(\uparrow\)
& WER (\%) \(\downarrow\)
& SIM \(\uparrow\)
& 
& 
& Seg1 \(\uparrow\)
& Seg2 \(\uparrow\)
& Seg3 \(\uparrow\)
& Seg4 \(\uparrow\)
& Seg5 \(\uparrow\) \\
\midrule

\multirow{3}{*}{\(768/\times\)}
& L & 1.31 & 0.683 & 1.90 & 0.622 & 7.82 & 0.716 & 0.732 & 0.694 & 0.670 & 0.644 & 0.639 \\
& M & 1.12 & 0.683 & 1.85 & 0.624 & 8.36 & 0.710 & 0.731 & 0.696 & 0.669 & 0.628 & 0.624 \\
& S & 2.04 & 0.691 & 2.15 & 0.625 & 25.35 & 0.687 & 0.743 & 0.677 & 0.623 & 0.522 & 0.448 \\
\midrule

\multirow{3}{*}{\(256/\times\)}
& L & 1.11 & 0.682 & 1.86 & 0.614 & 5.81 & 0.729 & 0.737 & 0.705 & 0.687 & 0.675 & \bf{0.680} \\
& M & 1.04 & 0.686 & 1.88 & 0.618 & 4.84 & 0.725 & 0.742 & 0.701 & 0.687 & \bf{0.677} & 0.667 \\
& S & 1.46 & 0.692 & 5.23 & 0.620 & 15.94 & 0.714 & 0.747 & 0.696 & 0.643 & 0.568 & 0.512 \\
\midrule

\multirow{3}{*}{\(64/\times\)}
& L & 1.09 & 0.683 & 1.94 & 0.620 & 4.91 & 0.717 & 0.726 & 0.702 & 0.688 & 0.669 & 0.660 \\
& M & 1.50 & 0.686 & 1.82 & 0.627 & 4.69 & 0.716 & 0.728 & 0.697 & 0.676 & 0.662 & 0.666 \\
& S & 1.56 & 0.691 & 2.26 & 0.625 & 9.68 & 0.713 & 0.738 & 0.690 & 0.654 & 0.611 & 0.566 \\
\midrule

\multirow{3}{*}{\(32/\times\)}
& L & 1.07 & 0.675 & 2.08 & 0.608 & 4.64 & 0.708 & 0.721 & 0.683 & 0.671 & 0.656 & 0.642 \\
& M & 1.12 & 0.679 & 2.16 & 0.615 & 5.44 & 0.705 & 0.723 & 0.688 & 0.656 & 0.630 & 0.621 \\
& S & 1.52 & 0.683 & 2.47 & 0.615 & 20.80 & 0.674 & 0.734 & 0.664 & 0.583 & 0.487 & 0.419 \\
\midrule

\multirow{3}{*}{\(16/\times\)}
& L & 1.25 & 0.672 & 1.95 & 0.603 & 6.23 & 0.714 & 0.726 & 0.696 & 0.680 & 0.648 & 0.646 \\
& M & 1.30 & 0.678 & 1.93 & 0.609 & 9.49 & 0.702 & 0.729 & 0.686 & 0.658 & 0.633 & 0.594 \\
& S & 2.58 & 0.677 & 2.98 & 0.607 & 34.77 & 0.630 & 0.740 & 0.661 & 0.510 & 0.413 & 0.273 \\
\midrule

\multirow{3}{*}{\(64/\checkmark\)}
& L & 1.11 & 0.680 & 2.00 & 0.616 & 4.96 & 0.726 & 0.734 & 0.705 & 0.690 & 0.673 & 0.668 \\
& M & 1.12 & 0.683 & 1.87 & 0.620 & 4.78 & 0.720 & 0.729 & 0.697 & 0.684 & 0.667 & 0.663 \\
& S & 1.54 & 0.695 & 2.05 & \bf{0.630} & 9.63 & \bf{0.748} & \bf{0.762} & \bf{0.720} & 0.683 & 0.628 & 0.574 \\
\midrule

\multirow{3}{*}{\(32/\checkmark\)}
& L & \bf{0.95} & 0.687 & 1.87 & 0.615 & 4.61 & 0.731 & 0.740 & 0.707 & \bf{0.694} & 0.673 & 0.672 \\
& M & 0.99 & 0.691 & 1.80 & 0.622 & 4.79 & 0.727 & 0.740 & 0.712 & 0.684 & 0.670 & 0.659 \\
& S & 1.24 & 0.697 & 2.09 & 0.628 & 9.73 & 0.724 & 0.758 & 0.716 & 0.667 & 0.627 & 0.578 \\
\midrule

\multirow{3}{*}{\(16/\checkmark\)}
& L & 1.66 & 0.688 & 1.94 & 0.619 & 4.80 & 0.726 & 0.732 & 0.700 & 0.686 & 0.674 & 0.676 \\
& M & 1.29 & 0.691 & \bf{1.75} & 0.624 & \bf{4.55} & 0.718 & 0.734 & 0.702 & 0.684 & 0.666 & 0.670 \\
& S & 1.81 & \bf{0.698} & 2.47 & \bf{0.630} & 10.18 & 0.725 & 0.746 & 0.708 & 0.680 & 0.649 & 0.601 \\

\bottomrule
\end{tabular}
\end{table}

\begin{table}[!t]
\centering
\small
\caption{Generation results on Seed-TTS-eval. For MP-ELD, we report the result under its best CFG configuration from Table~\ref{tab:generation_tokenizer_compare}. All benchmark systems are AR models, and we select the version without post-training when applicable. The ``Tok./LM rate'' column denotes the native token frame rate and the effective frame rate seen by the LM, respectively.}
\label{tab:generation_seedtts}
\begin{tabular}{l|c|cc|cc}
\toprule
\multirow{2}{*}{Model}
& \multirow{2}{*}{Tok./LM rate (Hz)}
& \multicolumn{2}{c|}{ZH}
& \multicolumn{2}{c}{EN} \\
\cline{3-6}
&
& WER (\%) \(\downarrow\)
& SIM \(\uparrow\)
& WER (\%) \(\downarrow\)
& SIM \(\uparrow\) \\
\midrule
Human
& \textemdash
& 1.25 & 0.755
& 2.14 & 0.734 \\
\midrule
CosyVoice 2~\cite{du2024cosyvoice}
& 25/25
& 1.45 & 0.748
& 2.57 & 0.652 \\
CosyVoice 3-1.5B~\cite{du2025cosyvoice}
& 25/25
& 1.12 & 0.781
& 2.21 & 0.720 \\
FireRedTTS~\cite{guo2024fireredtts}
& 25/25
& 1.51 & 0.635
& 3.82 & 0.460 \\
FireRedTTS-2~\cite{xie2025fireredtts}
& 12.5/12.5
& 1.14 & 0.736
& 1.95 & 0.665 \\
Spark TTS~\cite{wang2025spark}
& 50/50
& 1.20 & 0.672
& 1.98 & 0.584 \\
VibeVoice~\cite{peng2025vibevoice}
& 7.5/7.5
& 1.16 & 0.744
& 3.04 & 0.689 \\
DiTAR~\cite{jia2025ditar}
& 40/10
& 1.02 & 0.753
& 1.69 & 0.735 \\
VoxCPM2~\cite{zhou2026voxcpm2}
& 25/6.25
& 0.97 & 0.795
& 1.84 & 0.753 \\
dots. tts~\cite{lian2026dots}
& 25/6.25
& 0.96 & \bf{0.805}
& \bf{1.34} & \bf{0.768} \\
\midrule
MP-ELD, \(64/\checkmark\)
& \(8/8\)
& 1.54 & 0.695
& 2.05 & 0.630 \\
MP-ELD, \(32/\checkmark\)
& \(8/8\)
& \bf{0.95} & 0.687
& 1.87 & 0.615 \\
MP-ELD, \(16/\checkmark\)
& \(8/8\)
& 1.29 & 0.691
& 1.75 & 0.624 \\
\bottomrule
\end{tabular}
\end{table}

The preceding CFG analyses are based on the \(32/\checkmark\) tokenizer configuration. We now extend the comparison to the other seven token configurations and examine whether the above observations remain consistent. Table~\ref{tab:generation_tokenizer_compare} compares the eight Locodec token configurations under three representative CFG settings: CFG-L is the relatively most stable long-form configuration, CFG-M is the best overall configuration selected on the medium-length CFG search set, and CFG-S is the configuration that favors short-horizon SIM.

Several observations follow. First, the high-dimensional configurations without a very low-dimensional core are still learnable. In particular, the \(768/\times\) and \(256/\times\) tokenizers already obtain competitive short-form WER on Seed-TTS-eval, and the \(256/\times\) configuration remains reasonably stable on the long-form set under CFG-L and CFG-M. This suggests that the Locodec design itself, including spherical normalization, strong angular corruption, and decoder robustness training, already imposes useful structure on the high-dimensional token space. The low-dimensional core and PDD further improve predictability and stability, but they are not the only source of learnability.

Second, all token configurations achieve relatively strong WER on Seed-TTS-eval, and this may be partly explained by the native low frame rate of Locodec. At \(8\) Hz, each token covers roughly \(125\) ms of audio, which is close to the duration scale of phoneme-level content. Thus, it seems that a low-frame-rate token naturally aggregates information over a content-relevant temporal span even without external SSL or ASR supervisions. This is consistent with the restricted-reconstruction results in Table~\ref{tab:locodec_low_prefix_reconstruction}, where a \(64\)-dimensional core can already support low WER, while its SIM remains much lower than full-dimensional token reconstruction. This suggests that the core representation is relatively content-centric and less complete acoustically. From this perspective, native low frame rate itself may act as a \emph{semantic inductive bias}, reducing the need for externally imposed semantic alignment losses.

Third, the short-form and long-form results differ substantially. The effective generated durations on Seed-TTS-eval are only \(5.28\pm0.96\) s for ZH and \(4.18\pm1.16\) s for EN, whereas the long-form set has an effective duration of \(52.14\pm3.04\) s. The Seed-TTS-eval results therefore mainly measure short-form quality and do not fully expose AR error accumulation. As shown in Figure~\ref{fig:ar_drift_spectrum}, except for the two examples that exhibit severe error accumulation almost from the beginning, the remaining failure cases can still maintain stable synthesis for at least the first \(6\) seconds, corresponding to roughly \(50\) tokens, before visible drift emerges. This delayed failure mode is also reflected in Table~\ref{tab:generation_tokenizer_compare}, where many token and CFG configurations obtain similar WER on Seed-TTS-eval, but their long-form behavior differs sharply. Moreover, there also appears to be a dataset effect in the absolute SIM values: the first-segment SIM on the long-form set is typically about \(0.05\) higher than the ZH SIM on Seed-TTS-eval, and the best SIM observed on the medium-length CFG-selection set in Figure~\ref{fig:cfg_sweep} is nearly \(0.07\) higher. This may partly reflect a domain mismatch rather than only a duration effect. Since our training data are dominated by real-world recordings, whereas Seed-TTS-eval is constructed from curated benchmark datasets, differences in recording conditions, speaker characteristics, content style, and reference-audio distribution may therefore affect SIM evaluation. This provides one possible explanation for why the model appears stronger on internal real-world evaluation sets while showing a larger SIM gap on the standard benchmark.

Fourth, a moderate core dimension gives the best overall balance. Very small core dimensions provide stronger constraints and can help WER, but may reduce acoustic capacity; larger core dimensions preserve more information but are less constrained. Across training loss, short-form generation, and long-form stability, \(32/\checkmark\) is the most balanced configuration: it achieves the best ZH WER on Seed-TTS-eval, \(0.95\%\), and strong long-form performance under CFG-L, with WER \(4.61\%\), gSIM \(0.731\), and a stable segment-SIM curve ending at \(0.672\). This agrees with the prediction-loss curves in Figure~\ref{fig:generation_loss_curves}, where \(32/\checkmark\) also reaches the lowest final direction loss.

Finally, PDD consistently improves the generation behavior, especially for long-form stability and SIM. Its effect is most visible under the more aggressive CFG-S setting. For example, changing \(32/\times\) to \(32/\checkmark\) reduces long-form WER from \(20.80\%\) to \(9.73\%\), increases gSIM from \(0.674\) to \(0.724\), and improves the fifth-segment SIM from \(0.419\) to \(0.578\). Similarly, changing \(16/\times\) to \(16/\checkmark\) under CFG-S improves the fifth-segment SIM from \(0.273\) to \(0.601\). These results indicate that the PDD-induced coordinate hierarchy improves not only the optimization loss but also the robustness of AR rollout, supporting the proposed identifiability mechanism.

Table~\ref{tab:generation_seedtts} further compares MP-ELD with prior AR TTS systems on Seed-TTS-eval. The proposed system is highly competitive in WER, while its SIM scores remain lower than those of the strongest prior systems. One possible explanation is that the token frame rate acts not only as an efficiency parameter, but also as a \emph{temporal-resolution knob}, loosely analogous to the analysis-window length in time--frequency analysis. A lower token frame rate means that each token summarizes a longer span of audio, similar to how a larger FFT window provides a longer analysis context and higher frequency resolution, but becomes less sensitive to rapid temporal variation. Such longer-span aggregation can make phonetic or content-level structure more stable and can shorten the AR horizon, which may help WER and long-horizon stability. At the same time, it may reduce sensitivity to fine-grained local acoustic variations, such as micro-prosody, transient spectral details, and short-time speaker-specific cues, which are important for speaker similarity. Conversely, higher-rate tokens, or product-structured local representations that preserve multiple short-range sub-tokens within each LM step, can retain more local acoustic detail and therefore often achieve stronger SIM, but they usually require longer native sequences or additional local prediction modules.

However, VibeVoice suggests that this temporal-resolution interpretation might be incomplete, as it uses an even lower native token frame rate of \(7.5\) Hz but achieves substantially stronger SIM than MP-ELD. We hypothesize that this difference may come from its explicit semantic--acoustic tokenizer design. In such a factorized representation, the low-rate semantic component can still provide the long-span content aggregation and AR-stability benefits associated with low frame rate, while a more independent acoustic component can preserve speaker-specific and fine-grained acoustic information. By contrast, Locodec uses a single reconstruction-first continuous token space, so content, speaker identity, and local acoustic detail must be organized within the same low-rate high-dimensional token. Under AR prediction, this unified space may naturally favor content-stable components, helping WER, while making fine-grained acoustic similarity harder to model. This suggests a potential direction for future tokenizer design: introducing semantic--acoustic factorization into a low-rate continuous token space, possibly still without relying on explicit SSL or ASR supervision, may preserve the content clustering and stability benefits of native low-frame-rate tokens while improving acoustic fidelity and speaker similarity.
\section{Conclusion and Future Work}
\label{sec:conclusion}

In this paper, we studied whether low-frame-rate, high-dimensional continuous tokens can serve as stable targets for autoregressive speech generation. Our main finding is that such tokens are viable when the representation space and the generative framework are designed jointly. We proposed Locodec, a locally encoded tokenizer which shapes a spherical high-dimensional token space around a lower-dimensional core manifold and induces a coordinate-wise energy hierarchy through PDD, improving predictability without noticeably degrading full-dimensional token reconstruction quality. We further proposed MP-ELD, which separates local-continuity, self-consistency, and alignment-consistency information pathways, allowing residual CFG to control acoustic consistency and content alignment more explicitly. The experiments show that reconstruction quality alone is not sufficient to characterize a generative representation, and token-space geometry strongly affects generator training loss, CFG behavior, and long-form stability. A moderate core dimension combined with PDD provides the best overall balance, and the resulting \(8\)-Hz, \(768\)-dimensional continuous tokens support competitive WER without external SSL/ASR models, pretrained text LMs, or post-training stages. At the same time, the remaining SIM gap to the strongest prior systems suggests a semantic--acoustic trade-off: low frame rate appears to favor content aggregation and AR stability, but may make fine-grained acoustic and speaker-similarity modeling harder. We therefore view token frame rate not only as an efficiency parameter, but also as a semantic--acoustic resolution knob.

Several extensions are natural. First, although Locodec is formulated as a general stereo-audio tokenizer framework and can be instantiated for higher sampling rates such as \(48\) kHz, the experiments in this paper focus primarily on \(24\)-kHz speech reconstruction and the TTS task. We plan to further evaluate the Locodec framework on broader audio understanding and generation tasks and even more general multimodal settings. The same principles may also be relevant beyond audio: low-rate high-dimensional continuous tokens with shaped latent geometry could be useful for image, video, or other sequence generation problems where reconstruction capacity and AR stability must be balanced.

Second, our deliberate avoidance of pretrained SSL, ASR, and language models is a design choice for isolating the effects of token-space shaping and information routing, not a restriction of the method. Locodec and MP-ELD are naturally compatible with pretrained components. For example, the alignment-consistency LM could be initialized from or replaced by a pretrained text LM, which may strengthen the semantic bias of the alignment path and thereby alter the role of the self-consistency path. Similarly, the token encoders for different MP-ELD pathways could be initialized differently, or one pathway could be guided by a pretrained SSL or ASR representation. More generally, information routing can emerge from training dynamics, as shown in this paper, but it can also be strengthened by architectural and initialization priors. Designing better information pathways, both inside the tokenizer and inside the generator, can potentially be an important direction to further mitigate or even eliminate AR error accumulation.

Third, improving speaker similarity and acoustic fidelity remains a central challenge. The current results suggest that low-frame-rate tokens naturally favor content aggregation, which helps WER but may make fine-grained acoustic modeling harder. One possible direction is to introduce semantic--acoustic factorization into the token space, still without relying on explicit SSL or ASR supervision. More broadly, for general audio generation, the relevant factors may go beyond semantics and acoustics to include spatial attributes, source identity, or temporal event dynamics. Developing tokenizers that can organize these factors in a controllable way, while preserving reconstruction fidelity and AR predictability, is a key open problem.

Fourth, scaling the individual components of the proposed framework is a natural next step. In particular, since the decoder is the actual next-token prediction module that estimates the pathwise velocity field, understanding its scaling behavior is particularly important. Compared with product-structured token approaches, where each LM step may require an additional local DiT to decode a group of short-range tokens, the FFN decoder in MP-ELD requires significantly fewer MACs to predict one native high-dimensional token. This leaves substantial computational headroom for scaling the decoder or replacing it with more expressive architectures while still keeping the overall inference cost competitive.

Finally, the long-form experiments in this paper are still limited relative to truly long streaming scenarios. Minute-scale generation already reveals clear AR error accumulation, but hour-scale streaming, multi-speaker interaction, podcast-level synthesis, and complex acoustic scenes may introduce additional failure modes. Future work should therefore study MP-ELD under longer and more diverse generation regimes. In addition, the experiments show that the optimal CFG configuration depends on utterance length, token representation, and the desired trade-off between WER, SIM, and stability. This suggests that CFG may not necessarily be time-invariant during AR generation. Dynamic guidance schedules that vary not only with bridge time \(\tau\), but also with AR generation time, may provide a more flexible way to maintain both short-horizon quality and long-horizon stability.

\clearpage

\bibliographystyle{plainnat}
\bibliography{main}

@article{chen2024diffusion,
  title={Diffusion forcing: Next-token prediction meets full-sequence diffusion},
  author={Chen, Boyuan and Mart{\'\i} Mons{\'o}, Diego and Du, Yilun and Simchowitz, Max and Tedrake, Russ and Sitzmann, Vincent},
  journal={Advances in Neural Information Processing Systems},
  volume={37},
  pages={24081--24125},
  year={2024}
}

@inproceedings{schmidt2019generalization,
  title={Generalization in generation: A closer look at exposure bias},
  author={Schmidt, Florian},
  booktitle={Proceedings of the 3rd Workshop on Neural Generation and Translation},
  pages={157--167},
  year={2019}
}

@article{van2017neural,
  title={Neural discrete representation learning},
  author={Van Den Oord, Aaron and Vinyals, Oriol and others},
  journal={Advances in neural information processing systems},
  volume={30},
  year={2017}
}

@inproceedings{liu2022convnet,
  title={A convnet for the 2020s},
  author={Liu, Zhuang and Mao, Hanzi and Wu, Chao-Yuan and Feichtenhofer, Christoph and Darrell, Trevor and Xie, Saining},
  booktitle={Proceedings of the IEEE/CVF conference on computer vision and pattern recognition},
  pages={11976--11986},
  year={2022}
}

@inproceedings{leng2025repa,
  title={{{REPA}}-{E}: Unlocking {VAE} for end-to-end tuning of latent diffusion transformers},
  author={Leng, Xingjian and Singh, Jaskirat and Hou, Yunzhong and Xing, Zhenchang and Xie, Saining and Zheng, Liang},
  booktitle={Proceedings of the IEEE/CVF International Conference on Computer Vision},
  pages={18262--18272},
  year={2025}
}

@article{chen2023flow,
  title={Flow matching on general geometries},
  author={Chen, Ricky TQ and Lipman, Yaron},
  journal={arXiv preprint arXiv:2302.03660},
  year={2023}
}

@article{anastassiou2024seed,
  title={Seed-{TTS}: A family of high-quality versatile speech generation models},
  author={Anastassiou, Philip and Chen, Jiawei and Chen, Jitong and Chen, Yuanzhe and Chen, Zhuo and Chen, Ziyi and Cong, Jian and Deng, Lelai and Ding, Chuang and Gao, Lu and others},
  journal={arXiv preprint arXiv:2406.02430},
  year={2024}
}

@article{zheng2025diffusion,
  title={Diffusion transformers with representation autoencoders},
  author={Zheng, Boyang and Ma, Nanye and Tong, Shengbang and Xie, Saining},
  journal={arXiv preprint arXiv:2510.11690},
  year={2025}
}

@article{tong2026scaling,
  title={Scaling Text-to-Image Diffusion Transformers with Representation Autoencoders},
  author={Tong, Shengbang and Zheng, Boyang and Wang, Ziteng and Tang, Bingda and Ma, Nanye and Brown, Ellis and Yang, Jihan and Fergus, Rob and LeCun, Yann and Xie, Saining},
  journal={arXiv preprint arXiv:2601.16208},
  year={2026}
}

@inproceedings{peebles2023scalable,
  title={Scalable diffusion models with transformers},
  author={Peebles, William and Xie, Saining},
  booktitle={Proceedings of the IEEE/CVF international conference on computer vision},
  pages={4195--4205},
  year={2023}
}

@article{jia2025ditar,
  title={Ditar: Diffusion transformer autoregressive modeling for speech generation},
  author={Jia, Dongya and Chen, Zhuo and Chen, Jiawei and Du, Chenpeng and Wu, Jian and Cong, Jian and Zhuang, Xiaobin and Li, Chumin and Wei, Zhen and Wang, Yuping and others},
  journal={arXiv preprint arXiv:2502.03930},
  year={2025}
}

@article{ning2023input,
  title={Input perturbation reduces exposure bias in diffusion models},
  author={Ning, Mang and Sangineto, Enver and Porrello, Angelo and Calderara, Simone and Cucchiara, Rita},
  journal={arXiv preprint arXiv:2301.11706},
  year={2023}
}

@inproceedings{
higgins2017betavae,
title={beta-{VAE}: Learning Basic Visual Concepts with a Constrained Variational Framework},
author={Irina Higgins and Loic Matthey and Arka Pal and Christopher Burgess and Xavier Glorot and Matthew Botvinick and Shakir Mohamed and Alexander Lerchner},
booktitle={International Conference on Learning Representations},
year={2017},
url={https://openreview.net/forum?id=Sy2fzU9gl}
}

@inproceedings{yin2025slow,
  title={From slow bidirectional to fast autoregressive video diffusion models},
  author={Yin, Tianwei and Zhang, Qiang and Zhang, Richard and Freeman, William T and Durand, Fredo and Shechtman, Eli and Huang, Xun},
  booktitle={Proceedings of the IEEE/CVF Conference on Computer Vision and Pattern Recognition},
  pages={22963--22974},
  year={2025}
}

@article{ke2025hyperspherical,
  title={Hyperspherical latents improve continuous-token autoregressive generation},
  author={Ke, Guolin and Xue, Hui},
  journal={arXiv preprint arXiv:2509.24335},
  year={2025}
}

@article{davidson2018hyperspherical,
  title={Hyperspherical variational auto-encoders},
  author={Davidson, Tim R and Falorsi, Luca and De Cao, Nicola and Kipf, Thomas and Tomczak, Jakub M},
  journal={arXiv preprint arXiv:1804.00891},
  year={2018}
}

@article{ruhe2024rolling,
  title={Rolling diffusion models},
  author={Ruhe, David and Heek, Jonathan and Salimans, Tim and Hoogeboom, Emiel},
  journal={arXiv preprint arXiv:2402.09470},
  year={2024}
}

@article{lee2023sequential,
  title={Sequential data generation with groupwise diffusion process},
  author={Lee, Sangyun and Lee, Gayoung and Kim, Hyunsu and Kim, Junho and Uh, Youngjung},
  journal={arXiv preprint arXiv:2310.01400},
  year={2023}
}

@article{huang2018music,
  title={Music transformer},
  author={Huang, Cheng-Zhi Anna and Vaswani, Ashish and Uszkoreit, Jakob and Shazeer, Noam and Simon, Ian and Hawthorne, Curtis and Dai, Andrew M and Hoffman, Matthew D and Dinculescu, Monica and Eck, Douglas},
  journal={arXiv preprint arXiv:1809.04281},
  year={2018}
}

@inproceedings{luo2020dual,
  title={Dual-path {RNN}: Efficient long sequence modeling for time-domain single-channel speech separation},
  author={Luo, Yi and Chen, Zhuo and Yoshioka, Takuya},
  booktitle={ICASSP 2020-2020 IEEE International Conference on Acoustics, Speech and Signal Processing (ICASSP)},
  pages={46--50},
  year={2020},
  organization={IEEE}
}

@inproceedings{shi2021emformer,
  title={Emformer: Efficient memory transformer based acoustic model for low latency streaming speech recognition},
  author={Shi, Yangyang and Wang, Yongqiang and Wu, Chunyang and Yeh, Ching-Feng and Chan, Julian and Zhang, Frank and Le, Duc and Seltzer, Mike},
  booktitle={ICASSP 2021-2021 IEEE International Conference on Acoustics, Speech and Signal Processing (ICASSP)},
  pages={6783--6787},
  year={2021},
  organization={IEEE}
}

@article{van2016wavenet,
  title={Wavenet: A generative model for raw audio},
  author={Van Den Oord, Aaron and Dieleman, Sander and Zen, Heiga and Simonyan, Karen and Vinyals, Oriol and Graves, Alex and Kalchbrenner, Nal and Senior, Andrew and Kavukcuoglu, Koray and others},
  journal={arXiv preprint arXiv:1609.03499},
  volume={12},
  number={1},
  year={2016}
}

@article{mehri2016samplernn,
  title={Sample{RNN}: An unconditional end-to-end neural audio generation model},
  author={Mehri, Soroush and Kumar, Kundan and Gulrajani, Ishaan and Kumar, Rithesh and Jain, Shubham and Sotelo, Jose and Courville, Aaron and Bengio, Yoshua},
  journal={arXiv preprint arXiv:1612.07837},
  year={2016}
}

@article{mao2020speech,
  title={Speech recognition and multi-speaker diarization of long conversations},
  author={Mao, Huanru Henry and Li, Shuyang and McAuley, Julian and Cottrell, Garrison},
  journal={arXiv preprint arXiv:2005.08072},
  year={2020}
}

@inproceedings{narayanan2019recognizing,
  title={Recognizing long-form speech using streaming end-to-end models},
  author={Narayanan, Arun and Prabhavalkar, Rohit and Chiu, Chung-Cheng and Rybach, David and Sainath, Tara N and Strohman, Trevor},
  booktitle={2019 IEEE automatic speech recognition and understanding workshop (ASRU)},
  pages={920--927},
  year={2019},
  organization={IEEE}
}

@inproceedings{raj2021integration,
  title={Integration of speech separation, diarization, and recognition for multi-speaker meetings: System description, comparison, and analysis},
  author={Raj, Desh and Denisov, Pavel and Chen, Zhuo and Erdogan, Hakan and Huang, Zili and He, Maokui and Watanabe, Shinji and Du, Jun and Yoshioka, Takuya and Luo, Yi and others},
  booktitle={2021 IEEE spoken language technology workshop (SLT)},
  pages={897--904},
  year={2021},
  organization={IEEE}
}

@inproceedings{battenberg2020location,
  title={Location-relative attention mechanisms for robust long-form speech synthesis},
  author={Battenberg, Eric and Skerry-Ryan, RJ and Mariooryad, Soroosh and Stanton, Daisy and Kao, David and Shannon, Matt and Bagby, Tom},
  booktitle={ICASSP 2020-2020 IEEE International Conference on Acoustics, Speech and Signal Processing (ICASSP)},
  pages={6194--6198},
  year={2020},
  organization={IEEE}
}

@inproceedings{han2021continuous,
  title={Continuous Speech Separation Using Speaker Inventory for Long Recording.},
  author={Han, Cong and Luo, Yi and Li, Chenda and Zhou, Tianyan and Kinoshita, Keisuke and Watanabe, Shinji and Delcroix, Marc and Erdogan, Hakan and Hershey, John R and Mesgarani, Nima and others},
  booktitle={Interspeech},
  pages={3036--3040},
  year={2021}
}

@article{evans2024long,
  title={Long-form music generation with latent diffusion},
  author={Evans, Zach and Parker, Julian D and Carr, CJ and Zukowski, Zack and Taylor, Josiah and Pons, Jordi},
  journal={arXiv preprint arXiv:2404.10301},
  year={2024}
}

@article{yuan2025yue,
  title={Yue: Scaling open foundation models for long-form music generation},
  author={Yuan, Ruibin and Lin, Hanfeng and Guo, Shuyue and Zhang, Ge and Pan, Jiahao and Zang, Yongyi and Liu, Haohe and Liang, Yiming and Ma, Wenye and Du, Xingjian and others},
  journal={arXiv preprint arXiv:2503.08638},
  year={2025}
}

@article{dhariwal2020jukebox,
  title={Jukebox: A generative model for music},
  author={Dhariwal, Prafulla and Jun, Heewoo and Payne, Christine and Kim, Jong Wook and Radford, Alec and Sutskever, Ilya},
  journal={arXiv preprint arXiv:2005.00341},
  year={2020}
}

@article{ning2025diffrhythm,
  title={Diffrhythm: Blazingly fast and embarrassingly simple end-to-end full-length song generation with latent diffusion},
  author={Ning, Ziqian and Chen, Huakang and Jiang, Yuepeng and Hao, Chunbo and Ma, Guobin and Wang, Shuai and Yao, Jixun and Xie, Lei},
  journal={arXiv preprint arXiv:2503.01183},
  year={2025}
}

@article{jiang2025diffrhythm,
  title={DiffRhythm 2: Efficient and High Fidelity Song Generation via Block Flow Matching},
  author={Jiang, Yuepeng and Chen, Huakang and Ning, Ziqian and Yao, Jixun and Han, Zerui and Wu, Di and Meng, Meng and Luan, Jian and Fu, Zhonghua and Xie, Lei},
  journal={arXiv preprint arXiv:2510.22950},
  year={2025}
}

@article{zeghidour2021soundstream,
  title={Soundstream: An end-to-end neural audio codec},
  author={Zeghidour, Neil and Luebs, Alejandro and Omran, Ahmed and Skoglund, Jan and Tagliasacchi, Marco},
  journal={IEEE/ACM Transactions on Audio, Speech, and Language Processing},
  volume={30},
  pages={495--507},
  year={2021},
  publisher={IEEE}
}

@inproceedings{mentzer2024finite,
  title={Finite scalar quantization: {VQ}-{VAE} made simple},
  author={Mentzer, Fabian and Minnen, David and Agustsson, Eirikur and Tschannen, Michael},
  booktitle={International Conference on Learning Representations},
  volume={2024},
  pages={51772--51783},
  year={2024}
}

@article{copet2023simple,
  title={Simple and controllable music generation},
  author={Copet, Jade and Kreuk, Felix and Gat, Itai and Remez, Tal and Kant, David and Synnaeve, Gabriel and Adi, Yossi and D{\'e}fossez, Alexandre},
  journal={Advances in neural information processing systems},
  volume={36},
  pages={47704--47720},
  year={2023}
}

@inproceedings{lee2022autoregressive,
  title={Autoregressive image generation using residual quantization},
  author={Lee, Doyup and Kim, Chiheon and Kim, Saehoon and Cho, Minsu and Han, Wook-Shin},
  booktitle={Proceedings of the IEEE/CVF conference on computer vision and pattern recognition},
  pages={11523--11532},
  year={2022}
}

@article{liu2024autoregressive,
  title={Autoregressive diffusion transformer for text-to-speech synthesis},
  author={Liu, Zhijun and Wang, Shuai and Inoue, Sho and Bai, Qibing and Li, Haizhou},
  journal={arXiv preprint arXiv:2406.05551},
  year={2024}
}

@inproceedings{meng2025autoregressive,
  title={Autoregressive speech synthesis without vector quantization},
  author={Meng, Lingwei and Zhou, Long and Liu, Shujie and Chen, Sanyuan and Han, Bing and Hu, Shujie and Liu, Yanqing and Li, Jinyu and Zhao, Sheng and Wu, Xixin and others},
  booktitle={Proceedings of the 63rd Annual Meeting of the Association for Computational Linguistics (Volume 1: Long Papers)},
  pages={1287--1300},
  year={2025}
}

@article{wang2023neural,
  title={Neural codec language models are zero-shot text to speech synthesizers},
  author={Wang, Chengyi and Chen, Sanyuan and Wu, Yu and Zhang, Ziqiang and Zhou, Long and Liu, Shujie and Chen, Zhuo and Liu, Yanqing and Wang, Huaming and Li, Jinyu and others},
  journal={arXiv preprint arXiv:2301.02111},
  year={2023}
}

@article{chen2024vall,
  title={Vall-e 2: Neural codec language models are human parity zero-shot text to speech synthesizers},
  author={Chen, Sanyuan and Liu, Shujie and Zhou, Long and Liu, Yanqing and Tan, Xu and Li, Jinyu and Zhao, Sheng and Qian, Yao and Wei, Furu},
  journal={arXiv preprint arXiv:2406.05370},
  year={2024}
}

@inproceedings{zhang2024speechtokenizer,
  title={Speechtokenizer: Unified speech tokenizer for speech language models},
  author={Zhang, Xin and Zhang, Dong and Li, Shimin and Zhou, Yaqian and Qiu, Xipeng},
  booktitle={International Conference on Learning Representations},
  volume={2024},
  pages={31798--31818},
  year={2024}
}

@article{kumar2023high,
  title={High-fidelity audio compression with improved {RVQGAN}},
  author={Kumar, Rithesh and Seetharaman, Prem and Luebs, Alejandro and Kumar, Ishaan and Kumar, Kundan},
  journal={Advances in Neural Information Processing Systems},
  volume={36},
  pages={27980--27993},
  year={2023}
}

@article{defossez2022high,
  title={High fidelity neural audio compression},
  author={D{\'e}fossez, Alexandre and Copet, Jade and Synnaeve, Gabriel and Adi, Yossi},
  journal={arXiv preprint arXiv:2210.13438},
  year={2022}
}

@article{ju2024naturalspeech,
  title={Naturalspeech 3: Zero-shot speech synthesis with factorized codec and diffusion models},
  author={Ju, Zeqian and Wang, Yuancheng and Shen, Kai and Tan, Xu and Xin, Detai and Yang, Dongchao and Liu, Yanqing and Leng, Yichong and Song, Kaitao and Tang, Siliang and others},
  journal={arXiv preprint arXiv:2403.03100},
  year={2024}
}

@inproceedings{eskimez2024e2,
  title={E2 tts: Embarrassingly easy fully non-autoregressive zero-shot tts},
  author={Eskimez, Sefik Emre and Wang, Xiaofei and Thakker, Manthan and Li, Canrun and Tsai, Chung-Hsien and Xiao, Zhen and Yang, Hemin and Zhu, Zirun and Tang, Min and Tan, Xu and others},
  booktitle={2024 IEEE spoken language technology workshop (SLT)},
  pages={682--689},
  year={2024},
  organization={IEEE}
}

@inproceedings{chen2025f5,
  title={F5-tts: A fairytaler that fakes fluent and faithful speech with flow matching},
  author={Chen, Yushen and Niu, Zhikang and Ma, Ziyang and Deng, Keqi and Wang, Chunhui and JianZhao, JianZhao and Yu, Kai and Chen, Xie},
  booktitle={Proceedings of the 63rd Annual Meeting of the Association for Computational Linguistics (Volume 1: Long Papers)},
  pages={6255--6271},
  year={2025}
}

@article{yue2026matters,
  title={What matters for diffusion-friendly latent manifold? prior-aligned autoencoders for latent diffusion},
  author={Yue, Zhengrong and Hu, Taihang and Chen, Mengting and Zhang, Haiyu and Pan, Zihao and Liu, Tao and Wang, Zikang and Lan, Jinsong and Zhu, Xiaoyong and Zheng, Bo and others},
  journal={arXiv preprint arXiv:2605.07915},
  year={2026}
}

@article{xu2026making,
  title={Making Reconstruction {FID} Predictive of Diffusion Generation {FID}},
  author={Xu, Tongda and He, Mingwei and Abu-Hussein, Shady and Hernandez-Lobato, Jose Miguel and Zheng, Chunhang and Zhao, Kai and Zhou, Chao and Zhang, Ya-Qin and Wang, Yan},
  journal={arXiv preprint arXiv:2603.05630},
  year={2026}
}

@article{daly2022variational,
  title={Variational autoencoders without the variation},
  author={Daly, Gregory A and Fieldsend, Jonathan E and Tabor, Gavin},
  journal={arXiv preprint arXiv:2203.00645},
  year={2022}
}

@article{de2024pullback,
  title={Pullback flow matching on data manifolds},
  author={de Kruiff, Friso and Bekkers, Erik and {\"O}ktem, Ozan and Sch{\"o}nlieb, Carola-Bibiane and Diepeveen, Willem},
  journal={arXiv preprint arXiv:2410.04543},
  year={2024}
}

@article{arvanitidis2017latent,
  title={Latent space oddity: on the curvature of deep generative models},
  author={Arvanitidis, Georgios and Hansen, Lars Kai and Hauberg, S{\o}ren},
  journal={arXiv preprint arXiv:1710.11379},
  year={2017}
}

@inproceedings{wessels2025grounding,
  title={Grounding continuous representations in geometry: Equivariant neural fields},
  author={Wessels, David and Knigge, David and Valperga, Riccardo and Papa, Samuele and Vadgama, Sharvaree and Gavves, Efstratios and Bekkers, Erik},
  booktitle={International Conference on Learning Representations},
  volume={2025},
  pages={59774--59794},
  year={2025}
}

@article{arvanitidis2020geometrically,
  title={Geometrically enriched latent spaces},
  author={Arvanitidis, Georgios and Hauberg, S{\o}ren and Sch{\"o}lkopf, Bernhard},
  journal={arXiv preprint arXiv:2008.00565},
  year={2020}
}

@inproceedings{sun2024geometry,
  title={Geometry-aware autoencoders for metric learning and generative modeling on data manifolds},
  author={Sun, Xingzhi and Liao, Danqi and MacDonald, Kincaid and Zhang, Yanlei and Huguet, Guillaume and Wolf, Guy and Adelstein, Ian and Rudner, Tim GJ and Krishnaswamy, Smita},
  booktitle={ICML 2024 Workshop on Geometry-grounded Representation Learning and Generative Modeling},
  year={2024}
}

@article{altman2018curse,
  title={The curse (s) of dimensionality},
  author={Altman, Naomi and Krzywinski, Martin},
  journal={Nat Methods},
  volume={15},
  number={6},
  pages={399--400},
  year={2018}
}

@inproceedings{li2026back,
  title={Back to basics: Let denoising generative models denoise},
  author={Li, Tianhong and He, Kaiming},
  booktitle={Proceedings of the IEEE/CVF Conference on Computer Vision and Pattern Recognition},
  pages={36115--36125},
  year={2026}
}

@inproceedings{hoogeboom2025simpler,
  title={Simpler Diffusion: 1.5 FID on ImageNet512 with pixel-space diffusion},
  author={Hoogeboom, Emiel and Mensink, Thomas and Heek, Jonathan and Lamerigts, Kay and Gao, Ruiqi and Salimans, Tim},
  booktitle={Proceedings of the Computer Vision and Pattern Recognition Conference},
  pages={18062--18071},
  year={2025}
}

@inproceedings{yu2026pixeldit,
  title={Pixeldit: Pixel diffusion transformers for image generation},
  author={Yu, Yongsheng and Xiong, Wei and Nie, Weili and Sheng, Yichen and Liu, Shiqiu and Luo, Jiebo},
  booktitle={Proceedings of the IEEE/CVF Conference on Computer Vision and Pattern Recognition},
  pages={14273--14282},
  year={2026}
}

@article{xin2026longcat,
  title={Longcat-audiodit: High-fidelity diffusion text-to-speech in the waveform latent space},
  author={Xin, Detai and Hu, Shujie and Yang, Chengzuo and Huang, Chen and Yu, Guoqiao and Wan, Guanglu and Cai, Xunliang},
  journal={arXiv preprint arXiv:2603.29339},
  year={2026}
}

@article{chen2026wavtts,
  title={WavTTS: Towards High-Quality Zero-Shot TTS via Direct Raw Waveform Modeling},
  author={Chen, Wenxi and Jia, Dongya and Chen, Yushen and Niu, Zhikang and Liang, Yuzhe and Li, Xiquan and Yan, Ruiqi and Ma, Ziyang and Yang, Guanrou and Chen, Sanyuan and others},
  journal={arXiv preprint arXiv:2606.03455},
  year={2026}
}

@article{peng2025vibevoice,
  title={Vibevoice technical report},
  author={Peng, Zhiliang and Yu, Jianwei and Wang, Wenhui and Chang, Yaoyao and Sun, Yutao and Dong, Li and Zhu, Yi and Xu, Weijiang and Bao, Hangbo and Wang, Zehua and others},
  journal={arXiv preprint arXiv:2508.19205},
  year={2025}
}

@article{wang2026scaling,
  title={Scaling Speech Tokenizers with Diffusion Autoencoders},
  author={Wang, Yuancheng and Tang, Zhenyu and Wang, Yun and Hinsvark, Arthur and Liu, Yingru and Li, Yinghao and Peng, Kainan and Ao, Junyi and Ma, Mingbo and Seltzer, Mike and others},
  journal={arXiv preprint arXiv:2602.06602},
  year={2026}
}

@article{guo2024fireredtts,
  title={Fireredtts: A foundation text-to-speech framework for industry-level generative speech applications},
  author={Guo, Hao-Han and Hu, Yao and Liu, Kun and Shen, Fei-Yu and Tang, Xu and Wu, Yi-Chen and Xie, Feng-Long and Xie, Kun and Xu, Kai-Tuo},
  journal={arXiv preprint arXiv:2409.03283},
  year={2024}
}

@inproceedings{parker2025scaling,
  title={Scaling transformers for low-bitrate high-quality speech coding},
  author={Parker, Julian and Smirnov, Anton and Pons, Jordi and Carr, CJ and Zukowski, Zack and Evans, Zach and Liu, Xubo},
  booktitle={International Conference on Learning Representations},
  volume={2025},
  pages={51997--52021},
  year={2025}
}

@article{defossez2024moshi,
  title={Moshi: a speech-text foundation model for real-time dialogue},
  author={D{\'e}fossez, Alexandre and Mazar{\'e}, Laurent and Orsini, Manu and Royer, Am{\'e}lie and P{\'e}rez, Patrick and J{\'e}gou, Herv{\'e} and Grave, Edouard and Zeghidour, Neil},
  journal={arXiv preprint arXiv:2410.00037},
  year={2024}
}

@inproceedings{ji2025wavtokenizer,
  title={Wavtokenizer: an efficient acoustic discrete codec tokenizer for audio language modeling},
  author={Ji, Shengpeng and Jiang, Ziyue and Wang, Wen and Chen, Yifu and Fang, Minghui and Zuo, Jialong and Yang, Qian and Cheng, Xize and Li, Ruiqi and Zhang, Ziang and others},
  booktitle={International Conference on Learning Representations},
  volume={2025},
  pages={93809--93826},
  year={2025}
}

@article{luo2021group,
  title={Group communication with context codec for lightweight source separation},
  author={Luo, Yi and Han, Cong and Mesgarani, Nima},
  journal={IEEE/ACM Transactions on Audio, Speech, and Language Processing},
  volume={29},
  pages={1752--1761},
  year={2021},
  publisher={IEEE}
}

@inproceedings{chen2026sac,
  title={Sac: Neural speech codec with semantic-acoustic dual-stream quantization},
  author={Chen, Wenxi and Yan, Ruiqi and Chen, Yushen and Niu, Zhikang and Ma, Ziyang and Li, Xiquan and Liang, Yuzhe and Yin, Shunshun and Tao, Ming and Wang, Xinsheng and others},
  booktitle={Proceedings of the 64th Annual Meeting of the Association for Computational Linguistics (Volume 1: Long Papers)},
  pages={3030--3048},
  year={2026}
}

@article{liu2024semanticodec,
  title={Semanticodec: An ultra low bitrate semantic audio codec for general sound},
  author={Liu, Haohe and Xu, Xuenan and Yuan, Yi and Wu, Mengyue and Wang, Wenwu and Plumbley, Mark D},
  journal={IEEE Journal of Selected Topics in Signal Processing},
  volume={18},
  number={8},
  pages={1448--1461},
  year={2024},
  publisher={IEEE}
}

@article{mousavi2024should,
  title={How should we extract discrete audio tokens from self-supervised models?},
  author={Mousavi, Pooneh and Duret, Jarod and Zaiem, Salah and Della Libera, Luca and Ploujnikov, Artem and Subakan, Cem and Ravanelli, Mirco},
  journal={arXiv preprint arXiv:2406.10735},
  year={2024}
}

@article{borsos2023audiolm,
  title={Audiolm: a language modeling approach to audio generation},
  author={Borsos, Zal{\'a}n and Marinier, Rapha{\"e}l and Vincent, Damien and Kharitonov, Eugene and Pietquin, Olivier and Sharifi, Matt and Roblek, Dominik and Teboul, Olivier and Grangier, David and Tagliasacchi, Marco and others},
  journal={IEEE/ACM transactions on audio, speech, and language processing},
  volume={31},
  pages={2523--2533},
  year={2023},
  publisher={IEEE}
}

@inproceedings{ye2025codec,
  title={Codec does matter: Exploring the semantic shortcoming of codec for audio language model},
  author={Ye, Zhen and Sun, Peiwen and Lei, Jiahe and Lin, Hongzhan and Tan, Xu and Dai, Zheqi and Kong, Qiuqiang and Chen, Jianyi and Pan, Jiahao and Liu, Qifeng and others},
  booktitle={Proceedings of the AAAI Conference on Artificial Intelligence},
  volume={39},
  number={24},
  pages={25697--25705},
  year={2025}
}

@article{zhou2026voxcpm2,
  title={VoxCPM2 technical report},
  author={Zhou, Yixuan and Zeng, Guoyang and Liu, Xin and Li, Xiang and Yu, Renjie and Gui, Jiancheng and Wu, Jiaheng and Wang, Ziyang and Shen, Xudong and Ye, Runchuan and others},
  journal={arXiv preprint arXiv:2606.06928},
  year={2026}
}

@article{bai2024seed,
  title={Seed-music: A unified framework for high quality and controlled music generation},
  author={Bai, Ye and Chen, Haonan and Chen, Jitong and Chen, Zhuo and Deng, Yi and Dong, Xiaohong and Hantrakul, Lamtharn and Hao, Weituo and Huang, Qingqing and Huang, Zhongyi and others},
  journal={arXiv preprint arXiv:2409.09214},
  year={2024}
}

@article{hsu2021hubert,
  title={Hubert: Self-supervised speech representation learning by masked prediction of hidden units},
  author={Hsu, Wei-Ning and Bolte, Benjamin and Tsai, Yao-Hung Hubert and Lakhotia, Kushal and Salakhutdinov, Ruslan and Mohamed, Abdelrahman},
  journal={IEEE/ACM transactions on audio, speech, and language processing},
  volume={29},
  pages={3451--3460},
  year={2021},
  publisher={IEEE}
}

@inproceedings{ardila2020common,
  title={Common voice: A massively-multilingual speech corpus},
  author={Ardila, Rosana and Branson, Megan and Davis, Kelly and Kohler, Michael and Meyer, Josh and Henretty, Michael and Morais, Reuben and Saunders, Lindsay and Tyers, Francis and Weber, Gregor},
  booktitle={Proceedings of the twelfth language resources and evaluation conference},
  pages={4218--4222},
  year={2020}
}

@inproceedings{guo2021didispeech,
  title={Didispeech: A large scale mandarin speech corpus},
  author={Guo, Tingwei and Wen, Cheng and Jiang, Dongwei and Luo, Ne and Zhang, Ruixiong and Zhao, Shuaijiang and Li, Wubo and Gong, Cheng and Zou, Wei and Han, Kun and others},
  booktitle={ICASSP 2021-2021 IEEE International Conference on Acoustics, Speech and Signal Processing (ICASSP)},
  pages={6968--6972},
  year={2021},
  organization={IEEE}
}

@article{baevski2020wav2vec,
  title={wav2vec 2.0: A framework for self-supervised learning of speech representations},
  author={Baevski, Alexei and Zhou, Yuhao and Mohamed, Abdelrahman and Auli, Michael},
  journal={Advances in neural information processing systems},
  volume={33},
  pages={12449--12460},
  year={2020}
}

@article{chen2022wavlm,
  title={Wavlm: Large-scale self-supervised pre-training for full stack speech processing},
  author={Chen, Sanyuan and Wang, Chengyi and Chen, Zhengyang and Wu, Yu and Liu, Shujie and Chen, Zhuo and Li, Jinyu and Kanda, Naoyuki and Yoshioka, Takuya and Xiao, Xiong and others},
  journal={IEEE Journal of Selected Topics in Signal Processing},
  volume={16},
  number={6},
  pages={1505--1518},
  year={2022},
  publisher={IEEE}
}

@inproceedings{chiu2022self,
  title={Self-supervised learning with random-projection quantizer for speech recognition},
  author={Chiu, Chung-Cheng and Qin, James and Zhang, Yu and Yu, Jiahui and Wu, Yonghui},
  booktitle={International Conference on Machine Learning},
  pages={3915--3924},
  year={2022},
  organization={PMLR}
}

@article{chen2025minmo,
  title={Minmo: A multimodal large language model for seamless voice interaction},
  author={Chen, Qian and Chen, Yafeng and Chen, Yanni and Chen, Mengzhe and Chen, Yingda and Deng, Chong and Du, Zhihao and Gao, Ruize and Gao, Changfeng and Gao, Zhifu and others},
  journal={arXiv preprint arXiv:2501.06282},
  year={2025}
}

@inproceedings{radford2023robust,
  title={Robust speech recognition via large-scale weak supervision},
  author={Radford, Alec and Kim, Jong Wook and Xu, Tao and Brockman, Greg and McLeavey, Christine and Sutskever, Ilya},
  booktitle={International conference on machine learning},
  pages={28492--28518},
  year={2023},
  organization={PMLR}
}

@article{wang2025spark,
  title={Spark-tts: An efficient llm-based text-to-speech model with single-stream decoupled speech tokens},
  author={Wang, Xinsheng and Jiang, Mingqi and Ma, Ziyang and Zhang, Ziyu and Liu, Songxiang and Li, Linqin and Liang, Zheng and Zheng, Qixi and Wang, Rui and Feng, Xiaoqin and others},
  journal={arXiv preprint arXiv:2503.01710},
  year={2025}
}

@article{gorban2018blessing,
  title={Blessing of dimensionality: mathematical foundations of the statistical physics of data},
  author={Gorban, Alexander N and Tyukin, Ivan Yu},
  journal={Philosophical Transactions of the Royal Society A: Mathematical, Physical and Engineering Sciences},
  volume={376},
  number={2118},
  pages={20170237},
  year={2018},
  publisher={The Royal Society Publishing}
}

@incollection{chapelle2006discussion,
  title={A discussion of semi-supervised learning and transduction},
  author={Chapelle, Olivier and Sch{\"o}lkopf, Bernhard and Zien, Alexander},
  booktitle={Semi-supervised learning},
  pages={473--478},
  year={2006},
  publisher={MIT Press}
}

@article{campbell1968application,
  title={Application of Fourier analysis to the visibility of gratings},
  author={Campbell, Fergus W and Robson, John G},
  journal={The Journal of physiology},
  volume={197},
  number={3},
  pages={551},
  year={1968}
}

@book{barten1999contrast,
  title={Contrast sensitivity of the human eye and its effects on image quality},
  author={Barten, Peter GJ},
  year={1999},
  publisher={SPIE press}
}

@article{watson1997model,
  title={Model of visual contrast gain control and pattern masking},
  author={Watson, Andrew B and Solomon, Joshua A},
  journal={Journal of the optical society of America A},
  volume={14},
  number={9},
  pages={2379--2391},
  year={1997},
  publisher={Optical Society of America}
}

@book{zwicker2013psychoacoustics,
  title={Psychoacoustics: Facts and models},
  author={Zwicker, Eberhard and Fastl, Hugo},
  volume={22},
  year={2013},
  publisher={Springer Science \& Business Media}
}

@article{miller1947sensitivity,
  title={Sensitivity to changes in the intensity of white noise and its relation to masking and loudness},
  author={Miller, George A},
  journal={The Journal of the Acoustical Society of America},
  volume={19},
  number={4},
  pages={609--619},
  year={1947},
  publisher={Acoustical Society of America}
}

@article{chen2023importance,
  title={On the importance of noise scheduling for diffusion models},
  author={Chen, Ting},
  journal={arXiv preprint arXiv:2301.10972},
  year={2023}
}

@article{zhou2026wavflow,
  title={WavFlow: Audio Generation in Waveform Space},
  author={Zhou, Feiyan and Wang, Luyuan and Chen, Shoufa and Wang, Zhe and Liu, Zhiheng and Cong, Yuren and Zhang, Xiaohui and Yang, Fanny and Zeng, Belinda},
  journal={arXiv preprint arXiv:2605.18749},
  year={2026}
}

@article{fan2026barewave,
  title={BareWave: Waveform-Native Flow-Matching Text-to-Speech},
  author={Fan, Wei and Tan, Chao-Hong and Chen, Qian and Wang, Wen and Li, Xiangang and Chen, Kejiang and Zhang, Weiming and Yu, Nenghai},
  journal={arXiv preprint arXiv:2606.09048},
  year={2026}
}

@article{zhou2025voxcpm,
  title={Voxcpm: Tokenizer-free TTS for context-aware speech generation and true-to-life voice cloning},
  author={Zhou, Yixuan and Zeng, Guoyang and Liu, Xin and Li, Xiang and Yu, Renjie and Wang, Ziyang and Ye, Runchuan and Sun, Weiyue and Gui, Jiancheng and Li, Kehan and others},
  journal={arXiv preprint arXiv:2509.24650},
  year={2025}
}

@article{li2024autoregressive,
  title={Autoregressive image generation without vector quantization},
  author={Li, Tianhong and Tian, Yonglong and Li, He and Deng, Mingyang and He, Kaiming},
  journal={Advances in Neural Information Processing Systems},
  volume={37},
  pages={56424--56445},
  year={2024}
}

@article{yu2024representation,
  title={Representation alignment for generation: Training diffusion transformers is easier than you think},
  author={Yu, Sihyun and Kwak, Sangkyung and Jang, Huiwon and Jeong, Jongheon and Huang, Jonathan and Shin, Jinwoo and Xie, Saining},
  journal={arXiv preprint arXiv:2410.06940},
  year={2024}
}

@article{wang2025pixnerd,
  title={Pixnerd: Pixel neural field diffusion},
  author={Wang, Shuai and Gao, Ziteng and Zhu, Chenhui and Huang, Weilin and Wang, Limin},
  journal={arXiv preprint arXiv:2507.23268},
  year={2025}
}

@article{chu2023qwen,
  title={Qwen-audio: Advancing universal audio understanding via unified large-scale audio-language models},
  author={Chu, Yunfei and Xu, Jin and Zhou, Xiaohuan and Yang, Qian and Zhang, Shiliang and Yan, Zhijie and Zhou, Chang and Zhou, Jingren},
  journal={arXiv preprint arXiv:2311.07919},
  year={2023}
}

@article{ding2025kimi,
  title={Kimi-audio technical report},
  author={Ding, Ding and Ju, Zeqian and Leng, Yichong and Liu, Songxiang and Liu, Tong and Shang, Zeyu and Shen, Kai and Song, Wei and Tan, Xu and Tang, Heyi and others},
  journal={arXiv preprint arXiv:2504.18425},
  year={2025}
}

@article{zhang2025mimo,
  title={MiMo-Audio: Audio Language Models are Few-Shot Learners},
  author={Zhang, Dong and Wang, Gang and Xue, Jinlong and Fang, Kai and Zhao, Liang and Ma, Rui and Ren, Shuhuai and Liu, Shuo and Guo, Tao and Zhuang, Weiji and others},
  journal={arXiv preprint arXiv:2512.23808},
  year={2025}
}

@article{wu2025step,
  title={Step-audio 2 technical report},
  author={Wu, Boyong and Yan, Chao and Hu, Chen and Yi, Cheng and Feng, Chengli and Tian, Fei and Shen, Feiyu and Yu, Gang and Zhang, Haoyang and Li, Jingbei and others},
  journal={arXiv preprint arXiv:2507.16632},
  year={2025}
}

@article{xu2025qwen3,
  title={Qwen3-omni technical report},
  author={Xu, Jin and Guo, Zhifang and Hu, Hangrui and Chu, Yunfei and Wang, Xiong and He, Jinzheng and Wang, Yuxuan and Shi, Xian and He, Ting and Zhu, Xinfa and others},
  journal={arXiv preprint arXiv:2509.17765},
  year={2025}
}

@article{ghosh2026audio,
  title={Audio flamingo 3: Advancing audio intelligence with fully open large audio language models},
  author={Ghosh, Sreyan and Goel, Arushi and Kim, Jaehyeon and Kumar, Sonal and Kong, Zhifeng and Lee, Sang-gil and Yang, Chao-Han and Duraiswami, Ramani and Manocha, Dinesh and Valle, Rafael and others},
  journal={Advances in Neural Information Processing Systems},
  volume={38},
  pages={41819--41886},
  year={2026}
}

@article{yang2026moss,
  title={MOSS-Audio Technical Report},
  author={Yang, Chen and Yu, Chufan and Chen, Hanfu and Zhu, Jie and Chen, Jingqi and Chen, Ke and Wang, Wenxuan and Wang, Yang and Jiang, Yaozhou and Jiang, Yi and others},
  journal={arXiv preprint arXiv:2606.01802},
  year={2026}
}

@article{wu2023next,
  title={Next-gpt: Any-to-any multimodal llm},
  author={Wu, Shengqiong and Fei, Hao and Qu, Leigang and Ji, Wei and Chua, Tat-Seng},
  journal={arXiv preprint arXiv:2309.05519},
  year={2023}
}

@misc{gemmateam2026gemma4technicalreport,
      title={Gemma 4 Technical Report}, 
      author={Gemma Team},
      year={2026},
      eprint={2607.02770},
      archivePrefix={arXiv},
      primaryClass={cs.CL},
      url={https://arxiv.org/abs/2607.02770}, 
}

@article{hurst2024gpt,
  title={Gpt-4o system card},
  author={Hurst, Aaron and Lerer, Adam and Goucher, Adam P and Perelman, Adam and Ramesh, Aditya and Clark, Aidan and Ostrow, AJ and Welihinda, Akila and Hayes, Alan and Radford, Alec and others},
  journal={arXiv preprint arXiv:2410.21276},
  year={2024}
}

@article{lee2026geometry,
  title={Geometry-aware image flow matching},
  author={Lee, Junho and Kim, Kwanseok and Lee, Joonseok},
  journal={arXiv preprint arXiv:2605.25294},
  year={2026}
}

@article{meral2026aligning,
  title={Aligning Latent Geometry for Spherical Flow Matching in Image Generation},
  author={Meral, Tuna Han Salih and Oktay, Kaan and Yesiltepe, Hidir and Akan, Adil Kaan and Yanardag, Pinar},
  journal={arXiv preprint arXiv:2605.15193},
  year={2026}
}

@article{song2020score,
  title={Score-based generative modeling through stochastic differential equations},
  author={Song, Yang and Sohl-Dickstein, Jascha and Kingma, Diederik P and Kumar, Abhishek and Ermon, Stefano and Poole, Ben},
  journal={arXiv preprint arXiv:2011.13456},
  year={2020}
}

@article{albergo2025stochastic,
  title={Stochastic interpolants: A unifying framework for flows and diffusions},
  author={Albergo, Michael and Boffi, Nicholas M and Vanden-Eijnden, Eric},
  journal={Journal of Machine Learning Research},
  volume={26},
  number={209},
  pages={1--80},
  year={2025}
}

@article{albergo2022building,
  title={Building normalizing flows with stochastic interpolants},
  author={Albergo, Michael S and Vanden-Eijnden, Eric},
  journal={arXiv preprint arXiv:2209.15571},
  year={2022}
}

@article{luo2024gull,
  title={Gull: A generative multifunctional audio codec},
  author={Luo, Yi and Yu, Jianwei and Chen, Hangting and Gu, Rongzhi and Weng, Chao},
  journal={arXiv preprint arXiv:2404.04947},
  year={2024}
}

@article{atal1970adaptive,
  title={Adaptive predictive coding of speech signals},
  author={Atal, Bishnu S and Schroeder, Manfred R},
  journal={Bell System Technical Journal},
  volume={49},
  number={8},
  pages={1973--1986},
  year={1970},
  publisher={Wiley Online Library}
}

@inproceedings{schroeder1985code,
  title={Code-excited linear prediction ({CELP}): High-quality speech at very low bit rates},
  author={Schroeder, Manfred and Atal, B},
  booktitle={ICASSP'85. IEEE International Conference on Acoustics, Speech, and Signal Processing},
  volume={10},
  pages={937--940},
  year={1985},
  organization={IEEE}
}

@article{bessette2002adaptive,
  title={The adaptive multirate wideband speech codec ({AMR}-{WB})},
  author={Bessette, Bruno and Salami, Redwan and Lefebvre, Roch and Jelinek, Milan and Rotola-Pukkila, Jani and Vainio, Janne and Mikkola, Hannu and Jarvinen, Kari},
  journal={IEEE transactions on speech and audio processing},
  volume={10},
  number={8},
  pages={620--636},
  year={2002},
  publisher={IEEE}
}

@inproceedings{mao2017least,
  title={Least squares generative adversarial networks},
  author={Mao, Xudong and Li, Qing and Xie, Haoran and Lau, Raymond YK and Wang, Zhen and Paul Smolley, Stephen},
  booktitle={Proceedings of the IEEE international conference on computer vision},
  pages={2794--2802},
  year={2017}
}

@inproceedings{kubichek1993mel,
  title={Mel-cepstral distance measure for objective speech quality assessment},
  author={Kubichek, Robert},
  booktitle={Proceedings of IEEE pacific rim conference on communications computers and signal processing},
  volume={1},
  pages={125--128},
  year={1993},
  organization={IEEE}
}

@article{taal2011algorithm,
  title={An algorithm for intelligibility prediction of time--frequency weighted noisy speech},
  author={Taal, Cees H and Hendriks, Richard C and Heusdens, Richard and Jensen, Jesper},
  journal={IEEE Transactions on audio, speech, and language processing},
  volume={19},
  number={7},
  pages={2125--2136},
  year={2011},
  publisher={IEEE}
}

@inproceedings{chinen2020visqol,
  title={ViSQOL v3: An open source production ready objective speech and audio metric},
  author={Chinen, Michael and Lim, Felicia SC and Skoglund, Jan and Gureev, Nikita and O'Gorman, Feargus and Hines, Andrew},
  booktitle={2020 twelfth international conference on quality of multimedia experience (QoMEX)},
  pages={1--6},
  year={2020},
  organization={IEEE}
}

@article{shazeer2020glu,
  title={Glu variants improve transformer},
  author={Shazeer, Noam},
  journal={arXiv preprint arXiv:2002.05202},
  year={2020}
}

@article{su2024roformer,
  title={Roformer: Enhanced transformer with rotary position embedding},
  author={Su, Jianlin and Ahmed, Murtadha and Lu, Yu and Pan, Shengfeng and Bo, Wen and Liu, Yunfeng},
  journal={Neurocomputing},
  volume={568},
  pages={127063},
  year={2024},
  publisher={Elsevier}
}

@article{liao2026fish,
  title={Fish audio s2 technical report},
  author={Liao, Shijia and Wang, Yuxuan and Liu, Songting and Cheng, Yifan and Zhang, Ruoyi and Li, Tianyu and Li, Shidong and Zheng, Yisheng and Liu, Xingwei and Wang, Qingzheng and others},
  journal={arXiv preprint arXiv:2603.08823},
  year={2026}
}

@article{du2024cosyvoice,
  title={Cosyvoice: A scalable multilingual zero-shot text-to-speech synthesizer based on supervised semantic tokens},
  author={Du, Zhihao and Chen, Qian and Zhang, Shiliang and Hu, Kai and Lu, Heng and Yang, Yexin and Hu, Hangrui and Zheng, Siqi and Gu, Yue and Ma, Ziyang and others},
  journal={arXiv preprint arXiv:2407.05407},
  year={2024}
}

@article{du2025cosyvoice,
  title={Cosyvoice 3: Towards in-the-wild speech generation via scaling-up and post-training},
  author={Du, Zhihao and Gao, Changfeng and Wang, Yuxuan and Yu, Fan and Zhao, Tianyu and Wang, Hao and Lv, Xiang and Wang, Hui and Ni, Chongjia and Shi, Xian and others},
  journal={arXiv preprint arXiv:2505.17589},
  year={2025}
}

@article{xie2025fireredtts,
  title={Fireredtts-2: Towards long conversational speech generation for podcast and chatbot},
  author={Xie, Kun and Shen, Feiyu and Li, Junjie and Xie, Fenglong and Tang, Xu and Hu, Yao},
  journal={arXiv preprint arXiv:2509.02020},
  year={2025}
}

@article{lian2026dots,
  title={dots. tts Technical Report},
  author={Lian, Shi and Li, Changtao and Li, Bohan and Wang, Hankun and Zheng, Da and Tian, Junfeng and Ma, Yufeng and Zhang, Colin and Yu, Kai},
  journal={arXiv preprint arXiv:2606.07080},
  year={2026}
}

\clearpage

\end{document}